\documentclass[aps,reprint,superscriptaddress,letterpaper]{revtex4-2}

\usepackage{graphicx}
\usepackage{comment}
\usepackage{verbatim}
\usepackage{amsmath}
\usepackage{mathtools}
\usepackage{xfrac}
\usepackage{bm}
\usepackage{xcolor}
\usepackage{color}
\usepackage[hidelinks]{hyperref}
\usepackage[utf8]{inputenc}
\usepackage{siunitx}

\usepackage[inline]{enumitem}
\providecommand{\tabularnewline}{\\}
\usepackage{array}
\usepackage{color}
\usepackage[T1]{fontenc} 

\def \l {\left}
\def \r {\right}
\def\ket#1{\left| #1\right>}
\def\bra#1{\left< #1\right|}
\def\sp{\mathop{\rm Span}}

\newcommand{\ketbrad}[1]{|#1\rangle\!\langle #1|}
\newcommand{\ketbradt}[2]{|#1\rangle\!\langle #2|}

\definecolor{red}{rgb}{1,0,0}
\definecolor{green}{rgb}{0,1,0}
\definecolor{blue}{rgb}{0,0,1}
\definecolor{c1}{rgb}{0.178,0.63,0.17}
\definecolor{c2}{rgb}{0.09,0.740,0.81}
\definecolor{c3}{rgb}{0.12,0.46,0.7}
\definecolor{c4}{rgb}{0.58,0.40,0.74}

\newcommand{\nocontentsline}[3]{}
\newcommand{\tocless}[2]{\bgroup\let\addcontentsline=\nocontentsline#1{#2}\egroup}

\makeatletter
\let\saved@includegraphics\includegraphics
\AtBeginDocument{\let\includegraphics\saved@includegraphics}
\renewenvironment*{figure}{\@float{figure}}{\end@float}
\makeatother

\begin{document}

\title{Suppressing quantum errors by scaling a surface code logical qubit}
\author{Google Quantum AI}
\date{\today}

\email[Corresponding author (H.~Neven): ]{neven@google.com}

\begin{abstract}

Practical quantum computing will require error rates that are well below what is achievable with physical qubits.  
Quantum error correction \cite{shor1995scheme, gottesman1997stabilizer} offers a path to algorithmically-relevant error rates by encoding logical qubits within many physical qubits, where increasing the number of physical qubits enhances protection against physical errors.
However, introducing more qubits also increases the number of error sources, so the density of errors must be sufficiently low in order for logical performance to improve with increasing code size.
Here, we report the measurement of logical qubit performance scaling across multiple code sizes, and demonstrate that our system of superconducting qubits has sufficient performance to overcome the additional errors from increasing qubit number.
We find our distance-5 surface code logical qubit modestly outperforms an ensemble of distance-3 logical qubits on average, both in terms of logical error probability over 25 cycles and logical error per cycle (2.914\%±0.016\% compared to 3.028\%±0.023\%).
To investigate damaging, low-probability error sources, we run a distance-25 repetition code and observe a $1.7\times10^{-6}$ logical error per round floor set by a single high-energy event ($1.6\times10^{-7}$ when excluding this event).
We are able to accurately model our experiment, and from this model we can extract error budgets that highlight the biggest challenges for future systems. 
These results mark the first experimental demonstration where quantum error correction begins to improve performance with increasing qubit number, illuminating the path to reaching the logical error rates required for computation.
 
\end{abstract}

\maketitle
\vspace{6ex}
\tocless\section{Introduction}

Since Feynman's proposal to compute using quantum mechanics \cite{feynman1982simulating}, many potential applications have emerged, including factoring \cite{shor1999polynomial}, optimization \cite{farhi2001quantum}, machine learning \cite{biamonte2017quantum}, quantum simulation \cite{lloyd1996universal}, and quantum chemistry \cite{aspuru2005simulated}. These applications often require billions of quantum operations \cite{reiher2017elucidating, gidney2019factor, kivlichan2020improved} while state-of-the-art quantum processors typically have error rates around $10^{-3}$ per gate \cite{ballance2016high, huang2019fidelity, rol2019fast, jurcevic2020demonstration, foxen2020demonstrating, wu2021strong}, far too high to execute such large circuits. Fortunately, quantum error correction can exponentially suppress the operational error rates in a quantum processor, at the expense of physical qubit overhead \cite{knill1998resilient, aharonov2008fault}.

Several works have reported quantum error correction on small codes able to correct a single error \cite{ryan2021realization, egan2021fault, krinner2022realizing, sundaresan2022matching, zhao2021realizing, abobeih2022fault, satzinger2021realizing} and in continuous variable codes \cite{ofek2016extending, fluhmann2019encoding, campagne2020quantum, grimm2020stabilization}. However, a crucial question remains: will scaling up the error-correcting code size reduce logical error rates in a real device? In theory, logical errors should be reduced if physical errors are sufficiently sparse in the quantum processor. In practice, demonstrating reduced logical error requires scaling up a device to support a code which can correct at least two errors, without sacrificing state-of-the-art performance. In this work, we report a 72-qubit superconducting device supporting a 49-qubit distance-5 ($d=5$) surface code that narrowly outperforms its average subset 17-qubit distance-3 surface code, demonstrating a critical step towards scalable quantum error correction. 

\vspace{6ex}
\tocless\section{Surface codes with superconducting qubits}

\begin{figure}[t]
    \centering
    \includegraphics[width=89mm]{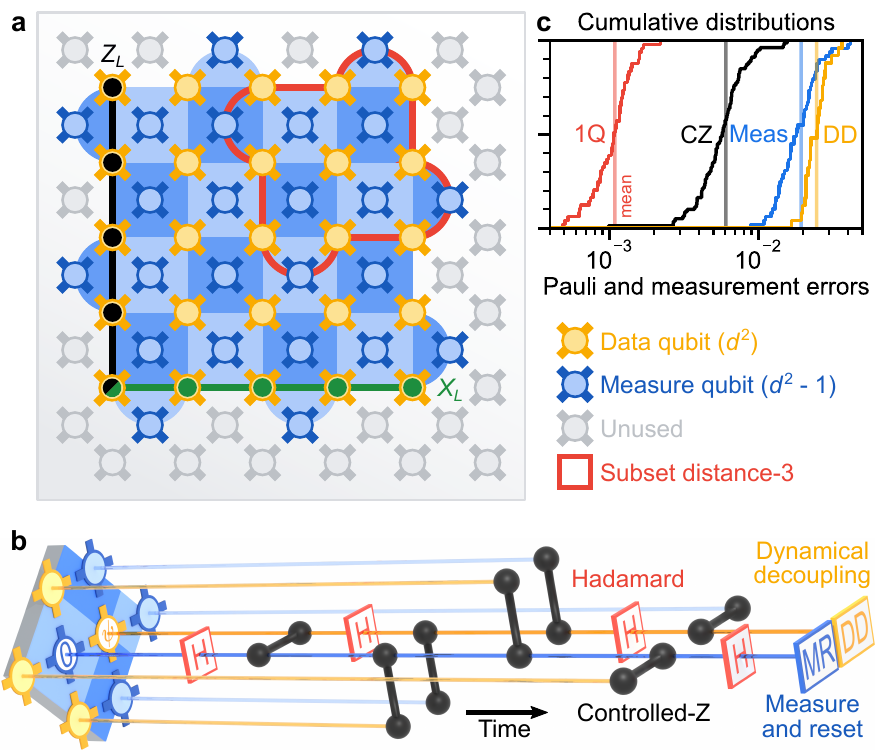}
    \caption{
    \textbf{Implementing surface code logical qubits.} 
    \textbf{a,} Schematic of a 72-qubit Sycamore device with a distance-5 surface code embedded, consisting of 25 data qubits (gold) and 24 measure qubits (blue). Each measure qubit is associated with a stabiliser (blue colored tile, dark: $X$, light: $Z$). Representative logical operators $Z_L$ (black) and $X_L$ (green) traverse the array, intersecting at the lower-left data qubit. The upper-right quadrant (red outline) is one of four subset distance-3 codes (the four quadrants) we compare to distance-5.
    \textbf{b,} Illustration of a stabiliser measurement, focusing on one data qubit (gold) and one measure qubit (blue), in perspective view with time progressing to the right. Each qubit participates in four controlled-Z (CZ) gates with its four nearest neighbours, interspersed with Hadamard gates (H), and finally, the measure qubit is measured and reset to $|0\rangle$. Data qubits perform dynamical decoupling (DD) while waiting for the measurement and reset. All stabilisers are measured in this manner concurrently. 
    Cycle duration is 921~ns, including 500~ns measurement and 160~ns reset.
    \textbf{c,} Cumulative distributions of errors for single-qubit gates, CZ gates, measurement, and data qubit DD (idle during measurement and reset). Benchmarked in simultaneous operation using random circuit techniques, on the 49 qubits used in distance-5 and the four CZ layers from the stabiliser circuit \cite{emerson2005scalable, arute2019quantum}. Vertical lines are means.
    }
\end{figure}

Surface codes \cite{kitaev2003, dennis2002topological, raussendorf2007fault, fowler2012surface, satzinger2021realizing} are a family of quantum error-correcting codes which encode a logical qubit into the joint entangled state of a $d\times d$ square of physical qubits, referred to as \emph{data qubit}s. The logical qubit states are defined by a pair of anti-commuting logical observables $X_L$ and $Z_L$. For the example shown in Fig.~1a, a $Z_L$ observable is encoded in the joint $Z$-basis parity of a line of qubits which traverse the lattice from top-to-bottom, and likewise an $X_L$ observable is encoded in the joint $X$-basis parity traversing left-to-right. 
This non-local encoding of information protects the logical qubit from local physical errors, provided we can detect and correct them.

To detect errors, we periodically measure $X$ and $Z$ parities of adjacent clusters of data qubits with the aid of $d^2 - 1$ \emph{measure qubits} interspersed throughout the lattice. As shown in Fig.~1b, each measure qubit interacts with its neighbouring data qubits in order to map the joint data qubit parity onto the measure qubit state, which is then measured. These parity measurements, or \emph{stabilisers}, are laid out so that each one commutes with the logical observables of the encoded qubit as well as every other stabiliser. Consequently, we can detect errors when parity measurements change unexpectedly, without disturbing the logical qubit state.

A \emph{decoder} uses the history of stabiliser measurement outcomes to infer likely configurations of physical errors on the device. We can then determine the overall effect of these inferred errors on the logical qubit, thus preserving the logical state.  Most surface code logical gates can be implemented by maintaining logical memory and executing different sequences of measurements on the code boundary \cite{horsman2012, fowler2019}. Thus, we focus on preserving logical memory, the core technical challenge in operating the surface code.

We implement the surface code on an expanded Sycamore device \cite{arute2019quantum} with 72 transmon qubits \cite{koch2007charge} and 121 tunable couplers \cite{neill2017path, yan2018tunable}. Each qubit is coupled to four nearest neighbours except on the boundaries, with mean qubit coherence times $T_1$ = 20~$\mu$s and $T_{2,\textrm{CPMG}}$ = 30~$\mu$s. As in Ref.\,\cite{chen2021exponential}, we implement single-qubit rotations, controlled-Z (CZ) gates, reset, and measurement, demonstrating similar or improved simultaneous performance as shown in Fig.~1c.

The distance-5 surface code logical qubit is encoded on a 49-qubit subset of the device, with 25 data qubits and 24 measure qubits. Each measure qubit corresponds to one stabiliser, classified by its basis ($X$ or $Z$) and the number of data qubits involved (weight, 2 or 4). Ideally, to assess how logical performance scales with code size, we would compare distance-5 and distance-3 logical qubits under identical noise.  Although device inhomogeneity makes this comparison difficult, we can compare the distance-5 logical qubit to the average of four distance-3 logical qubit subgrids, each containing 9 data qubits and 8 measure qubits. These distance-3 logical qubits cover the four quadrants of the distance-5 code with minimal qubit overlap, capturing the average performance of the full distance-5 grid.

In a single instance of the experiment, we initialise the logical qubit state, run several cycles of error correction, then measure the final logical state. We show an example in Fig.~2a. 
To prepare a $Z_L$ eigenstate, we first prepare the data qubits in $|0\rangle$’s and $|1\rangle$’s, an eigenstate of the $Z$ stabilisers. The first cycle of stabiliser measurements then projects the data qubits into an entangled state that is also an eigenstate of the $X$ stabilisers. 
Each cycle contains CZ and Hadamard gates sequenced to extract $X$ and $Z$ stabilisers simultaneously, and ends with the measurement and reset of the measure qubits. In the final cycle, we also measure the data qubits in the $Z$ basis, yielding both parity information and a measurement of the logical state. 
Preparing and measuring $X_L$ eigenstates proceeds analogously.
The instance succeeds if the corrected logical measurement agrees with the known initial state; otherwise, a logical error has occurred.

Our stabiliser circuits contain a few modifications \cite{supplement} to the standard gate sequence described above, including phase corrections to correct for unintended qubit frequency shifts \cite{kelly2016scalable} and dynamical decoupling gates during qubit idles. We also remove certain Hadamard gates to implement the \emph{ZXXZ} variant of the surface code \cite{wen2003quantum,bonilla2021xzzx}, helping to symmetrise the $X$- and $Z$-basis logical error rates. Finally, during initialization, the data qubits are prepared into randomly selected bitstrings. This ensures that we do not preferentially measure even parities in the first few rounds of the code, which could artificially lower logical error rates due to bias in measurement error.

\vspace{6ex}
\tocless\section{Error detectors}

\begin{figure*} [t]
    \centering
    \includegraphics[width=183mm]{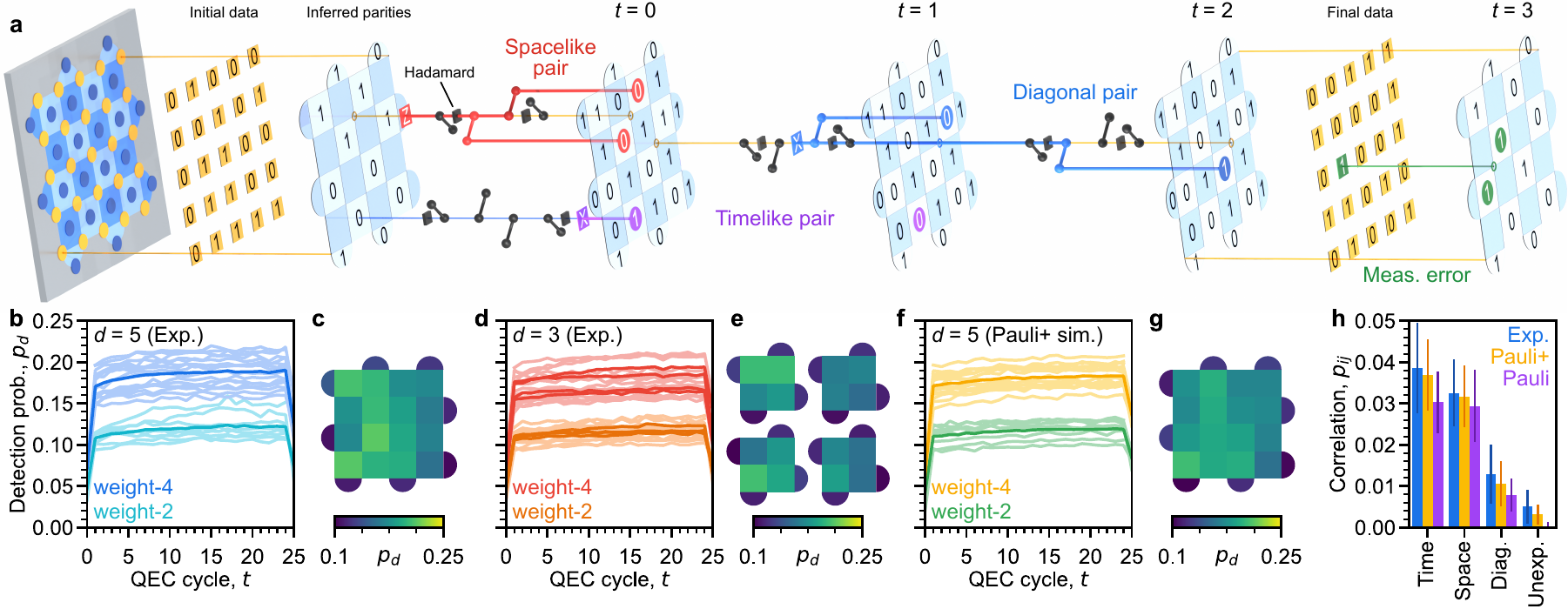}
    \caption{
    \textbf{Error detection in the surface code.}
    \textbf{a,} Illustration of a surface code experiment, in perspective view with time progressing to the right. We begin with an initial data qubit state which has known parities in one stabiliser basis (here, $Z$). We show example errors that manifest in detection pairs: a $Z$ error (red) on a data qubit (spacelike pair), a measurement error (purple) on a measure qubit (timelike pair), an $X$ error (blue) during the CZ gates (diagonal pair), and a measurement error (green) on a data qubit (detected in the final inferred $Z$ parities).
    \textbf{b,} Detection probability for each stabiliser over a 25-cycle distance-5 experiment (50,000 repetitions). Darker lines: average over all stabilisers with the same weight. 
    There are fewer detections at $t=0$ because there is no preceding data idling, and at $t=25$ because the final parities are calculated from data qubit measurements.
    \textbf{c,} Detection probability heatmap, averaging over $t=1$ to 24.
    \textbf{d-e,} Similar to \textbf{b-c} for four separate distance-3 experiments covering the four quadrants of the distance-5 code.
    \textbf{f-g,} Similar to \textbf{b-c} using a simulation with Pauli errors plus leakage, crosstalk, and stray interactions (\emph{Pauli+}).
    \textbf{h,} Bar chart summarising the detection correlation matrix $p_{ij}$, comparing the distance-5 experiment from \textbf{b} to the simulation in \textbf{f} (\emph{Pauli+}) and a simpler simulation with only Pauli errors. We aggregate four groups of correlations: timelike pairs; spacelike pairs; diagonal pairs expected for Pauli noise; and diagonal pairs unexpected for Pauli noise (``Unexp.''), including correlations over two timesteps. Each bar shows a mean and standard deviation of correlations from a 25-cycle, 50,000-repetition dataset.
    }
\end{figure*}

After initialisation, parity measurements should produce the same value in each cycle, up to known flips applied by the circuit. If we compare a parity measurement to the corresponding measurement in the preceding cycle and their values are inconsistent, a \emph{detection event} has occurred, indicating an error. We refer to these comparisons as \emph{detectors}.

The detection event probabilities for each detector indicate the distribution of physical errors in space and time while running the surface code.  In Fig.~2, we show the detection event probabilities in the distance-5 code (b, c) and the distance-3 codes (d, e) running for 25 rounds, as measured over 50,000 experimental instances. For the weight-4 stabilisers, the average detection probability is $0.185\pm 0.018$ ($1\sigma$) in the distance-5 code and $0.175 \pm 0.017$ averaged over the distance-3 codes. The weight-2 stabilisers interact with fewer qubits and hence detect fewer errors.  Correspondingly, they yield a lower average detection probability of $0.119 \pm 0.012$ in the distance-5 code and $0.115 \pm 0.008$ averaged over the distance-3 codes. The relative consistency between code distances suggests that growing the lattice does not substantially increase the component error rates during error correction.

The average detection probabilities exhibit a relative rise of 12\% for distance-5 and 8\% for distance-3 over 25 cycles, with a typical characteristic risetime of roughly 5 cycles \cite{supplement}.
We attribute this rise to data qubits leaking into non-computational excited states and anticipate that the inclusion of leakage-removal techniques on data qubits would help to mitigate this rise \cite{aliferis2005fault, suchara2015leakage, mcewen_removing_2021, chen2021exponential}. We hypothesise that the greater increase in detection probability in the distance-5 code is due to increased stray interactions or leakage from simultaneously operating more gates and measurements.

We test our understanding of the physical noise in our system by comparing the experimental data to a simulation. We begin with a Pauli simulation based on the component error information in Fig.~1c in a probabilistic Clifford simulator, then incorporate coherence information, transitions to leaked states, and crosstalk errors from unwanted coupling (\emph{Pauli+}). Fig.~2f shows that this second simulator accurately predicts the average detection probabilities, finding $0.180 \pm 0.013$ for the weight-4 stabilisers and $0.116 \pm 0.011$ for the weight-2 stabilisers, with average detection probabilities increasing 7\% over 25 rounds (distance-5).
\vspace{6ex}
\tocless\section{Understanding errors through correlations}

We next examine pairwise correlations between detection events, which give us fine-grained information about which types of errors are occurring during error correction.  Fig.~2a illustrates a few examples of pairwise detections which are generated by $X$ or $Z$ errors in the surface code. Measurement and reset errors are detected by the same stabiliser in two consecutive rounds, which we classify as a \emph{timelike} pair.  Data qubits may experience an $X$ ($Z$) error while idling during measurement which is detected by its neighbouring $Z$ ($X$) stabilisers in the same round, forming a \emph{spacelike} pair. Errors during CZ gates may cause a variety of pairwise detections to occur, including \emph{diagonal} pairs which are separated in both space and time. More complex clusters of detection events arise when a $Y$ error occurs, which generates detection events for both $X$ and $Z$ errors.

To estimate probabilities of detection event pairs from our data, we compute an appropriately normalised correlation $p_{ij}$ between detection events occurring on any two detectors $i$ and $j$ \cite{chen2021exponential}. In Fig.~2h, we show the estimated probabilities for experimental and simulated distance-5 data, aggregated and averaged according to the different classes of pairs.  In addition to the expected pairs, we also quantify how often detection pairs occur which are unexpected from component Pauli errors. 
Overall, the Pauli simulation systematically underpredicts these probabilities compared to experimental data, while the \emph{Pauli+} simulation is closer and predicts the presence of unexpected pairs, 
which we surmise are related to leakage and stray interactions.  
These errors can be especially harmful to the surface code because they can generate multiple detection events distantly separated in space or time, which a decoder might wrongly interpret as multiple independent component errors. We expect that mitigating leakage and stray interactions will become increasingly important as error rates decrease.
\vspace{6ex}
\tocless\section{Decoding and logical error probabilities}

\begin{figure}[t]
    \includegraphics[width=89mm]{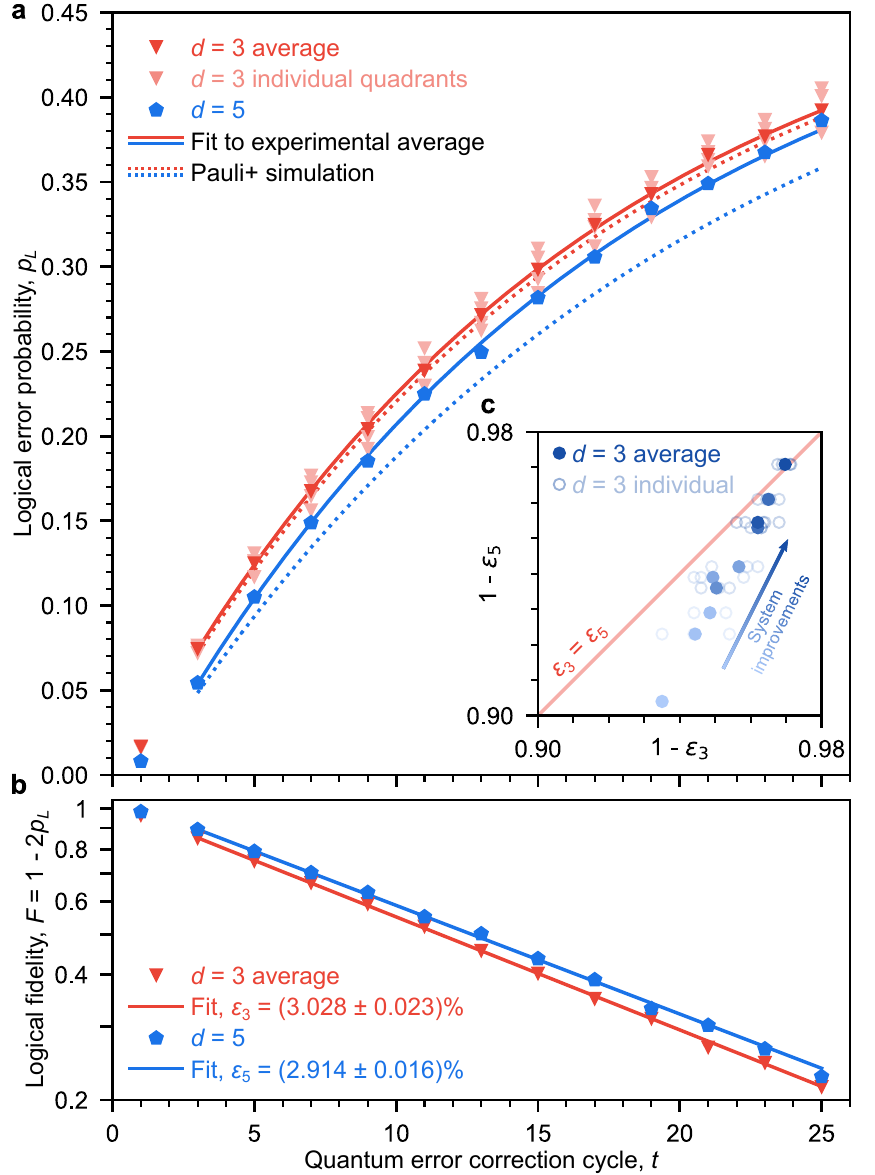}
    \caption{
    \textbf{Logical error reduction.} 
    \textbf{a,} Logical error probability $p_L$ vs. cycle comparing distance-5 (blue) to distance-3 (pink: four separate quadrants, red: average), all averaged over $Z_L$ and $X_L$. 
    Each individual datapoint represents 100,000 repetitions.
    Solid line: fit to experimental average, $t=3$ to 25 (see main text).
    Dotted line: comparison to \emph{Pauli+} simulation.
    \textbf{b,} Logical fidelity $F=1-2p_L$ vs. cycle, semilog plot. Same experimental averages and fits from \textbf{a}. 
    \textbf{c,} Summary of experimental progression comparing logical error per round $\varepsilon_d$ (specifically plotting $1-\varepsilon_d$) between distance-3 and $5$, where system improvements lead to faster improvement for distance-5 (see main text). Each open circle is a comparison to a specific distance-3 code, and filled circles average over multiple distance-3 codes measured in the same session. Markers are colored chronologically from light to dark. Typical $1\sigma$ statistical and fit uncertainty is $0.02\%$, smaller than the points.
    }
\end{figure}

We next examine the logical performance of our surface code qubits.  To infer the error-corrected logical measurement, the decoder requires a probability model for physical error events.  This information may be expressed as an \emph{error hypergraph}: detectors are vertices, physical error mechanisms are hyperedges connecting the detectors they trigger, and each hyperedge is assigned its corresponding error mechanism probability.  We use a generalisation of $p_{ij}$ to determine these probabilities \cite{chen2021exponential, chen2022calibrated}.

Given the error hypergraph, we implement two different decoders: belief-matching, an efficient combination of belief propagation and minimum-weight perfect matching \cite{higgott2022fragile}; and tensor network decoding, a slow but accurate approximate maximum-likelihood decoder.  
The belief-matching decoder first runs belief propagation on the error hypergraph to update hyperedge error probabilities based on nearby detection events \cite{criger2018multi, higgott2022fragile}.  The updated error hypergraph is then decomposed into a pair of disjoint error graphs, one each for $X$ and $Z$ errors \cite{dennis2002topological}. These graphs are decoded efficiently using minimum-weight perfect matching \cite{fowler2012towards} to select a single probable set of errors.

By contrast, a maximum-likelihood decoder considers all possible sets of errors consistent with the detection events, splits them into two groups based on whether they flip the logical measurement, and chooses the group with the greater total likelihood.
The two likelihoods are each expressed as a tensor network contraction \cite{BSV2014,chubb2021statistical, higgott2022fragile} that exhaustively sums the probabilities of all sets of errors within each group.  We can contract the network approximately, and verify that the approximation converges.  This yields a decoder which is nearly optimal given the hypergraph error priors, but is considerably slower. 
Further improvements could come from a more accurate prior, or incorporating more fine-grained measurement information \cite{suchara2015leakage, pattison2021improved}.

Figure~3 compares the logical error performance of the distance-3 and distance-5 codes using the approximate maximum-likelihood decoder. Because the \emph{ZXXZ} variant of the surface code symmetrises the $X$ and $Z$ bases, differences between the two bases’ logical error per round are small and attributable to spatial variations in physical error rates. Thus, for visual clarity, we report logical error probabilities averaged between the $X$ and $Z$ basis; the full data set may be found in the supplement. Note that we do not post-select on leakage or high-energy events in order to capture the effects of realistic non-idealities on logical performance. Over all 25 cycles of error correction, the distance-5 code realises lower logical error probabilities $p_L$ than the average of the subset distance-3 codes.   

We fit the logical fidelity $F=1-2p_L$ to an exponential decay, starting at $t=3$ to avoid
time-boundary effects that are advantageous to the distance-5 code.
We obtain a logical error per cycle $\varepsilon_5 = (2.914 \pm 0.016)$\% ($1\sigma$ statistical and fit uncertainty) for the distance-5 code, compared to an average of $\varepsilon_3 = (3.028 \pm 0.023)$\% for the subset distance-3 codes, a relative error reduction of about 4\%.  When decoding with the faster belief-matching decoder, we fit a logical error per cycle of $(3.056 \pm 0.015)$\% for the distance-5 code, compared to an average of $(3.118 \pm 0.025)$\% for the distance-3 codes, a relative error reduction of about 2\%. 
We note that the distance-5 logical error per cycle is slightly higher than two of the distance-3 codes individually, and that leakage accumulation may cause distance-5 performance to degrade faster than distance-3 as logical error probability approaches 50\%. 

In principle, the logical performance of a distance-5 code should improve faster than a distance-3 code as physical error rates decrease \cite{fowler2012surface}.
Over time, we improved our physical error rates, for example by optimising single and two qubit gates, measurement, and data qubit idling \cite{supplement}.
In Fig.~3c, we show the corresponding performance progression of distance-5 and distance-3 codes.
The larger code improved about twice as fast until finally overtaking the smaller code, validating the benefit of increased-distance protection in practice.

To understand the contributions of individual components to our logical error performance, we follow Ref.\,~\cite{chen2021exponential} and simulate the distance-5 and distance-3 codes while varying the physical error rates of the various circuit components. Because the logical error-suppression factor $\Lambda_{d/(d+2)} = \varepsilon_d / \varepsilon_{d+2}$ is approximately inversely proportional to the physical error rate, we can budget how much each physical error mechanism contributes to $1/\Lambda$ as shown in Fig.~4a to assess scaling. This error budget shows that CZ error and data qubit decoherence during measurement and reset are dominant contributors. 
\vspace{6ex}
\tocless\section{Algorithmically-relevant error rates with repetition codes}

\begin{figure} [t]
    \centering
    \includegraphics[width=89mm]{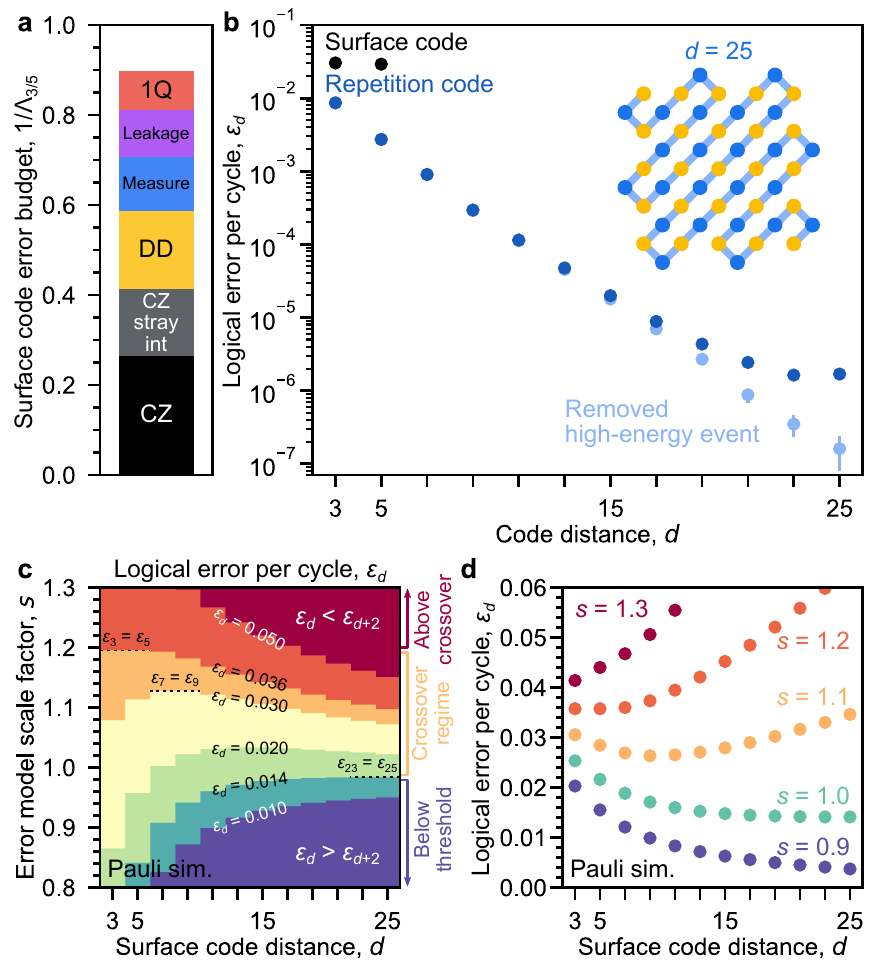}
    \caption{
    \textbf{Towards algorithmically-relevant error rates.}
    \textbf{a,} Estimated error budgets for the surface code, based on operation errors (see Fig.~1c) and \emph{Pauli+} simulations.
    $\Lambda_{3/5} = \varepsilon_3/\varepsilon_5$.
    CZ: contributions from CZ error (excluding leakage and stray interactions).
    CZ stray int: CZ error from unwanted interactions.
    DD: dynamical decoupling, data qubit idle error during measurement and reset.
    Measure: measurement and reset error.
    Leakage: leakage during CZs and due to heating.
    1Q: single-qubit gate error.
    \textbf{b,} Logical error for repetition codes. 
    Inset: schematic of the distance-25 repetition code, using the same data and measure qubits as the distance-5 surface code. Smaller codes are subsampled from the same distance-25 data~\cite{chen2021exponential}. 
    A high-energy event resulted in an apparent error floor around $10^{-6}$. After removing the instances nearby (light blue), error decreases more rapidly with code distance.
    50 cycles, $5\times 10^5$ repetitions.
    We also plot the surface code error per cycle from Fig.~3b in black.
    \textbf{c,} Contour plot of simulated surface code logical error per round $\varepsilon_d$ as a function of code distance $d$ and a scale factor $s$ on the error model in Fig.~1c (Pauli simulation, $s=1.0$ corresponds to the current device error model).
    \textbf{d,} Horizontal slices from c, each for a value of error model scale factor $s$. $s=1.3$ is above threshold (larger codes are worse), while $s=1.2$ to $1.0$ represent the crossover regime, where progressively larger codes get better until a turnaround. $s=0.9$ is below threshold (larger codes are better).
}
\end{figure}

Even as known error sources are suppressed in future devices, new dominant error mechanisms may arise as lower logical error rates are realised.  To test the behaviour of codes with substantially lower error rates, we employ the bit-flip repetition code, a 1D version of the surface code.  The bit-flip repetition code does not correct for phase flip errors and is thus unsuitable for quantum algorithms. However, correcting only bit-flip errors allows it to achieve much lower logical error probabilities.

Without post-selection, we achieve a logical error per cycle of $(1.7 \pm 0.3) \times 10^{-6}$ using a distance-25 repetition code decoded with minimum-weight perfect matching.  We attribute many of these high-distance logical errors to a high-energy impact, which can temporarily impart widespread correlated errors to the system \cite{mcewen2022resolving}.  These events may be identified by spikes in detection-event counts \cite{chen2021exponential}, and such error mechanisms must be mitigated for scalable quantum error correction to succeed.  In this case, there was one such event; after removing it (0.15\% of trials), we observe a logical error per cycle of $(1.6 \pm 0.8) \times 10^{-7}$ \cite{supplement}. The repetition code results demonstrate that low logical error rates are possible in a superconducting system, but finding and mitigating highly correlated errors such as cosmic ray impacts will be an important area of research moving forward.
\vspace{6ex}
\tocless\section{Towards large-scale quantum error correction}

To understand how our surface code results project forward to future devices, we simulate the logical error performance of surface codes ranging from distance-3 to 25, while also scaling the physical error rates shown in Fig.~1c. For efficiency, the simulation considers only Pauli errors. Figures 4c-d illustrate the contours of this parameter space, which has three distinct regions. When the physical error rate is high (for example, the initial runs of our surface code in the inset of Fig.~3), logical error probability increases with increasing system size ($\varepsilon_{d+2} > \varepsilon_d$). On the other hand, low physical error rates show the desired exponential suppression of logical error ($\varepsilon_{d+2} < \varepsilon_d$). Our experiment lies in a crossover regime where, due to finite-size effects, increasing system size initially suppresses the logical error rate before later increasing it.

While our device is close to threshold, reaching algorithmically-relevant logical error rates with manageable resources will require an error-suppression factor $\Lambda \gg 1$. Based on the error budget and simulations in Fig.~4, we estimate that component performance must improve by at least 20\% to move below threshold, and significantly improve beyond that to achieve practical scaling. However, these projections rely on simplified models and must be validated experimentally, testing larger code sizes with longer durations to eventually realise the desired logical performance. This work demonstrates the first step in that process, suppressing logical errors by scaling a quantum error-correcting code -- the foundation of a fault-tolerant quantum computer.
\vspace{6ex}
\tocless\section{Author Contributions}
The Google Quantum AI team conceived and designed the experiment. The theory and experimental teams at Google Quantum AI developed data analysis, modeling and metrological tools that enabled the experiment, built the system, performed the calibrations, and collected the data. The modeling was done jointly with collaborators outside Google Quantum AI. All authors wrote and revised the manuscript and the Supplementary Information.
\vspace{6ex}
\tocless\section{Acknowledgements}
We are grateful to S. Brin, S. Pichai, R. Porat, J. Dean, E. Collins, and J. Yagnik for their executive sponsorship of the Google Quantum AI team, and for their continued engagement and support. A portion of this work was performed in the UCSB Nanofabrication Facility, an open access laboratory. J. Marshall  acknowledges support from NASA Ames Research Center (NASA-Google SAA 403512), NASA Advanced Supercomputing Division for access to NASA HPC systems, and NASA Academic Mission Services (NNA16BD14C). D. Bacon is a CIFAR Associate Fellow in the Quantum Information Science Program.
\vspace{6ex}
\tocless\section{Data availability}
The data that support the plots within this paper and other findings of this study are available 
upon reasonable request, or at \href{https://doi.org/10.5281/zenodo.6804040}{10.5281/zenodo.6804040}.

\newpage
\onecolumngrid

\vspace{1em}
\begin{flushleft}
{\small Google Quantum AI: }

\bigskip
{\small
\renewcommand{\author}[2]{#1$^\textrm{\scriptsize #2}$}
\renewcommand{\affiliation}[2]{$^\textrm{\scriptsize #1}$ #2 \\}

\newcommand{\xGoogle}{\affiliation{1}{Google Research}}

\newcommand{\xCU}{\affiliation{2}{Department of Physics, Columbia University, New York, NY}}

\newcommand{\xUMass}{\affiliation{3}{Department of Electrical and Computer Engineering, University of Massachusetts, Amherst, MA}}

\newcommand{\xAU}{\affiliation{4}{Department of Electrical and Computer Engineering, Auburn University, Auburn, AL}}

\newcommand{\xUTSydney}{\affiliation{5}{Centre for Quantum Computation and Communication Technology, Centre for Quantum Software and Information, Faculty of Engineering and Information Technology, University of Technology Sydney, NSW 2007, Australia}}

\newcommand{\xCaltech}{\affiliation{6}{Department of Physics, Institute for Quantum Information and Matter, and Walter Burke Institute for Theoretical Physics, California Institute of Technology, Pasadena, CA}}

\newcommand{\xUCR}{\affiliation{7}{Department of Electrical and Computer Engineering, University of California, Riverside, CA}}

\newcommand{\xUSRA}{\affiliation{8}{USRA Research Institute for Advanced Computer Science, Mountain View, CA}}

\newcommand{\xQUAIL}{\affiliation{9}{QuAIL, NASA Ames Research Center, Moffett Field, CA}}

\newcommand{\xUCSB}{\affiliation{10}{Department of Physics, University of California, Santa Barbara, CA}}

\newcommand{\xUCRPhys}{\affiliation{11}{Department of Physics and Astronomy, University of California, Riverside, CA}}

\newcommand{\Google}{1}
\newcommand{\CU}{2}
\newcommand{\UMass}{3}
\newcommand{\AU}{4}
\newcommand{\UTSydney}{5}
\newcommand{\Caltech}{6}
\newcommand{\UCR}{7}
\newcommand{\USRA}{8}
\newcommand{\QUAIL}{9}
\newcommand{\UCSB}{10}
\newcommand{\UCRPhys}{11}

\author{Rajeev Acharya}{\Google},
\author{Igor Aleiner}{\Google,\CU},
\author{Richard Allen}{\Google},
\author{Trond I. Andersen}{\Google},
\author{Markus Ansmann}{\Google},
\author{Frank Arute}{\Google},
\author{Kunal Arya}{\Google},
\author{Abraham Asfaw}{\Google},
\author{Juan Atalaya}{\Google},
\author{Ryan Babbush}{\Google},
% ...

\bigskip

\xGoogle
\xCU
\xUMass
\xAU
\xUTSydney
\xCaltech
\xUCR
\xUSRA
\xQUAIL
\xUCSB
\xUCRPhys
}

\end{flushleft}

\twocolumngrid

\clearpage
\bibliographystyle{naturemag}
\vspace{6ex}
\let\oldaddcontentsline\addcontentsline
\renewcommand{\addcontentsline}[3]{}
\bibliography{refs}
\let\addcontentsline\oldaddcontentsline

\clearpage 

\title{Supplemental materials for \emph{Suppressing quantum errors by scaling a surface code logical qubit}}
\author{Google Quantum AI}
\date{\today}

\newpage
\null
\onecolumngrid
\vskip 2em%
\begin{center}%
\let \footnote \thanks
{\large \textbf{Supplemental materials for \emph{Suppressing quantum errors by scaling a surface code logical qubit}} \par}%
\vskip 1.5em%
{\large
  \lineskip .5em%
  \begin{tabular}[t]{c}%
    Google Quantum AI\\
    (Dated: \today)
  \end{tabular}\par}%
\vskip 1em%
\end{center}%
\par
\vskip 1.5em

\twocolumngrid
\tableofcontents

\renewcommand{\thefigure}{S\arabic{figure}}
\setcounter{page}{11}

\begin{figure}[htbp]
    \centering
    \includegraphics[width=3.5in]{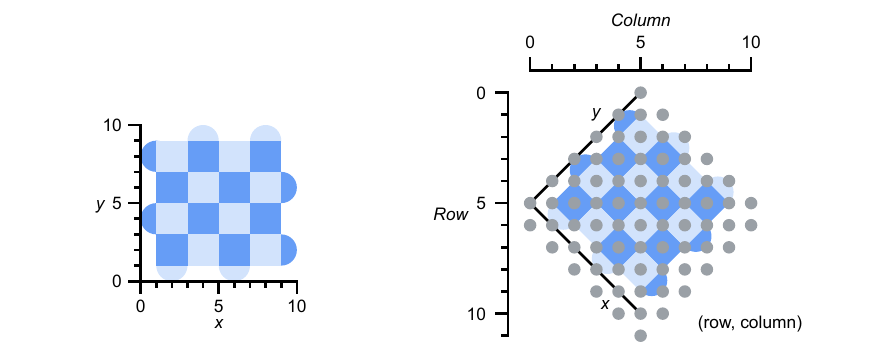}
    \caption{
        \textbf{Qubit coordinates.}
        Left: Distance-5 surface code as presented in the main text, with an $(x, y)$ coordinate system indicated.
        Right: Rotated coordinate system used for heatmap plots in the supplement, using (row, column).
    }
    \label{fig:sup_ksatz_coordinates}
\end{figure}

\section{Experimental details}

\subsection{Idling errors}

A significant contribution to the logical error budget is data qubit decoherence during the readout and reset of the measure qubits. The primary decoherence mechanism is dephasing induced by low frequency flux noise. We mitigate dephasing through dynamical decoupling with XY-4 phase cycling, which compensates first-order pulse errors and protects arbitrary quantum states equally. As mentioned in the main text, the specific dynamical decoupling sequence is chosen on a per-qubit basis in order to tailor the sequence filter function to each qubit's unique noise environment.

\subsection{Optimizing gate parameters}
\label{sec:snake}

\begin{figure*}
    \centering
    \includegraphics[width=\linewidth]{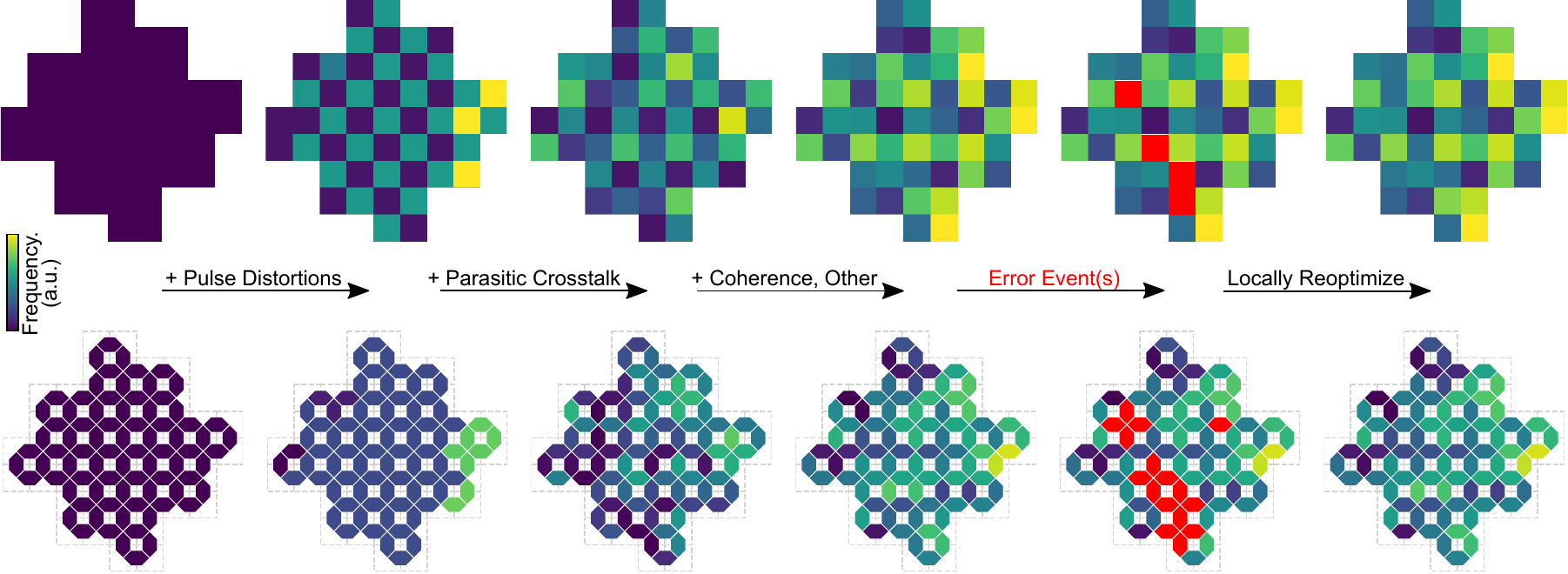}
    \caption{Idle (top) and interaction (bottom) frequencies found by our Snake optimizer as error mechanisms are progressively enabled. Readout parameters are not shown. Starting from an unconstrained processor, enabling pulse distortions results in an idle checkerboard with nearly-degenerate interactions. This configuration minimizes frequency-sweeps during CZ gates. Enabling crosstalk results in idle and interaction patterns that resemble multi-layer checkerboards. This configuration minimizes frequency collisions over the surface-code. Finally, enabling dephasing, relaxation, and other error mechanisms leads to frequency configurations with no obvious structure. After optimization, random error events can render arbitrary gates unusable (red). In such scenarios, we locally reoptimize parameters and stitch solutions. Color scales were chosen to maximize contrast.}
    \label{fig:snake}
\end{figure*}

Our quantum processor employs frequency-tunable qubits on a two-dimensional lattice with coupler-mediated nearest-neighbor coupling and dispersive readout. Quantum logic is implemented via single-qubit (SQ), two-qubit (CZ), and readout (RO) gates. SQ gates are implemented via resonant microwave pulses at qubits’ respective $f_{10}$ \textit{idle} frequencies. CZ gates are implemented by sweeping qubit pairs into $f_{10}/f_{21}$ resonance at respective \textit{interaction} frequencies and actuating respective couplers. RO gates are implemented by sweeping qubits' $f_{10}$ frequencies to respective \textit{readout} frequencies and measuring them via variable-amplitude and -length readout pulses. Most error mechanisms depend strongly on gate frequencies and readout-pulse parameters, which we refer to collectively as \textit{gate parameters}. Optimizing gate parameters is a critical error mitigation strategy necessary for state-of-the-art surface-code performance \cite{arute2019quantum, chen2021exponential}. 

To optimize gate parameters, we developed a surface-code objective through benchmarking and machine learning. It includes error contributions from relaxation, dephasing, crosstalk, and pulse distortion, along with various heuristics. Furthermore, it embeds the surface code circuit and it’s mapping onto our processor. To offer a sense of scale, the distance-5 objective incorporates $O(10^4)$ error terms and is defined over 49 idle, 49 readout, and 80 interaction frequencies, and 49 readout pulse amplitudes and 49 lengths. It is noisy, non-convex, and all parameters are explicitly or implicitly intertwined due to engineered interactions and/or crosstalk. Furthermore, since each parameter is constrained to $\sim10^2$ values by the control electronics, processor circuit, and gate parameters, the search-space is $\sim10^{552}$. This space is intractable to search exhaustively and traditional global optimizers do not perform well on the objective. Therefore, we invented the Snake optimizer to address it \cite{klimov2020snake}. 

The Snake leverages concepts in graph optimization and dynamic programming to split complex high-dimensional optimization problems into simpler lower-dimensional subproblems. One key hyperparameter of the Snake is the subproblem dimension, which enables us to trade optimization complexity for accessible solutions. We operate the Snake in an intermediate dimensional regime that benefits from the speed of local search and non-locality of global search. This strategy outperforms local and global optimization in convergence rate and error on the distance-5 objective. Furthermore, we believe it will scale towards fault-tolerant processors. 

To illustrate how the Snake trades between error mechanisms, we plot optimized idle and interaction frequencies as error mechanisms are progressively enabled in Fig.~\ref{fig:snake}. Readout parameters experience similar tradeoffs but are not shown. After optimization, some gates unexpectedly experience uncharacteristically large error rates. These events happen randomly and are often due to spurious resonances, for example due to two-level-system defects, moving into the path of SQ and/or CZ gates \cite{klimov2018tls}. Reoptimizing all gates after such failures is unscalable due to considerations including calibration runtime. Instead, we use the Snake to locally reoptimize failing gates and stitch solutions. To improve the quality and longevity of our solutions, we employ higher dimensional optimization and embed historical data into our objective.

\subsection{Single qubit gates}
\begin{figure*}[htbp]
    \centering
    \includegraphics[width=\linewidth]{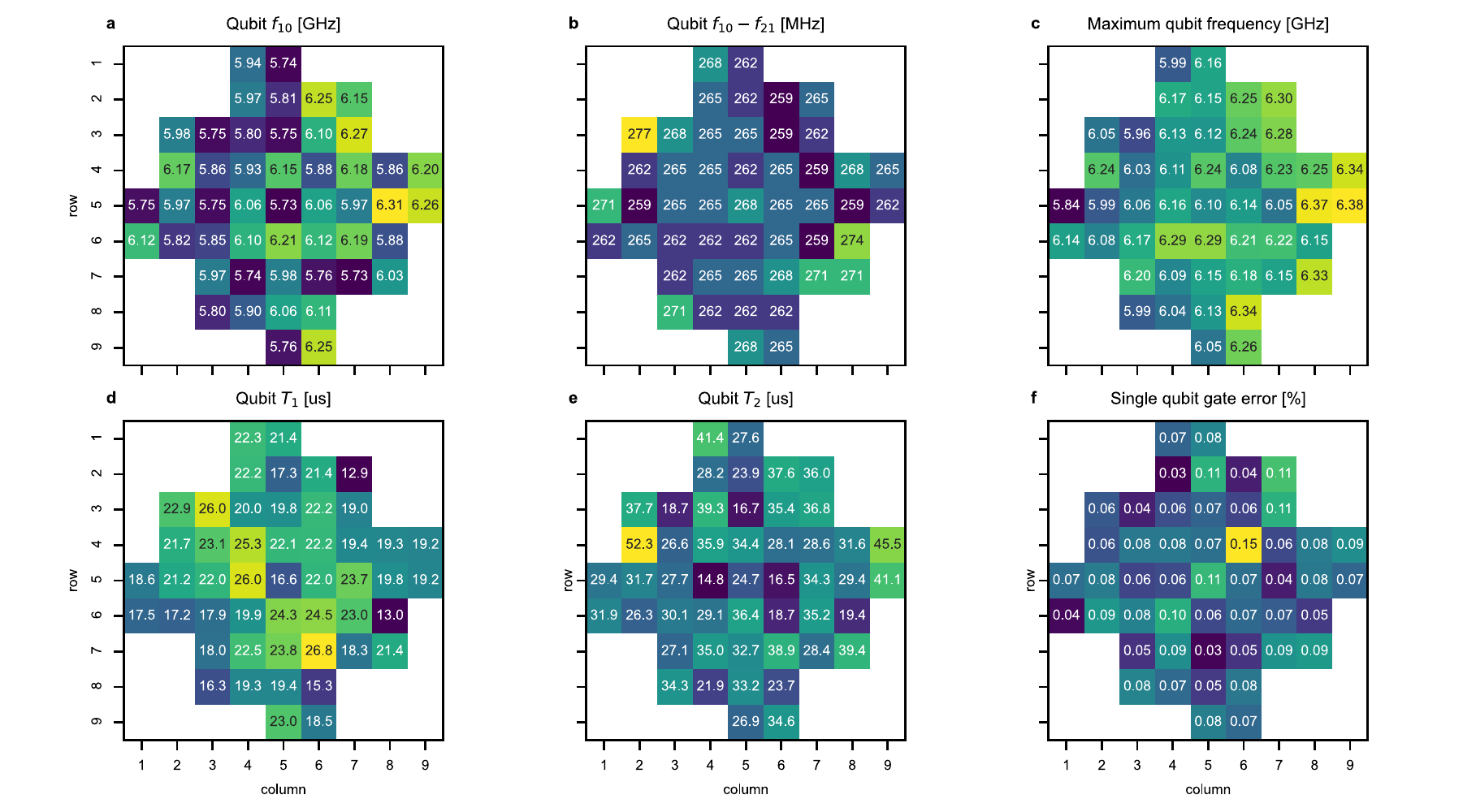}
    \caption{Heatmaps of various single qubit parameters:
    \textbf{a,} ground to first excited state transition frequency.
    \textbf{b,} Qubit anharmonicity at the operating frequency.
    \textbf{c,} Maximum achievable qubit frequencies.
    \textbf{d,} Qubit relaxation time.
    \textbf{e,} Qubit decoherence time as measured by CPMG.
    \textbf{f,} Simultaneous single-qubit gate error.
    }
    \label{fig:sq_parameters}
\end{figure*}

Single qubit gates are implemented through a combination of microwave $XY$ rotations and virtual $Z$ rotations. All microwave pulses utilize DRAG pulse shaping to mitigate leakage from off-resonant driving of the $|1\rangle\leftrightarrow|2\rangle$ transition, as detailed in \cite{chen2021exponential}. We assess our single qubit gate performance through randomized benchmarking, acting on all qubits simultaneously. A summary of our single qubit device parameters, including operating frequencies, coherence times, and gate errors can be found in \ref{fig:sq_parameters}.

\subsection{Two qubit gates}
\begin{figure*}[htbp]
    \centering
    \includegraphics[width=\linewidth]{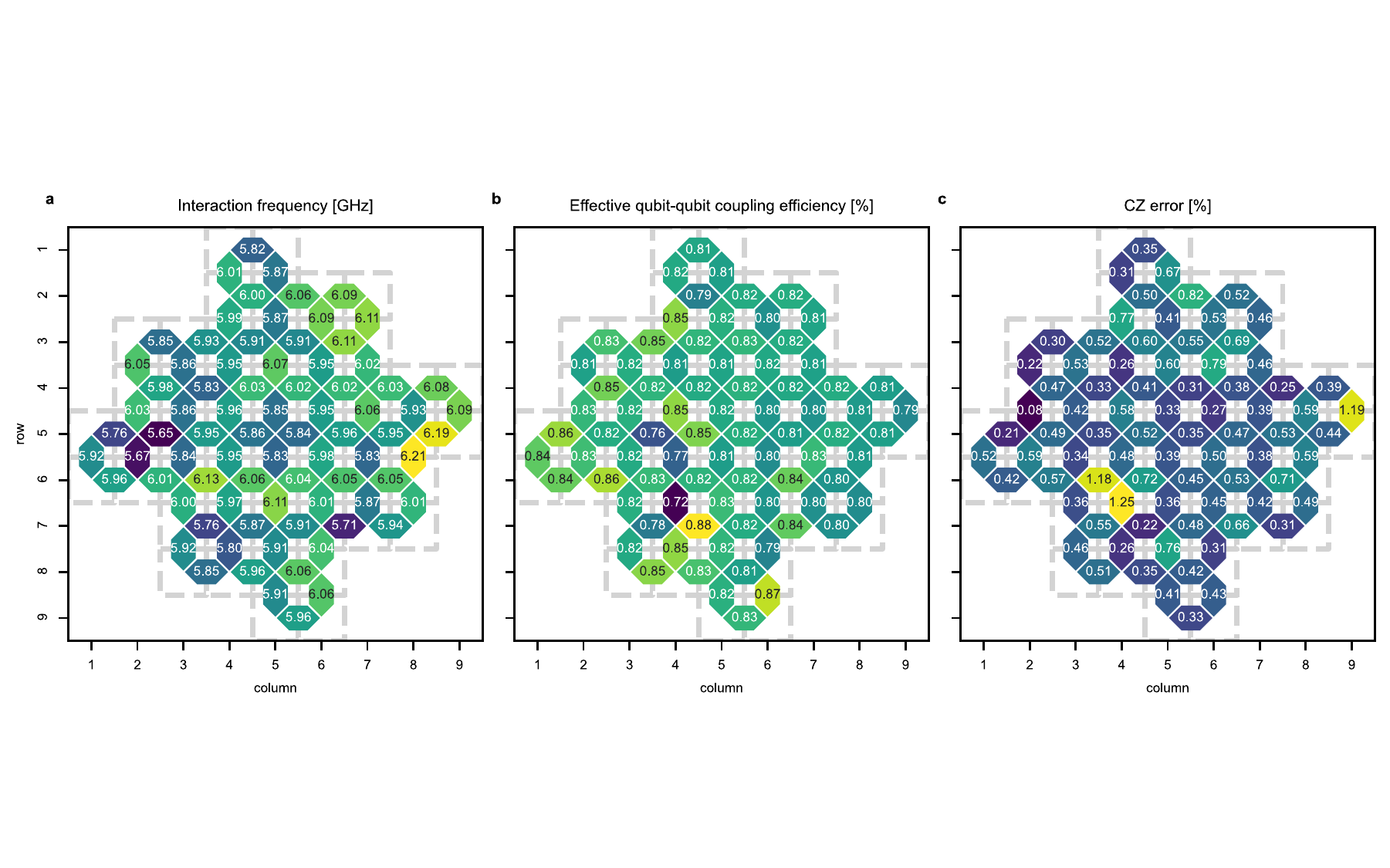}
    \caption{Heatmaps of various two qubit gate parameters: 
    \textbf{a,} frequency where two coupled qubits interact to perform a CZ gate.
    \textbf{b,} The effective coupling efficiency between two qubits.
    \textbf{c,} CZ gate error as measured by XEB and in the surface code layers.
    }
    \label{fig:cz_parameters}
\end{figure*}

The two qubit gate utilized in this work is the controlled-Z (CZ) gate. CZ gates in the sycamore architecture arise from a state-selective dispersive shift on the two qubit state $|11\rangle$, which is introduced through a tunable coupling in the two excitation manifold, as detailed in \cite{foxen2020demonstrating}. We assess our CZ gate performance by performing cross-entropy benchmarking (XEB). In a single round of error correction in the surface code, CZ gates are applied in layers which correspond to their position in a given stabilizer. Therefore, to more accurately represent the CZ performance in the surface code, we perform simultaneous XEB on pairs using their respective CZ layers. We summarize our CZ gate parameters and measured errors in \ref{fig:cz_parameters}

\subsection{Measurement and reset}
\begin{figure*}[htbp]
    \centering
    \includegraphics[width=\linewidth]{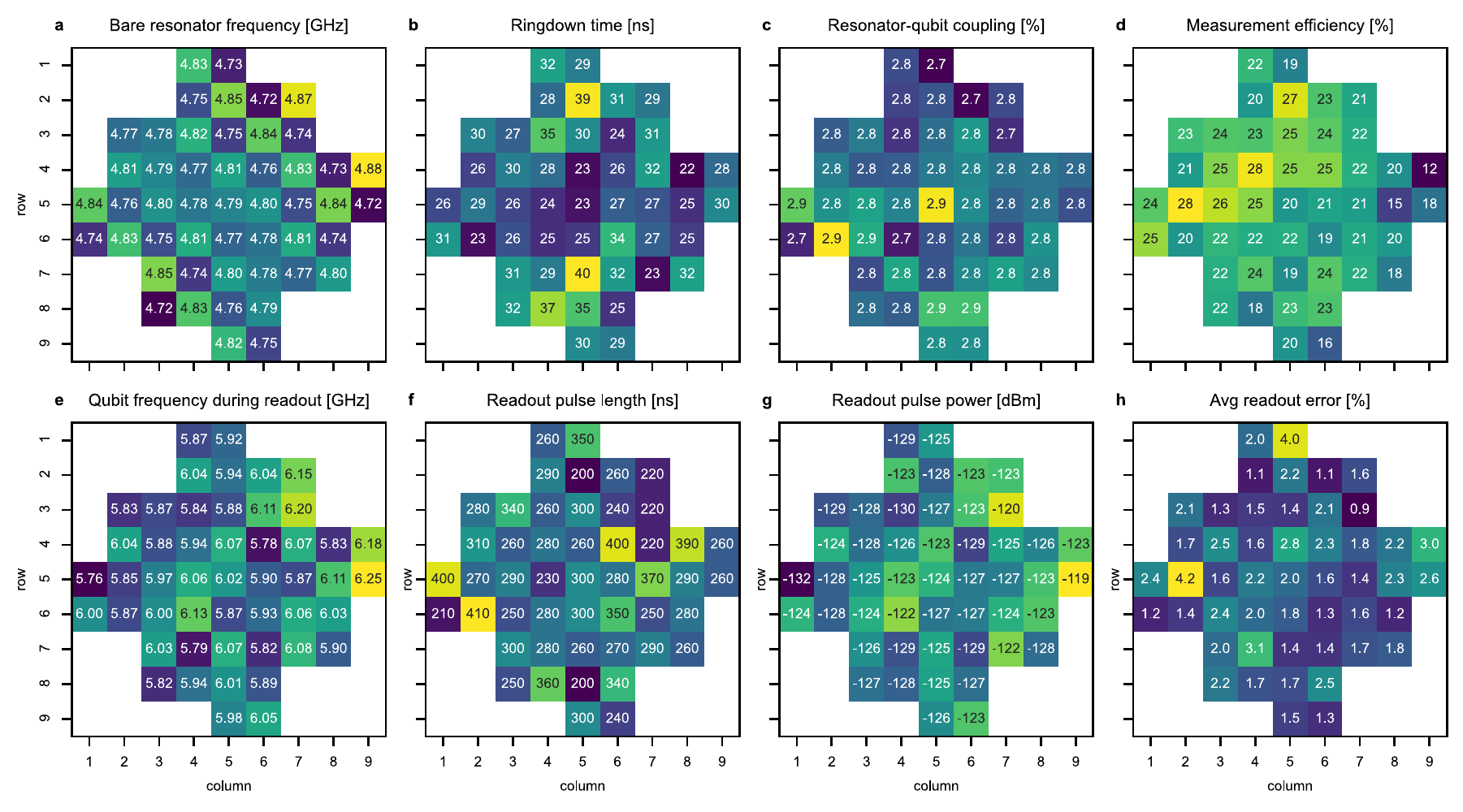}
    \caption{Heatmaps of various readout parameters:
    \textbf{a,} bare frequency of each readout resonator.
    \textbf{b,} Ringdown time (equal to $1/\kappa$) of each resonator.
    \textbf{c,} The coupling efficiency between resonator and qubit.
    \textbf{d,} Efficiency of the measurement chain close to the readout frequency.
    \textbf{e,} Optimized qubit frequency during readout (in the absence of a readout drive).
    \textbf{f,} Optimized readout pulse lengths.
    \textbf{g,} Optimized readout pulse power referenced to the input of the readout resonator and calibrated using the AC-Stark shift.
    \textbf{h,} Measured average readout error over $\ket{0}$ and $\ket{1}$ when preparing a random set of qubit states across all 49 qubits.
    }
    \label{fig:readout_parameters}
\end{figure*}

Every qubit on the Sycamore processor has a dedicated resonator used for measurement and reset of the qubit state, see details in Refs.~\cite{arute2019quantum, chen2021exponential, mcewen_removing_2021}. In \ref{fig:readout_parameters} (a-d) we summarize various parameters related to the resonators and the measurement chain. Feeding those parameters into models for different readout and reset error mechanisms, we optimize three readout parameters (qubit frequenecy during readout, readout pulse length, and readout pulse power) individually for each qubit. The total readout time is kept at 500 ns for each qubit; however for measure qubits we need the respective resonators to be empty before reset starts so we include time (equal to 500 ns minus the readout pulse length) for the resonators to ring down. The optimization is done using the Snake optimizer, see \ref{sec:snake}. For instance, we model errors due to finite signal-to-noise ratio, qubit relaxation during readout, swapping between neighboring qubits, and qubit state transition due to the resonator drive \cite{sank2016measurement}. The optimized values are found in \ref{fig:readout_parameters} (e-g). In the end, we benchmark readout by preparing and measuring a random set of qubit states, summarized in \ref{fig:readout_parameters} (h).

\subsection{Microwave crosstalk induced leakage mitigation}

\begin{figure}[htbp]
    \centering
    \includegraphics[width=3.5in]{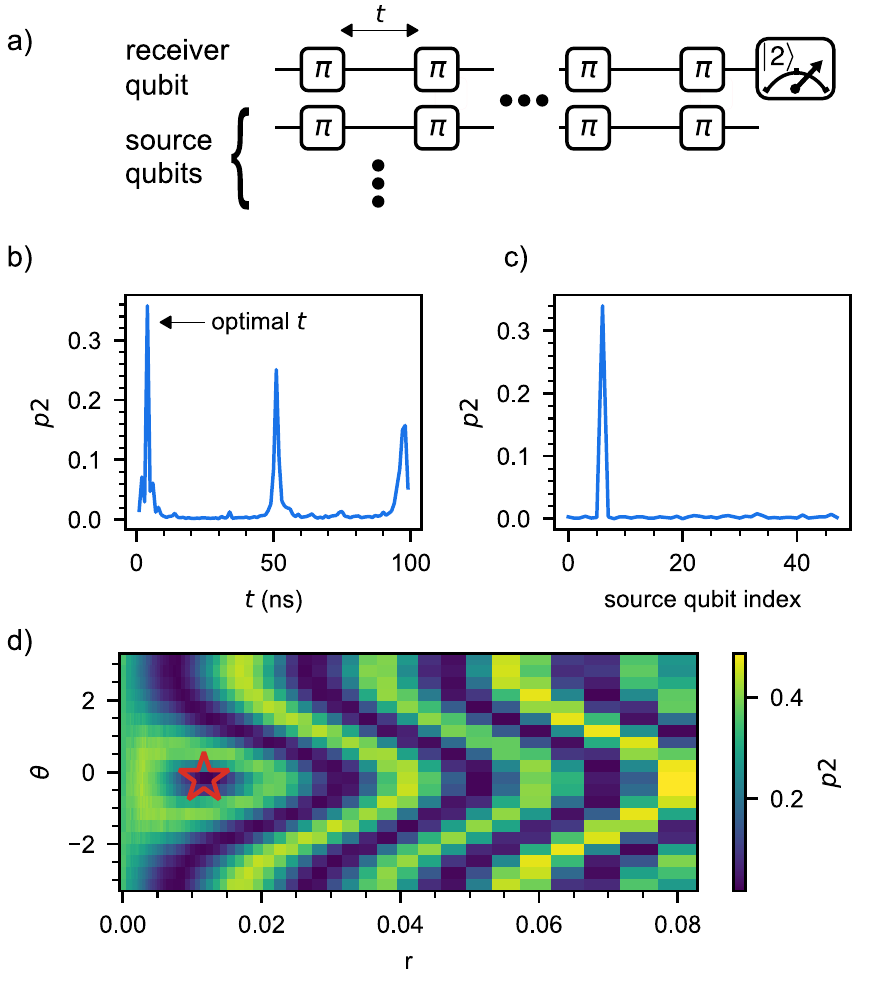}
    \caption{Mitigation of microwave crosstalk induced leakage. \textbf{a,} The Ramsey error filter pulse sequence consists of $\pi$ rotations separated by a variable delay $t$.  We repeat this several times to amplify the coherent leakage, which we observe by direct measurement of $\ket{2}$. \textbf{b,} Leakage probability $p2$ vs $t$ during the Ramsey error filter.  This data is taken with all sources operating in parallel.  The $t$ that maximizes leakage is held fixed throughout the remainder of the calibration sequence. \textbf{c,} $p2$ vs source qubit during pairwise operation at optimal $t$. The peak identifies the dominant source. \textbf{d,} $p2$ after Ramsey error filter during pairwise operation with the dominant source vs r and $\theta$, the magnitude and phase of the compensation tone used to null the crosstalk leakage.  The red star indicates the optimal $r$ and $\theta$.}
    \label{fig:xtalk_leakage}
\end{figure}

Leakage out of the computational subspace is an important error class as it leads to correlated error patterns during error detection.  Microwave crosstalk is one mechanism that causes leakage during single qubit gates.  If the drive frequency is near f21 of the receiver qubit, a parasitic $\ket{1} \rightarrow \ket{2}$ leakage error can result.

To mitigate this leakage channel we apply a compensation pulse, out of phase with the parasitic drive, directly to the receiver qubit to null this leakage.  We apply a signal $V' \left( t \right) = V \left( t \right) r e^{i \theta}$ for $V'$ the applied compensation, $V$ the source signal.  In this section we outline our procedure for calibrating the parameters $r$ and $\theta$.

We use a procedure based on the Ramsey error filter pulse sequence, Fig.~\ref{fig:xtalk_leakage}a, to calibrate the magnitude and phase of the compensation tone.  For certain delay times $t$, this sequence amplifies the coherent crosstalk leakage to facilitate the calibration.  This amplification can be understood as a result of constructive interference between the leakage amplitudes induced by successive pulses. 

Fig.~\ref{fig:xtalk_leakage}b shows representative data from the Ramsey error filter vs delay time $t$.  This data is taken while driving all qubits in parallel to minimize data acquisition time.  In the following calibration steps, $t$ is held fixed at the value that maximizes the leakage population $p2$.  In Fig.~\ref{fig:xtalk_leakage}c, we show the output of the Ramsey error filter at optimal $t$ during pairwise operation for all possible pairs that include the receiver qubit.  This data identifies the source of the parasitic drive.

Fig.~\ref{fig:xtalk_leakage}d
shows $p2$ on the receiver qubit during pairwise operation with the dominant source qubit at optimal $t$ vs the amplitude $r$ and phase $\theta$ of the compensation tone.  The red star indicates the optimal $r$ and $\theta$ to mitigate crosstalk induced leakage.

\subsection{Surface code experimental details}

\subsubsection{Surface code circuits}
\begin{figure*}[htbp]
    \centering
    \includegraphics[width=7.2in]{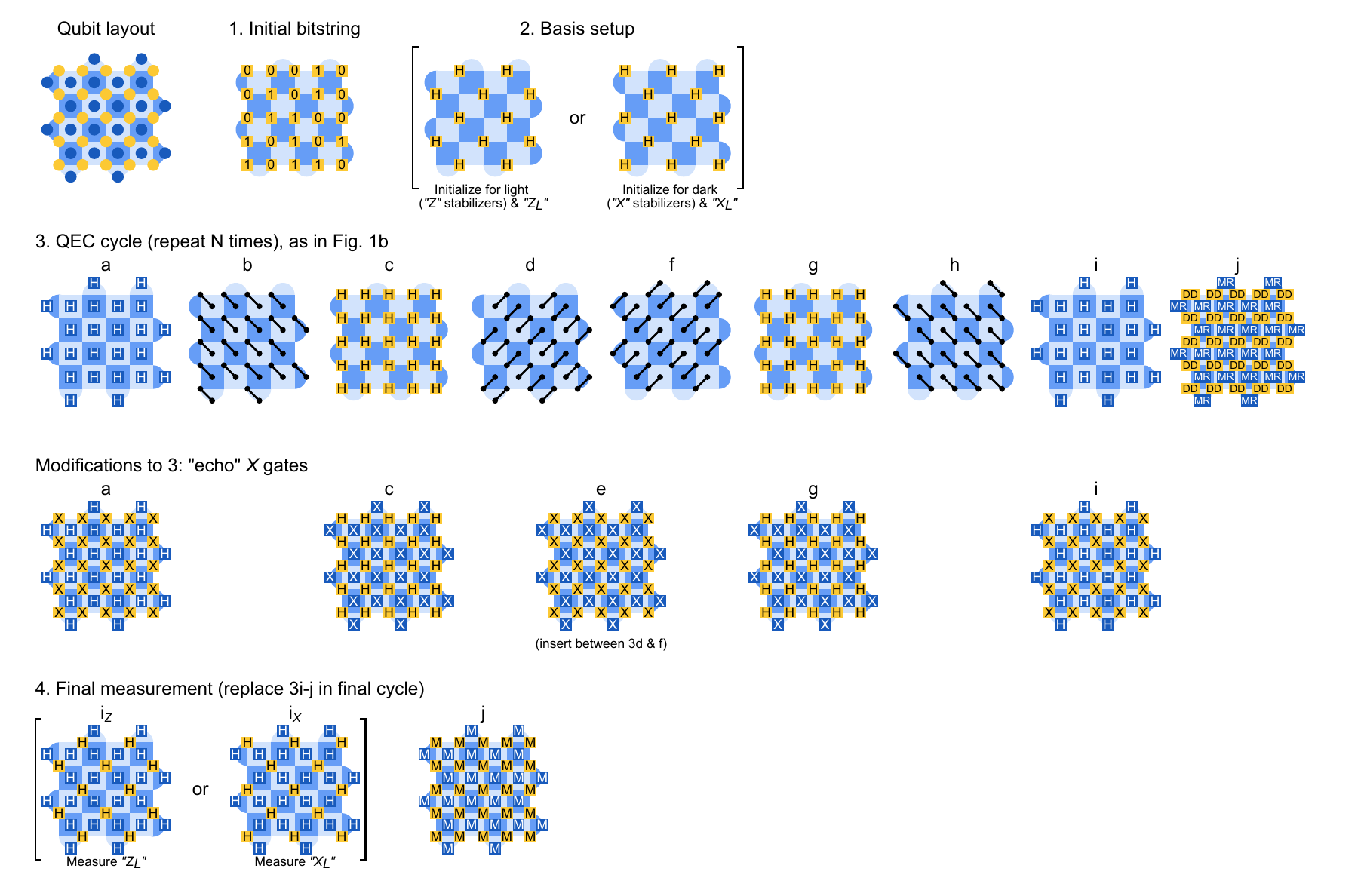}
    \caption{
        \textbf{Distance-5 surface code circuits.}
        Expanded version of Fig.~1b (main text) showing the full sequence of operations.
        \textbf{1.} We reset all qubits to $|0\rangle$ and then prepare an initial bitstring state on the data qubits using $X$ gates.
        \textbf{2.} We apply $H$ (Hadamard) gates to some of the data qubits to convert the initial bitstring (eigenstate of all $Z$ operators) into an eigenstate of half the \textit{ZXXZ} stabilisers, the half matching to logical operator of interest ($X_L$ or $Z_L$) for the specific experiment. Note steps (1) and (2) could be combined into a single moment of Clifford gates, but for simplicity we execute them in two moments as shown.
        \textbf{3.} QEC cycle, as in Fig.~1b (main text), showing the explicit gate patterns. Note although all stabilisers measure \textit{ZXXZ}, the "$X$" stabilizers and "$Z$" stabilizers apply their CZs in a different order (specifically in 3d and f). This pattern is carefully designed to manage ``hook'' errors.
        As mentioned in the main text, we modify this gate sequence with additional ``echo'' or ``dynamical decoupling'' gates \emph{within} the unitary part of the circuit. This amounts to adding $X$ gates to the data qubits in 3a and 3i, to the measure qubits in 3c and 3g, and to all qubits in 3e (inserted between the middle two CZ moments).
        \textbf{4.} For the final measurement, we replace 3i-j with a different Hadamard pattern (transforming the data qubit state to the appropriate basis for logical measurement) and measure all qubits simultaneously. This final measurement is used for detectors for the penultimate round (measure qubit results) and the final round (data qubit results, converted to parities in the relevant basis).
    }
    \label{fig:sup_ksatz_circuit1}
\end{figure*}

\begin{figure}
    \centering
    \includegraphics[width=3.5in]{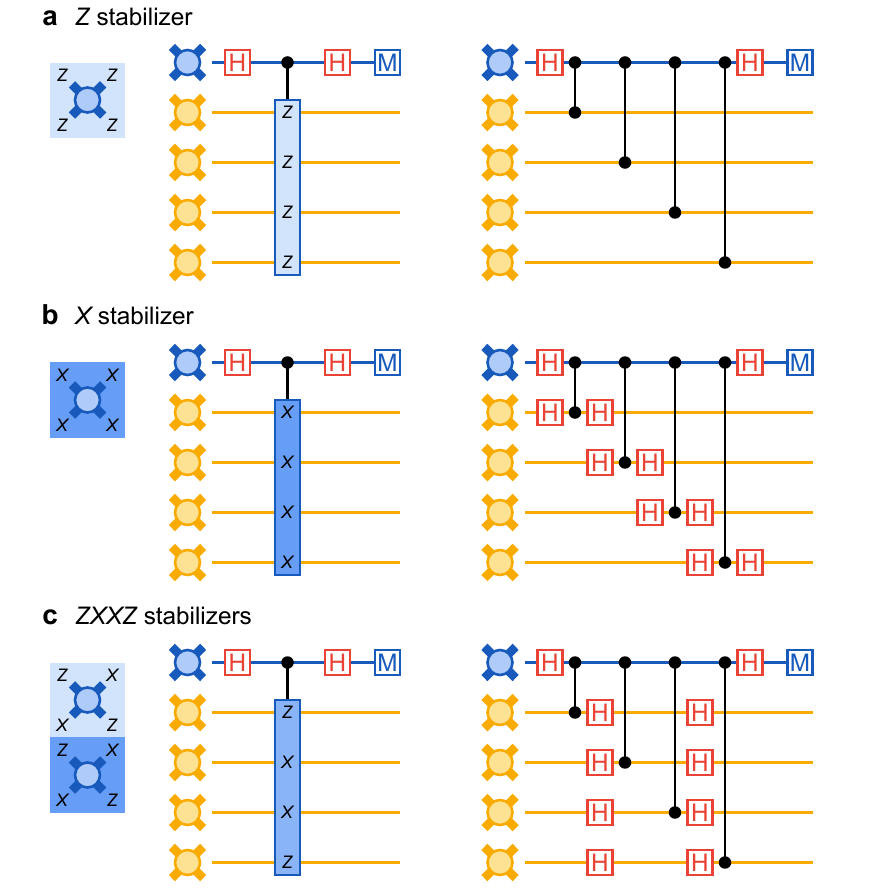}
    \caption{
    \textbf{\textit{ZXXZ} stabilisers.}
    \textbf{a,} Illustration of measuring a \textit{ZZZZ} stabiliser including compilation into CZ gates.
    \textbf{b,} Similar for \textit{XXXX}.
    \textbf{c,} Similar for \textit{ZXXZ} which we use for all stabilisers in our experiments.
    }
    \label{fig:sub_ksatz_zxxz}
\end{figure}

We present the surface code circuit in Fig.~1b (main text), focusing primarily on two qubits. The circuit generalises across the surface code, which we show here for clarity in Fig.~\ref{fig:sup_ksatz_circuit1}. We also clarify the \textit{ZXXZ} stabilisers and how they relate to other stabilisers in Fig.~\ref{fig:sub_ksatz_zxxz}.

In addition to individual-gate calibrations to track how qubit phases are affected by each gate, we add additional corrections within the surface code circuit, empirically optimized to minimize detection probabilities \cite{kelly2015state, chen2021exponential}. These catch-all corrections allow us to mitigate the impact of a variety of noise sources which all effectively act as a single qubit phase shift. We place a ``virtual $Z^\alpha$'' correction preceding each Hadamard gate, and then tune the value of $\alpha$ for each correction to minimize the detection rate of the detectors that the correction impacts.

The measure qubit case is simpler because they are reset to $|0\rangle$ to start each round, meaning that the initial Hadamard does not need a correction. As a result, each measure qubit only has a single correction per error correction cycle. Data qubits, which are assumed to be in a non-trivial state throughout, require two corrections per cycle. In case there are ``warmup'' effects, we give each measure qubit a special correction value for the first cycle, and then each uses another value for the rest of the cycles, while data qubit corrections have unique values for the first cycle and the bulk cycle for each correction. Data qubits also have a single extra correction in the last round, to correct for errors which occur right before logical measurement.

These corrections $\{\alpha\}$, typically small (< 0.1), are optimized using separate surface code experiments. We can divide the corrections into ten groupings of compatible corrections for any distance of surface code, meaning that the corrections impact different detectors and can therefore be optimized simultaneously. We then find the value for each correction which minimized the detection rate in parallel. Since the number of groupings is constant for any distance of surface code, we expect this method to scale well as we increase the sizes of our system.

\subsubsection{Choice of $d=3$ grids}
\begin{figure*}[htbp]
    \centering
    \includegraphics[width=\linewidth]{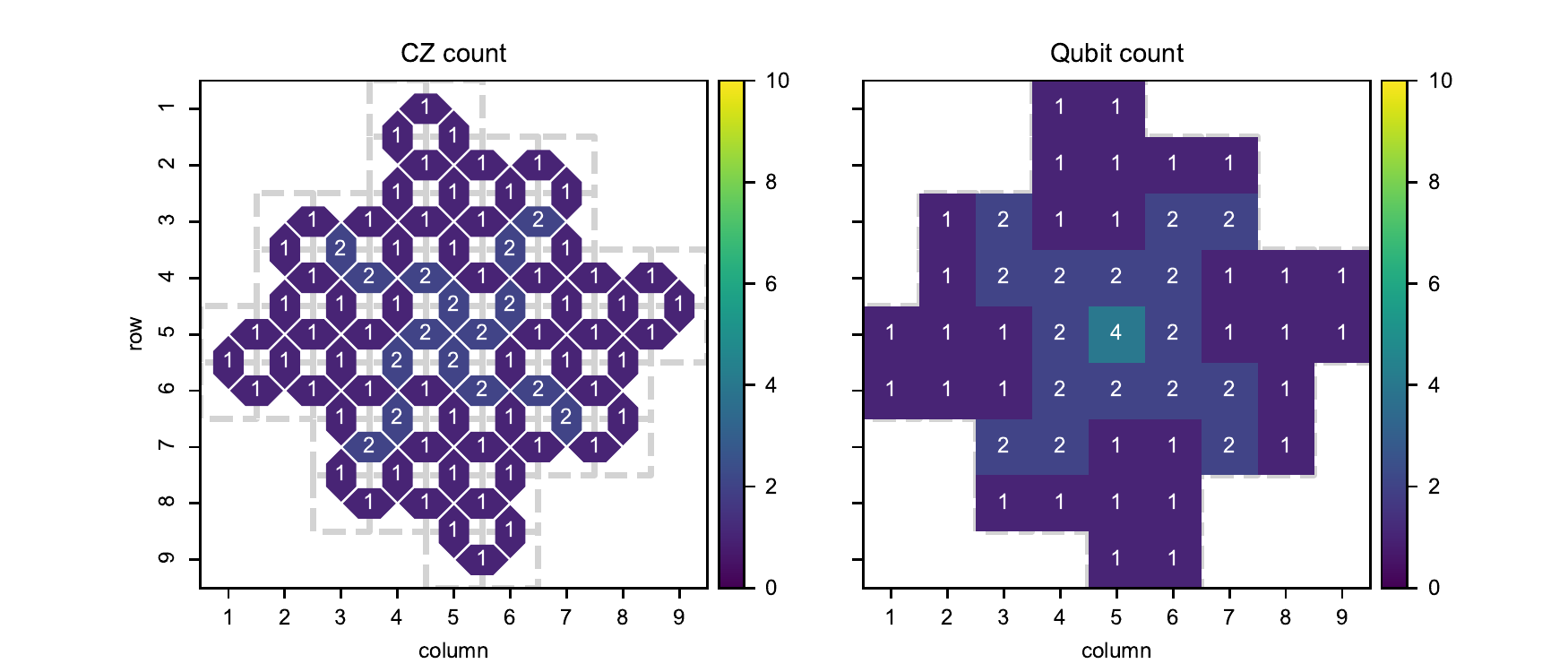}
    \caption{Heatmaps of CZ and qubit involvements in the four distance-3 grids considered in this work.}
    \label{fig:qubit heatmap}
\end{figure*}
In total, there are 10 possible distance-3 grids whose footprint lies within the footprint of the distance-5 surface code grid. The four grids chosen were picked such that their overlap was minimized, in order to reduce susceptibility to bias. There is still non-trivial overlap at the boundaries of the distance-3 grids, as well as the center qubit, which is included in all 4 of the smaller codes as seen in Fig.~\ref{fig:qubit heatmap}. Since the distance-3 codes are more susceptible to outlier qubits, the performance of the center qubit is essential to a fair measurement of $\Lambda_{35}$. 

\begin{figure}[htbp]
    \centering
    \includegraphics{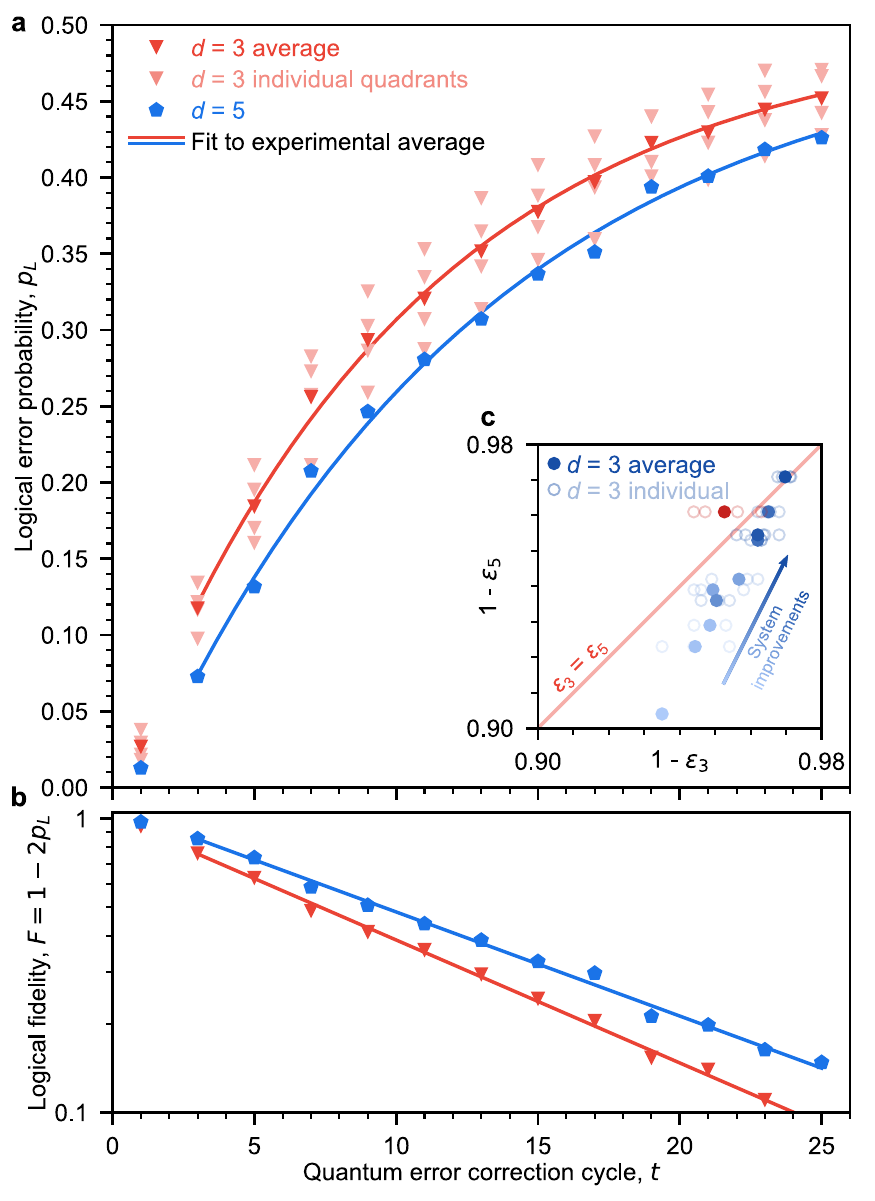}
    \caption{Plots in the style of Fig.~3 in the main text, but for a dataset where the center qubit in the grid was experiencing performance issues. In \textbf{a} and \textbf{b}, we present the logical error probability versus round, while in the inset \textbf{c}, we present the same progression shown in the main text, but with this outlier set plotted in red.}
    \label{fig:outlier}
\end{figure}

In Fig.~\ref{fig:outlier}, we show a dataset where the center qubit was experiencing excess phase errors, dramatically impacting the performance of the distance-3 codes. In the inset, we can see that although this dataset does land in the desired region, it is clear that it is due to outlier performance. We can also see this by comparing to our models and observing that the logical error per cycle does not match where our models would expect such a crossover to occur. By looking at the detection event fractions in Fig.~2 of the main text, we can see that the performance around the center of the chip was within the standard range for the experiment presented, so we have no reason to believe that our measurement of $\Lambda$ was skewed. 

\subsubsection{Order of experiments}
We interleave the experiments in Fig.~3 (main text). Specifically, we acquire one (number of cycles, basis, code) dataset at a time. Each dataset consists of 50000 total surface code runs (10 initial bitstring states times 5000 repetitions). 

We shuffle the order in which we acquire each (number of cycles) group of datasets. We begin with an extra 25-cycle experiment which we use for some decoder ``training'' (see decoding section) and then proceed with the other numbers of cycles in random order, in particular (25 (extra), 15, 17, 21, 13, 1, 19, 5, 23, 25, 7, 9, 3, 11). 

Additionally, after each 5 cycles, we run a ``frequency update'' calibration on all the qubits. This is a simultaneous Ramsey experiment to quickly check the qubit frequency and update the flux bias values to tune the qubits to the desired frequencies if needed.

The pseudocode below describes the specific order in which the (number of cycles, basis, code) datasets were acquired.

\verb||

\verb|num_cycles = (25, 15, 17, ...)  # see text|

\verb|codes = (|

\verb|    d5,|

\verb|    upper_left,|

\verb|    lower_right,|

\verb|    upper_right,|

\verb|    lower_left,|

\verb|)|

\verb|for idx, n in enumerate(num_cycles):|

\verb|    if should_run_frequency_update(idx):|

\verb|        run_frequency_update()|

\verb|    for basis in (Z, X):|

\verb|        for code in codes:|

\verb|            take_data(n, basis, code)|

\subsubsection{Initial states}
As discussed in the main text, we use random bitstrings as the initial data qubit states. This avoids a situation where we start with a bias to more $|0\rangle$ measurement outcomes: if all the data qubits start in $|0\rangle$, about 3/4 of the measure qubits will see $|0\rangle$ in the first cycle, as opposed to 1/2, and there would be an asymmetry in measurement fidelity. This would artificially lower the error for the first several rounds as the code ``warms up'' to the steady state.

For the data in Fig.~3 (main text), we choose to initialize with 10 different bitstrings, 5 with each logical value. To accomplish this, for $N$ qubits, we generate 5 integers between 0 (inclusive) and $2^{N-1}$ (exclusive) and then interleave them with their $N$-bit bitwise complements (for odd surface code distance, bitwise complements have opposite logical values).

Specifically, the bit-packed decimal integers we use for distance-5 are (1497382, 32057049, 12984827, 20569604, 10981887, 22572544, 7363158, 26191273, 7264790, 26289641), and for distance-3, (22, 489, 198, 313, 167, 344, 112, 399, 110, 401).

The 25-bit representation of 1497382 is 0b0000101101101100100100110. Referring to Fig.~\ref{fig:sup_ksatz_coordinates}, this is assigned to the data qubits in a big-endian fashion using the (row, column) coordinate system, with the most significant bit being (row=1, column=5), followed by (2, 4) and (2, 6) and ending with the least significant bit (9, 5). 1497382 is followed by its 25-bit complement, 32057049 or 0b1111010010010011011011001.

\section{Decoding}
\label{sec:decoding}

\subsection{Setting the prior distribution}
\label{sec:prior}
Each decoder requires a prior distribution on the set of errors occurring in the device.  However, there are many sets of errors which trigger the same set of detectors.  Because our decoder only depends on the detection events themselves, we group error probabilities collectively into bins according to the set of detectors they activate.  This is done using Stim's detector error model functionality \cite{gidney2021stim}, which tracks Pauli errors to perform this binning automatically.  Specifically, we use a simple depolarizing circuit noise model, derived from Stim circuit descriptions of the experimental circuits, to generate an initial set of bins of detectors and their collective probabilities.  This information represents an initial error hypergraph - detectors are vertices, bins act as hyperedges connecting those detectors they activate, and hyperedge weights are defined from the collective error probabilities of the errors contained in each bin.  Note that this is a slightly restricted model of decoding, and that more general models may be employed to increase the probability of success \cite{pattison2021improved}.

From here, we use a higher-order extension of the $p_{ij}$ method to assign probabilities to this initial ansatz of hyperedges using device-level data \cite{chen2021exponential, chen2022calibrated}.  We compute the highest-order correlations present in the standard depolarizing circuit model, which are weight-$4$.  We then continue to lower-order correlations, subtracting the probabilities of higher-order correlations that contain them to avoid double counting. However, this procedure can underestimate probabilities due to over-subtracting when the detection event data exhibits additional correlations that are not captured in the initial ansatz \cite{chen2022calibrated}.  This is most common in one- and two-body correlations, which may be contained in many three- or four-body correlations.  To account for this, we enforce a simple averaged depolarizing $T_1/T_2$ noise probability floor for one-body and spacelike two-body correlations to avoid unphysically small (or even negative) probabilities.

Because we use device-level data to calibrate the decoder, we must be careful not to fit the parameters of the decoder to the specific data that must be decoded.  This can happen e.g. when optimizing minimum-weight perfect matching edge weights using gradient descent with the logical error per cycle as an objective function \cite{kelly2015state}.  To avoid this issue, we decode an even subset of experimental trials by computing $p_{ij}$ on the odd subset, and vice versa, similar to \cite{kelly2015state}, and then average the two.  This sidesteps the potential concern of having computed $p_{ij}$ on the same data set as decoding (although in practice, we observe a negligible advantage when decoding with $p_{ij}$ computed from the decoded data set directly, as there is no optimization step). Without perfect statistics, this leads to a mild interdependence between the two individual decoding problems.

We validate the assumption that the interdependence is negligible by decoding half of a single experiment via this averaging method, and then decoding it again with $p_{ij}$ computed from another quarter of the data independent from the half we decode.  This ensures that each $p_{ij}$ is given the same amount of statistics, so that the only additional difference between the two decoding pathways are statistical fluctuations.  We perform an abbreviated (due to the computational cost of the tensor network) distance-3 decoding experiment using 5, 9, and 13 rounds with this method.  With the tensor network decoder and belief-matching decoder, averaged over round, we observe a \emph{relative} change of average logical error probability by $0.2\%$ and $0.1\%$, respectively, when using the independently computed $p_{ij}$ compared to averaging.  These relatively weak dependencies of performance on the precise edge weights (rather than their general features) provide evidence that any interdependence between the two data sets due to statistical fluctuation is small.  

For real-time decoding, it is also important that a decoder is robust to imperfections in the prior distribution e.g. caused by device drift.  Although the tensor network decoder will likely be too slow to keep up with the throughput of a surface code processor, belief-matching is a promising candidate for building such a real-time decoder.  To test belief-matching's sensitivity to an imperfect prior, we additionally use it to decode the experimental data with $p_{ij}$ computed from earlier data generated by the device.  We observe a very small increase in logical error per cycle from $3.118\%$ to $3.129\%$ for distance-3 and $3.056\%$ to $3.059\%$ for distance-5.  This robustness reinforces belief-matching as a promising candidate for real-time decoding.

\subsection{Correlated minimum-weight perfect matching}
For simulations involving larger scans through parameter space (e.g. for error budgeting), we employ a minimum-weight perfect matching decoder that uses a variant of the two-pass correlation strategy detailed in \cite{fowler2013optimal}.  Because the time cost of this decoder is essentially twice the time cost of minimum-weight perfect matching, we can leverage a fast custom matching engine to perform these larger scans quickly.  Matching is also used to decode the repetition code, which does not require a two-pass strategy.

\subsection{Belief-matching}

The minimum-weight perfect matching (MWPM) decoder often used for surface codes is efficient, but only considers edge-like fault mechanisms in the error hypergraph, ignoring hyperedge fault mechanisms that include more than two detectors~\cite{dennis2002topological}.
Like the correlated MWPM decoder, the belief-matching decoder exploits information about hyperedge fault mechanisms, but in a different way - by combining belief propagation (BP) with MWPM~\cite{higgott2022fragile, criger2018multi}.  In terms of efficiency, it has the same average and worst-case asymptotic running time as the conventional MWPM decoder~\cite{fowler2012towards}. 

\begin{figure}
    \centering
    \includegraphics[width=\linewidth]{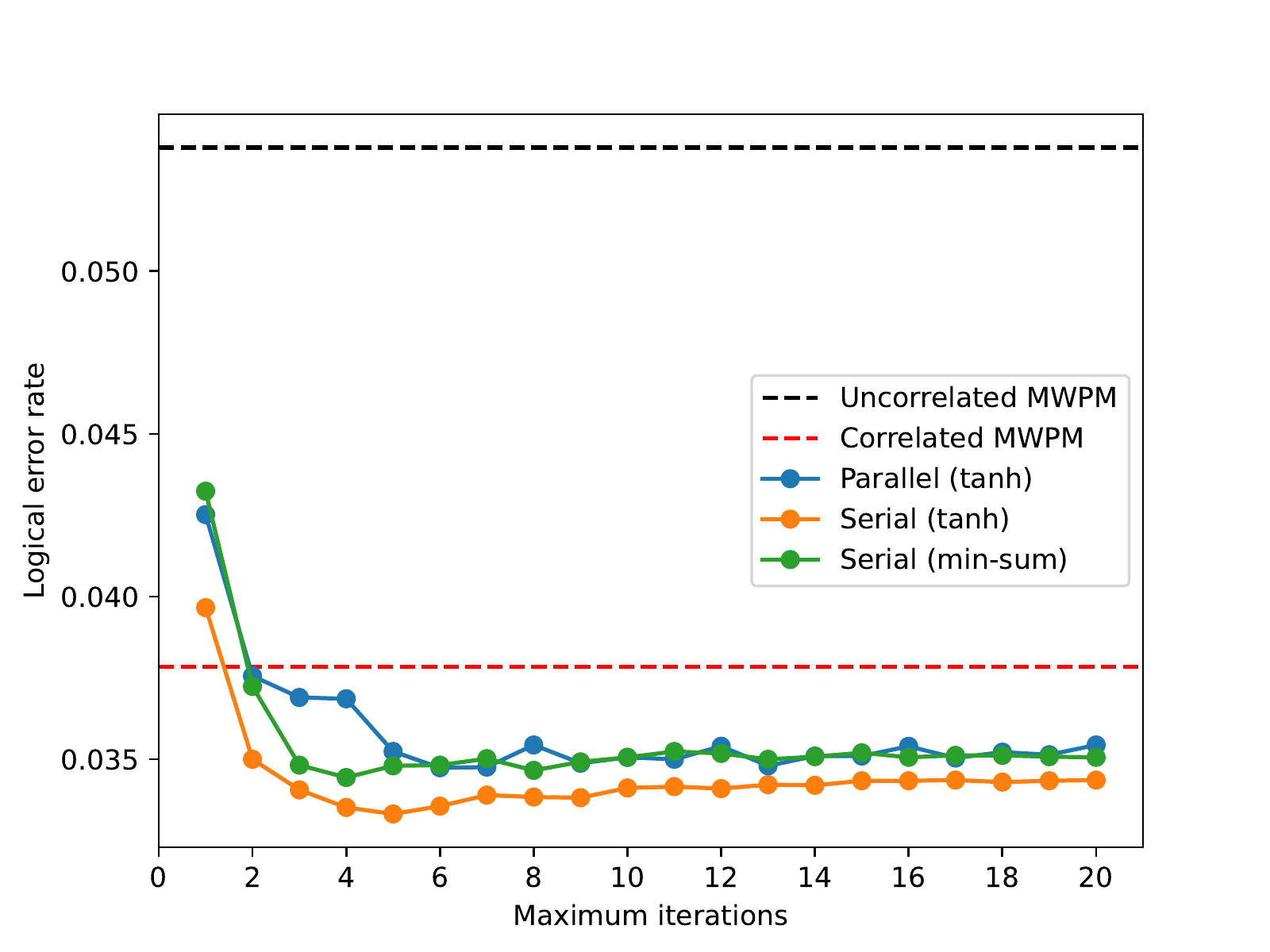}
    \caption{Comparison of different variants of BP used as a subroutine in the belief-matching decoder for a distance-5 surface code with simulated circuit-level depolarising noise over 5 rounds of syndrome measurement. We compare a parallel and serial schedule using the tanh update rule, as well as a serial schedule using the min-sum approximation of BP with a scaling factor of $
    0.7$. The $x$-axis shows the maximum number of iterations used for BP. The black and red dashed lines show the performance of an uncorrelated~\cite{dennis2002topological} and correlated~\cite{fowler2013optimal} MWPM decoder, respectively. All decoders were run on the same simulated data set of 50,000 shots.}
    \label{fig:belief_matching_serial_vs_parallel_schedule}
\end{figure}

In the first stage of belief-matching, BP is used to estimate the posterior marginal probability that each hyperedge fault mechanism has occurred, based on the prior probability of each hyperedge in the error hypergraph, as well as the set of detection events that have occurred in the experiment.
In the second stage, the error hypergraph is decomposed into two error graphs, using the BP posterior marginal probabilities to set the edge weights.
These two error graphs are then decoded using a MWPM decoder~\cite{dennis2002topological}.
We refer the reader to Ref.~\cite{higgott2022fragile} for a more detailed description.

The main difference between our implementation of belief-matching and that of Ref.~\cite{higgott2022fragile} is that we use a serial (rather than a parallel) schedule for BP and a maximum of 5 iterations.
This choice of parameters is motivated by our analysis of different variants of BP in belief-matching for simulated surface code data, shown in \ref{fig:belief_matching_serial_vs_parallel_schedule}.
The serial schedule increases the rate of convergence of BP by roughly a factor of two~\cite{zhang2005shuffled,Panteleev2021degeneratequantum,kuo2020refined}, and we also find that it improves the accuracy of belief-matching relative to using a parallel schedule, which we attribute to the serial schedule mitigating against the problem of split-beliefs caused by quantum degeneracy~\cite{poulin2008iterative}. 
We find that using more than 5 iterations for BP did not improve the accuracy of belief-matching for our simulations, and increasing the number of iterations can even degrade performance.
We expect this is due to short loops in the Tanner graph bounding information spread beyond some local region~\cite{higgott2022improved} and leading to unreasonably confident log-likelihood ratios when the number of iterations is increased beyond the length of these short loops.
Although we use the standard ``tanh'' rule for BP in our belief-matching implementation for the experiment, in \ref{fig:belief_matching_serial_vs_parallel_schedule} we also show the performance of belief-matching when using the min-sum approximation of BP.
The min-sum algorithm is an approximation of BP that is more efficient to implement in hardware.
Despite its relative simplicity, we find that the accuracy of the min-sum algorithm with a serial schedule is only slightly worse than that of BP using the tanh rule with the same serial schedule, and is comparable to that of the tanh rule with a parallel schedule.
A review of the tanh rule and min-sum algorithm can be found in Ref.~\cite{chen2005reduced}.

Practically, this decoder holds promise for scaling to the $\sim1$  $\mu$s per round throughput required by a real-time superconducting quantum computer.
In addition to using the min-sum approximation and a small number of iterations, the BP subroutine can be made even faster through the use of parallelisation.
Furthermore, by using belief-find, which uses weighted union-find~\cite{huang2020fault,Delfosse2021almostlineartime} instead of MWPM for post-processing, further speedups might be achieved with only a very small reduction in accuracy~\cite{higgott2022fragile}.

\subsection{Tensor network decoding}

We implement a close-to-optimal decoder mapping the prior distribution defined in Sec.~\ref{sec:prior} to a tensor network.
Given a configuration of detection events and a choice of logical frame change, the contraction of this tensor network estimates its probability.
It does so by summing the probabilities of all error configurations compatible with the set of detection events and logical frame change considered.
This decoder guesses the more plausible logical frame change by comparing the likelihood of both outcomes.
Unlike similar previous approaches~\cite{BSV2014, chubb2021statistical, higgott2022fragile}, our protocol takes as input device level noise specified by $p_{ij}$ correlations rather than gate-level noise.

\begin{figure}[tbp]
    \centering
    \includegraphics[width=\linewidth]{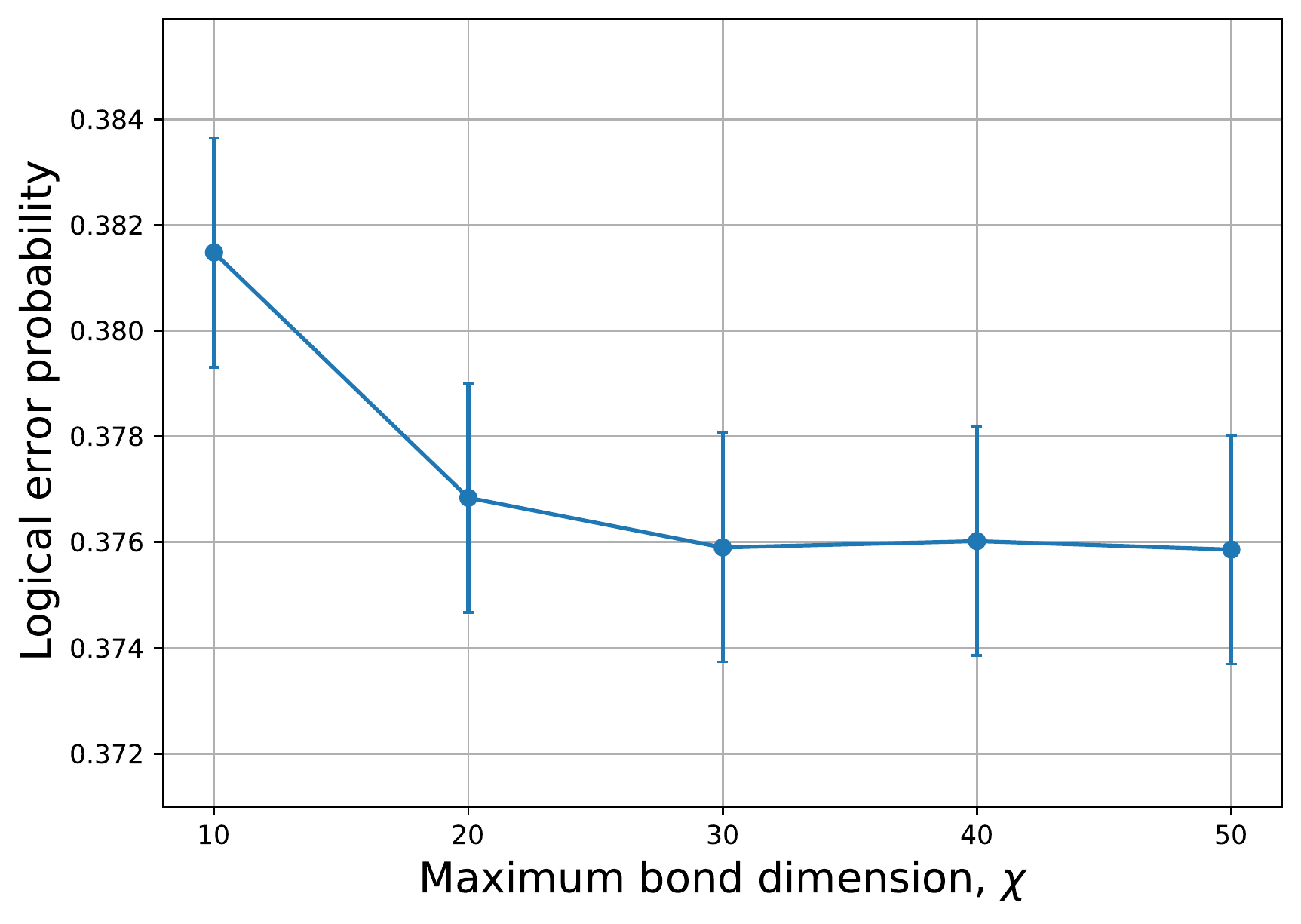}
    \caption{Logical error probability of a distance-5, $Z$-basis, 25-round experiment with 50,000 shots as a function of the maximum bond dimension $\chi$ used in the tensor network contraction.
    Error bars denote standard error of the mean.
    We observe that the logical error probability stabilizes at $\chi=30$.}
    \label{fig:lep_vs_bd}
\end{figure}

The contraction complexity of the resulting tensor network grows exponentially in $d^2$, where $d$ is the distance of the code.
In practice, we contract it approximately as a matrix product state evolution with a finite maximum bond dimension, $\chi$.
The complexity of this approximate contraction grows with $\chi^3$.
In order to guarantee that the value of $\chi$ used achieves convergence, we study the logical error probability of a distance-5 $Z$-basis experiment over 25 rounds - the largest experiment run.
We show in Fig.~\ref{fig:lep_vs_bd} that the logical error probability already converges at $\chi = 30$, which we use to decode.
Our implementation of this decoder uses the tensor network library quimb as a backend~\cite{gray2018quimb}.

\section{Fitting error versus rounds to extract logical error per cycle}

\begin{figure*}[htbp]
    \centering
    \includegraphics[width=7.2in]{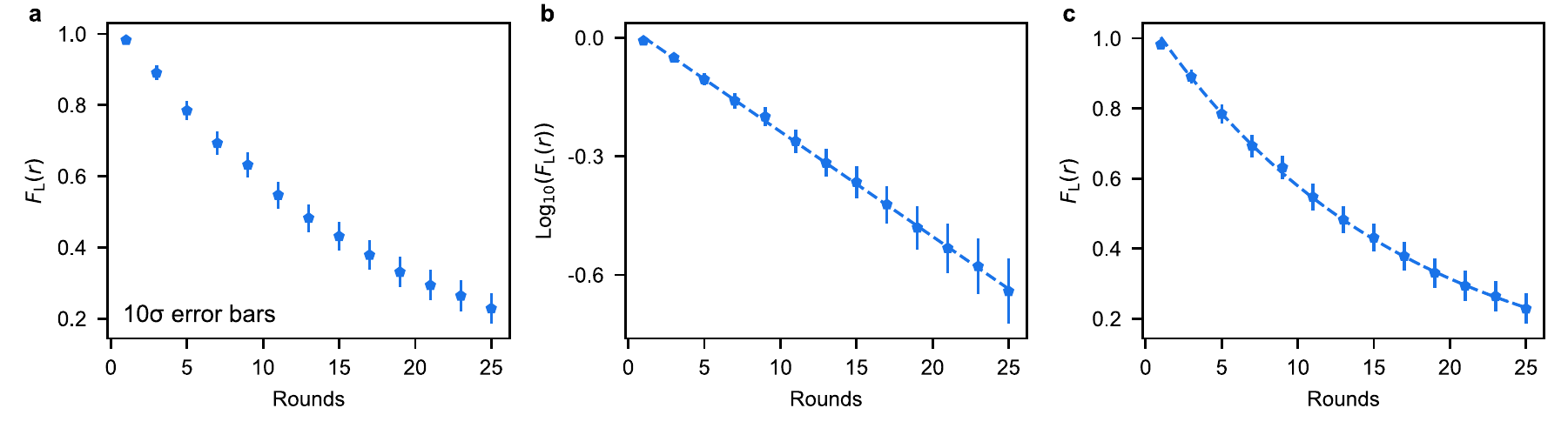}
    \caption{A figure demonstrating our fitting procedure on the Z-basis distance-5 data from the run presented in the main text, decoded with the tensor network decoder. 
    \textbf{a,} The logical fidelity data as a function of rounds, with error bars pulled from the binomial distribution (in this figure, all error bars shown at $10\sigma$ to help with visibility.) 
    \textbf{b,} The same data after a base-10 logarithm has been applied to it, with the error bars transformed appropriately, along with a two-parameter linear fit. \textbf{c,} The data transformed back into the original coordinates, along with the transformed fit.}
    \label{fig:fitting}
\end{figure*}
There are many ways to fit the data for error versus rounds to extract the logical error per cycle, which can yield slightly different results. In this section we lay out the procedure followed for this work.

We start with logical fidelities $F_\text{L}(r) = 1 - 2 P_\text{err}(r)$, discarding the first round as it decreases the quality of our fits due to the unique nature of its errors, which disproportionately favor distance-5.
Our goal is to find some logical error per cycle $\varepsilon$ such that
\begin{align}
\begin{split}
    P_{\text{err}}(r+1)
    &=
    (1 - \varepsilon)P_{\text{err}}(r)
    + \varepsilon\big(1 - P_{\text{err}}(r)\big)
    \\
    &=
    \varepsilon + (1 - 2\varepsilon)P_{\text{err}}(r)
\end{split}
\end{align}
which corresponds to exponential decay in the fidelity:
\begin{align}
    F_\text{L}(r+1)
    &=
    (1 - 2\varepsilon)F_\text{L}(r).
\end{align}
To do this, we first take our $F_\text{L}(r)$ data and append error bars from a binomial distribution, meaning that the variance at each point is given by
\begin{align}
    \label{eq:binomial-variance}
    \sigma^2
    &=
    \frac{P_{\text{err}}(r)\big(1 - P_{\text{err}}(r)\big)}{N}.
\end{align}
We then take the base-10 logarithm of each of these points, transforming the error bars as appropriate, before fitting to simple two-parameter linear fit using least squares, weighting the squared errors by the transformed variances to approximate a maximum-likelihood estimate. We then exponentiate this linear fit in order to get the fit in the original coordinates. In~\ref{fig:fitting} we show this process applied to the Z-basis distance-5 dataset presented in the main text. When combining data from multiple experiments, such as when considering data from multiple logical bases, this process is done on each dataset separately, and the final $\varepsilon$'s are averaged.

We additionally quantify out-of-model errors (such as leakage and device drift) by comparing the residuals of the fit to the residuals we would expect from binomial sampling errors alone. As the residuals are roughly three times larger than what we expect from sampling noise alone, we scale up the uncertainty calculated for sampling noise by this factor to account for the additional variance in our estimates from these out-of-model factors. As a final conservative measure, we upper bound our uncertainty by combining uncertainties from individual fits (from different initial-state bases and different distance-3 configurations) via averaging, which accounts for the possibility that these out-of-model errors might be correlated across different subsets of experiments.

\begin{figure}[htbp]
    \centering
    \includegraphics[width=\columnwidth]{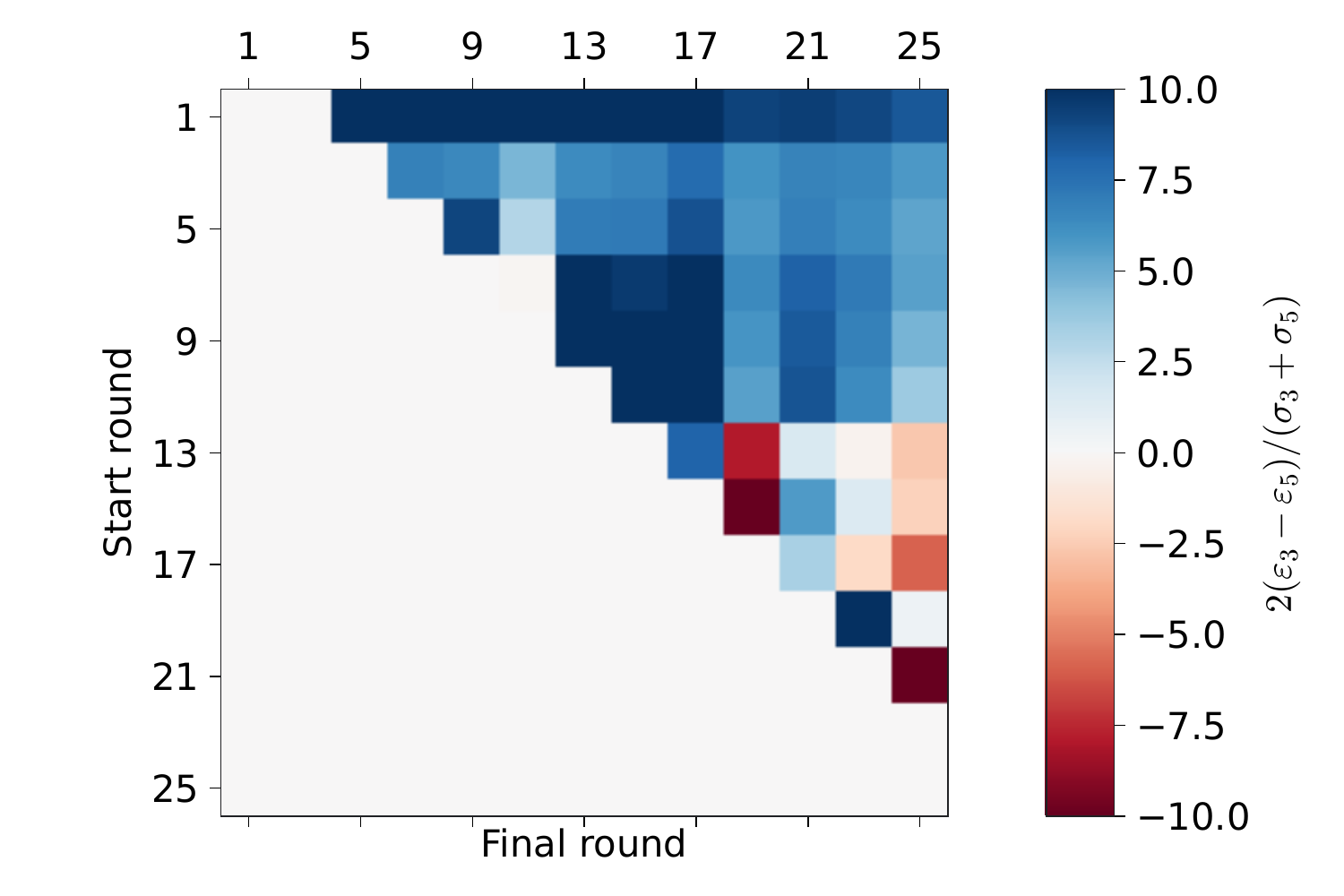}
    \caption{Results of fitting to an extended model with fluctuating $\varepsilon$. Fits are performed for all subintervals of rounds containing at least three points, and the result of the fit is presented as the difference in logical error per cycle between distance-3 and distance-5 measured in the average of the uncertainties of logical error per cycle for each distance given the extended model. Fits starting from round 1 clearly favor distance-5 disproportionately, and are not used for our quoted results. Fits starting from rounds 3--11 show consistent significant separation between the estimates for logical error per cycle. The fit from rounds 3--25, the range used for our quoted logical error per cycles, shows a separation of $5.7\sigma$ between distance-3 and distance-5, which is consistent with the $5.8\sigma$ separation we get from our original fits when scaling our error bars by the excess residual factor.}
    \label{fig:extended-model-fits}
\end{figure}

As a check that this can appropriately account for additional errors, we also fit an extended model in which the residuals in our fits come primarily from fluctuations in the logical error per cycle $\varepsilon$ itself instead of binomial sampling noise. This model results in near identical separation between the logical error per cycle for distance-3 and distance-5, as measured in standard deviations of the estimate. The results of this fit are illustrated in~\ref{fig:extended-model-fits} for all potential subintervals, demonstrating the robustness of our findings to out-of-model errors and fitting choices.

\section{Simulation of logical memory experiment}
\label{sec:simulation_Dvir}
We consider multiple simulation strategies for the the logical memory experiment. The simplest approach, labeled "Pauli" in the main text, is implemented as a standard Pauli frame simulation. This simulation is based on the \emph{Stim} open source library\cite{gidney2021stim}. The associated error model corresponds to adding one and two-qubit depolarizing channels each device operation, matched to experimentally characterized fidelities.

A more sophisticated simulation approach is labeled "Pauli$+$". This is still a classical simulation in the sense that the cost grows polynomially in the number of qubits, though its performance is not as heavily optimized. The advantage of this simulation is that it explicitly accounts for correlated errors such as qubit leakage and parallel gate crosstalk. 

Finally, we do a "brute force" quantum simulation of the distance-3 experiment. This is a quantum trajectories simulation \cite{tomita_low-distance_2014, isakov2021simulations}, meaning that each experimental sample corresponds to propagating the noisy evolution of the full quantum state vector. Unlike the previous approaches, the quantum simulation is "exact" in the sense that all noise is modeled using explicit quantum channels derived from physical considerations. The previous approaches must represent noise using Pauli channels, which can be derived from the more general quantum channels through approximations (detailed below). While the quantum simulation is not scalable to large code distances, we can use it to validate the approximations required for the Pauli$+$ simulator. 

Below we outline the general procedure to simulate the logical memory experiment. 
\begin{enumerate}
    \item The "ideal" experiment is represented as a noiseless quantum circuit. Each layer of gate, idle, measure, or reset Operations is encoded as a \emph{Circuit} object of the Cirq open source library \cite{isakov2021simulations}. Simultaneous Operations are organized into Moments. 
    \item The "ideal" circuit is mapped to a "noisy" circuit. This corresponds to inserting additional moments into the circuit composed of operations representing noise. (The noise models used are described below.) Each noise operation is a quantum noise channel encoded using Kraus operators. Such a representation is sufficient for the quantum simulator. Both noisy circuit processing and quantum trajectories simulation are implemented using the kraus\_sim library (Section \ref{sec:kraus_sim}).
    \item Additional processing of the noise is required for the "Pauli$+$" simulator. As detailed in section \ref{sec:twirling}, we apply a generalization of the Pauli Twirling Approximation \cite{geller2013efficient} to construct Generalized Pauli Channels (GPC's) capable of including leakage. 
    \item The chosen simulator generates many measurement samples of the experiment represented by the noisy circuit. The initial data qubit states are matched to those used in experiment, and the results are then processed in the same way as experimental measurement data. 
\end{enumerate}
In the following sections, we discuss details of the simulated noise models as well as the Pauli$+$ simulator and the Generalized Pauli Twirling Approximation. We conclude with a comparison of results between the "Pauli$+$" and quantum simulations for a distance-3 experiment.

\subsection{Noise models}
\label{sec:noise}
Below we give a brief summary of all noise models included in our simulations. These models are accounted for using Kraus operators, which define quantum channels that are applied to each qubit after their respective gate or idling operations. In the Pauli$+$ simulation, these channels are first converted into corresponding Generalized Pauli Channels (GPC's) so that the noisy experiment is amenable to efficient simulation. 

\subsubsection{Decay, dephasing, and leakage heating}
\label{sec:decoherence}
We consider the effects of qubit decay, white noise dephasing, and passive leakage to state $\ket{2}$ for an idling transmon qubit. We assume decay into a large, zero temperature environment with characteristic decay time $T_1$. Each transmon (both measure and data qubits) couples to the environment through its charge operator $q$. This induces a decay (in the energy eigenbasis) through the $\ketbradt{j}{j+1}$ elements of the charge operator, which we assume scale as $q_{j,j+1}\propto\sqrt{j+1}$.
Fermi's Golden rule predicts a $j+1\rightarrow j$ decay rate proportional
to $|q_{j+1,j}|^{2}\propto(j+1)$ , so we expect the state $\ket{2}$
to decay twice as fast as state $\ket{1}$. Similarly, one can derive
dephasing as arising from a weak measurement of the number operator
$n$, with information leaking at a rate $1/T_{\phi}$. Finally, we include a phenomenological leakage heating rate $\Gamma_{1\rightarrow2}$ described by the Linbladian operator $\sqrt{\Gamma_{1\rightarrow2}} \ketbradt{2}{1}$. The dynamics of an idle transmon in the presence of these processes is described through the master equation,
\begin{eqnarray}
    \label{eq:deco_lindblad}
            \partial_{t}\rho & = & -\frac{i}{\hbar}[H^{\text{idle}},\rho]+ \mathcal{L}[\rho]\nonumber\\
     &=&-\frac{i}{\hbar}[H^{\text{idle}},\rho]\nonumber\\
      & & +\frac{1}{T_{1}}\l(a\,\rho\:a^{\dagger}-\frac{1}{2}\{n,\rho\}\r)+\frac{2}{T_{\phi}}\l(n\rho\,n-\frac{1}{2}\{n^{2},\rho\}\r)\nonumber\\
    & & + \Gamma_{1\rightarrow2}\l(\ketbradt{2}{1}\rho \ketbradt{1}{2} - \frac{1}{2}\{\ketbrad{1},\rho\}\r),\\
    a  &=&\sum_{j\geq 0} \sqrt{j+1}\ketbradt{j}{j+1},\nonumber\\
    n  &=&\sum_{j\geq 0} j\ketbrad{j}=a^{\dagger}a.\nonumber
\end{eqnarray} 
The values for $T_1$ and $T_{\phi}$ used in our simulations are both qubit specific. They are extracted from experimental decay and CPMG data. For $\Gamma_{1\rightarrow2}$ we use a homogeneous value of $1 / (700 \mu \text{s})$, which is in agreement with typical numbers for both our current device and past experiments\cite{chen_measuring_2016}.

We use the dissipative dynamics in (\ref{eq:deco_lindblad}) to extract corresponding discrete quantum noise channels. These are applied on each qubit following every idle or gate operation. For an operation of duration $t$, evolution under the master equation can be approximately decomposed as $\mathcal{E}\circ \hat U_0$, where $\hat U_0$ represents the ideal unitary evolution and
\begin{equation}
    \mathcal{E} = \exp\l(t \mathcal{L}\r)
\end{equation}
represents the effect of decoherence. We solve for $\mathcal{E}$ by representing $\mathcal{L}$ as a linear operator and doing the direct matrix exponentiation; this ignores a correction due to weak non-linearity of the spectrum of $H^{\rm idle}$. From this we follow the standard prescription to extract the Kraus operators of $\mathcal{E}$\cite{wilde_2013_quantum}.

\subsubsection{Readout and reset error}
\label{sec:readout}
Readout error is assumed to be a classical process. Accordingly, our simulations in fact implement perfect readout in the standard basis (including leaked states $\ket{2}$ and $\ket{3}$) and errors are added using a Markov transition matrix in post-processing. To populate this matrix we use the calibrated readout fidelities in the device. Measure qubit fidelities are calibrated during parallel measurement of all measure qubits, while for data qubits we use fidelities calibrated during parallel measurement of \emph{all} qubits.  Since the experiment does not actually implement readout that distinguishes the leaked states, we assume that both state $\ket{2}$ and $\ket{3}$ are read out as $\ket{0}$ or $\ket{1}$ with equal probability. Additionally, we account for measure qubit reset error by adding this probability to the average readout error of the measure qubits. This is justified as a reset error on a measure qubit at the beginning of an error correction round has the same effect as a measurement error at the end of the round.

\subsubsection{Leakage}
\label{sec:leakage}
Leakage outside the computational subspace is caused by three mechanisms included in our models.
\begin{enumerate}
    \item Heating: Passive heating from state $\ket{1}$ to $\ket{2}$, as included in the decoherence model above.
    \item CZ gate dephasing: Since our CZ gates are diabatic (implemented as two complete Rabi swaps) dephasing processes may cause transitions between states $\ket{11}$ and $\ket{02}$. To model this process, after each CZ gate we include a channel with Kraus operators
    \begin{eqnarray}
    K_1&=& \sqrt{p_t}\l(\ketbradt{02}{11} + \ketbradt{11}{02}\r)\nonumber\\
    K_0&=& \sqrt{I-K_1^\dagger K_1}\,.
    \end{eqnarray}
    The value $p_t=8\times 10^{-4}$ used matches the median value observed in characterization of isolated CZ gates. The qubit ordering above matches the $f_{10}$ transition frequencies of the qubits, so that only the higher frequency qubit leaks \cite{chen2021exponential}.
    \item CZ crosstalk: As discussed below, coupling crosstalk between independent CZ pairs or idling qubits may lead to leakage transitions.
\end{enumerate}

Along with affecting measurement outcomes, we also account for the effect of leakage on our CZ gates. This is captured by two processes:
\begin{enumerate}
    \item Controlled phases: When one of two qubits is in a leaked state $\ket{j}$ during the CZ gate, the applied Z phase on the other, $\phi_j$, will typically not be $0$ or $\pi$. When the non-leaked qubit is a measure qubit, this will have a randomizing effect on the observed stabilizer measurement \cite{battistel2021hardware}. Under the twirling approximation, this effect is mapped to a $\sigma_z$ Pauli channel applied conditionally on the non-leaked qubit. 
    \item Higher level transitions: The physical implementation of the diabatic CZ gate leads to resonances between higher energy states (e.g. $\ket{22}$ and $\ket{13}$). Accordingly, this leads to coherent transitions between these states and effective transport of leakage between measure and data qubits\cite{mcewen_removing_2021}.
\end{enumerate}
In order to accurately capture the above processes, we numerically calibrated then simulated the CZ gate matching the experimental qubit and coupler control schedules. The observed values for $\phi_j$, as well as the transition amplitudes (specifically between states $\ket{22}\Longleftrightarrow\ket{13}$ and $\ket{21}\leftrightarrow\ket{03}$) were extracted. 

\subsubsection{Crosstalk} 
\label{sec:x-talk}

We model coupling crosstalk between qubits undergoing parallel CZ or idling operations. This crosstalk error is caused by two effects. First, the \emph{inactive} couplers (i.e those not implementing a CZ gate) used to mediate the qubit-qubit couplings at idle are unable to cancel all effective couplings over all fixed excitation subspaces\cite{youngkyu_coupler_2021}. Second, diagonal capacitive couplings (e.g., between next-nearest-neighbor qubits) are also present throughout the device, with $\ket{01}\leftrightarrow\ket{10}$ swap rates on the order of $200 $ kHz. To account for both of these effects, we consider all pairs of simultaneous gates (either single qubit idle or two-qubit CZ). For each pair of gates containing a diagonal or nearest neighbor, we simulate the parallel operation of both gates in the lab frame (while including the undesired couplings above). The full 3- or 4- qubit unitary $U_{lab}$ is then re-expressed in the "interaction picture"
\begin{equation}
\label{eq:xtalk_unitary}
    U_{lab} U_0 ^\dagger = U_I 
\end{equation}
where $U_0$ represents the "ideal" evolution in the absence of the unwanted couplings\footnote{Note that the ordering of operations is swapped compared to the typical interaction picture definition $U_{lab} = U_0 U_I$. This is to retain consistency with the other noise models, for which we append error channels after (instead of before) each ideal gate operation.}. (This ideal evolution can be computed as a tensor product of unitaries acting on the two subsystems involved in each gate.) Note that the simulations are carried out using 4-level qubits, so that leakage-related effects are explicitly included. Details of these simulations are deferred to a later work. The $U_I$ unitaries for all pairs of neighboring gates are included as error channels in representation of the noisy circuit. 

\subsubsection{Depolarizing errors}
\label{sec:depolarizing}
While the models above account for a majority of errors in the experiment, they typically under-predict the experimentally calibrated decay of fidelity. To compensate for this missing error, we append additional depolarizing errors after each gate. In all cases, the fidelity decay curves are converted to an inferred Pauli error. This error is compared to the sum of Pauli errors accumulated over the error models described above. The effect of leakage transitions on the Pauli error are explicitly accounted for as in ref.~\cite{pedersen2007fidelity}. Single qubit gates are characterized using parallel randomized benchmarking\cite{emerson2005scalable}, while CZ gates are characterized with parallel cross-entropy benchmarking (XEB)\cite{arute2019quantum}. Additionally, the data qubit idles (during measure qubit measurement and reset) are characterized using interleaved single qubit XEB.

\subsection{Simulator details}
\subsubsection{kraus\_sim - noisy circuit compiler and quantum simulator }
\label{sec:kraus_sim}
The kraus\_sim library serves as an intermediate layer between the Cirq (used to represent the circuit) and \emph{qsim} (used for high-performance qubit state evolution)\cite{arute2019quantum,quantum_ai_team_and_collaborators_2020_4023103} libraries. At the interface to Cirq, it processes the ideal circuit to generate the noisy circuit (as described in Section \ref{sec:noise}). These noisy circuits are then consumed by kraus\_sim's quantum trajectories simulator. A single quantum trajectory is equivalent to consecutively sampling the Kraus operators of each applied quantum channel using Born's rule for POVMs\cite{wilde_2013_quantum}, then applying the Kraus operator to the state vector and normalizing the result\cite{isakov2021simulations}. Below, we highlight unique features of the library, which necessitate a custom simulator backend based on the the \emph{qsim} library.

The kraus\_sim analog of Cirq's Operation is the KrausOperation. It is used to encode all quantum channels (both gates and errors) applied to the system. These channels are encoded as Kraus operators, each of which is represented as a multi-dimensional tensor. While typical Kraus operators are "square" matrices, KrausOperations allow these tensors to have an arbitrary shape. This means that channels can be defined which "create" or "destroy" degrees of freedom in the system. Accordingly, in many cases it is possible to reorder the application of commuting KrausOperations to greatly speed up calculations by decreasing the size of the state vector. Specifically, the structure of the surface code and bit flip code circuits means that operations can be ordered so that a "destructive" measurement of a single measure qubit is following by a "creative" reset of another measure qubit\cite{obrien_density-matrix_2017}. Thus, only a single measure qubit ever needs to be represented in memory. This allows for noisy simulations of up to a distance-5 surface code (25 data + 24 measure qubits) while keeping at most 25+1 qubits in memory. However, this technique is not possible in the presence of 3- and 4-local crosstalk errors, since if two measure qubits are affected by such error they must both exist concurrently in the simulation. We therefore focus only on distance-3 simulations (requiring only 9+8=17 qubits in memory).

kraus\_sim also implements functionality to represent classical degrees of freedom (registers), which are updated using stochastic matrices. KrausOperations can be conditioned on the values of these registers, which allows the registers to affect the quantum state evolution. Analogously, the index of the sample Kraus operator for a given KrausOperation can be used to apply conditional stochastic matrices to classical registers. This is relevant for simulations of leakage, as in our models we approximate all leakage transitions as incoherent. This means that no superposition is ever formed between the computational and leakage subspaces of any qubit, and we therefore are never required to keep more than two levels per qubit in our state vector. Instead we use classical registers to track the leakage status of each qubit, applying only the parts of our Kraus operators that act in the current computational subspace\footnote{Details of this approximation are deferred to a future work.}.

\subsubsection{Pauli$+$ simulator implementation}

Pauli$+$ carries out a classical simulation of the logical memory
experiment. The term ``classical'' here denotes the fact that the
simulation cost scales quadratically with the
number of qubits and linearly with the number of error correction
rounds, so it is amenable to classical computers. For a system of qubits, polynomial scaling of simulation cost for an
arbitrary Clifford circuit with Pauli-channel errors is a consequence
of the Gottesmann-Knill theorem \cite{Gottesman-Knill-1999}.  Related
methods have long been used for accuracy threshold calculations \cite{Steane-2003,Fowler-QEC-2004,raussendorf2007fault, Wang-Fowler-Stephens-Hollenberg-2010,Wang-Fowler-Hollenberg-2011,  Landahl-2011,Brown-Nickerson-Browne-2016,Tuckett-Bartlett-Flammia-2018}
and non-Pauli errors \cite{Gutierrez-Svec-Vargo-Brown-2013},
and even for simulating circuits with a small fraction of non-Clifford
gates\cite{Bravyi-Gosset-2016,Bravyi-etal-sim-2019}.

In the presence of leakage, we achieve such a scaling in a
Markov-chain simulation by using an approximate Generalized Pauli Channel (GPC)
error model similar to that used in Pauli-frame
simulations by Fowler\cite{fowler2013coping}. In our
case, to ensure correct correlations between measurement events in an
arbitrary Clifford circuit, the state of the system is described by a
vector of device leakage labels and an $m$-qubit stabilizer state for
the devices in the computational (qubit) subspace.

The noisy circuit is modified for Pauli$+$ simulations by decomposing each noisy gate as a product of an ideal Clifford gate and a sequence of residual error channels. The error channels are then approximated as a GPC.  As a result, the entire simulation consists of four operation types: 
\begin{enumerate}
\item A unitary that acts as a perfect Clifford gate in the
  computational subspace and identity on its complement.
\item A perfect measurement (all leakage levels resolved) in the
  computational basis.
\item A perfect reset to a single-device $\ket0$ state in the
  computational basis.
\item An error channel, represented as a Generalized Pauli
  (GP)  channel.
\end{enumerate}
(As noted previously, classical readout error is handled through post-processing of measurement data). We describe how each of these operations are handled by the simulator below.

The full Hilbert space of the $n$-device system is approximated as a
$q^n$-dimensional space, where each device has $q=4$ energy levels,
with $\ket0$ and $\ket1$ forming the computational subspace.  Further,
because the corresponding phases are not being tracked, the states
$\ket2$ and $\ket3$ are subject to rapid dephasing.  As a result, the
full Hilbert space is separated into mutually incoherent leakage
subspaces given by Kronecker products,
\begin{equation}
{\cal H}_{\bar a}={\cal H}_{a[1]}\otimes {\cal H}_{a[2]}\otimes \ldots
\otimes {\cal H}_{a[n]}, \label{eq:hilbert-space-decomp}
\end{equation}
where the local subspace ${\cal H}_{a[j]}$ of the $j$\,th device with
leakage state label $a[j]\in\{c,2,3\}$ is selected from
\begin{equation} {\cal H}_c = \sp\{\ket0 , \ket 1\},\quad {\cal H}_2 =
  \sp\{\ket2\},\quad {\cal H}_3 = \sp\{\ket3\}.
  \label{eq:leakage-decomposition}
\end{equation}
Each GP channel acting on $k$ devices in the index set
$J=\{j_1,j_2,\ldots, j_k\}$ is represented by a collection of possible transitions between leakage subspaces: 
\begin{equation}
  \bar a[J]\equiv \left(a[j_1], a[j_2], \ldots , a[j_k]\right) \to
  \bar b[J]\equiv \left(b[j_1], b[j_2], \ldots , b[j_k]\right).
  \label{eq:space-decomp}
\end{equation}
For each transition $\bar a \rightarrow \bar b$, the set of devices in $J$ are partitioned into four categories, depending on their initial and final leakage status. The devices which begin and \emph{remain} in the computational subspace are denoted
\begin{equation}
  R\equiv \{j\in J\mid a[j]=b[j]=c\}.\label{eq:qubits-remain}
\end{equation}
These devices will receive a standard Pauli channel. The devices that transition \emph{up} to leakage are denoted by
\begin{equation}
  U\equiv \{j\in J\mid a[j]=c, b[j]\neq c\}.\label{eq:up-leak}
\end{equation}
Leaked devices are not modified after the transition. Conversely, devices which transition \emph{down} to the computational subspace are labeled
\begin{equation}
  D\equiv \{j\in J\mid a[j]\neq b[j]= c\}.\label{eq:down-leak}
\end{equation}
We assume that such devices are fully disordered, meaning that they receive a maximally depolarizing Pauli channel.
Finally, the remaining devices are those which start and end \emph{leaked},
\begin{equation}
  L\equiv \{j\in J\mid a[j]\neq c, b[j]\neq c\}.\label{eq:stay-leak}
\end{equation}
Like the devices in $U$, no Pauli channel is applied for these devices. However, transitions between different leakage subspaces are allowed. As detailed below, approximating quantum error channels in terms of GP channels is premised on the assumption that the input is a
highly-entangled stabilizer state, that the output is eventually
projected to such a state, e.g., at the end of a measurement cycle, and that a ``large'' part of evolution during any gate is factored out as a Clifford rotation, while the remaining error
channel is in some sense close to identity.

Each transition (denoted by the initial and final subspace vectors
$\bar\imath$, $\bar f$ supported on $J$) is assigned a conditional
probability $P(\bar f\mid\bar \imath )\equiv P(\bar \imath \to \bar f)$.  Whenever the GP channel is applied to the system initially in subspace ${\cal H}_{\bar a}$, there is a unique subspace label $\bar\imath\equiv \bar a[J]$ which is updated according to the distribution $P( \bar f\mid\bar \imath )$.
Additionally, each transition is associated with a standard Pauli
channel acting on the qubits in $R$.  Initial state of the qubits in $U$ is traced over, while 
the final state for each qubit in $D$ is a randomly selected $\ket0$ or $\ket1$ with equal probability.  Overall, each GP channel acting on devices in
the index set $J$ is represented by compound data
\begin{equation}
  \left\{
    \left(P(\bar{\imath}\rightarrow\bar{f}),\,
      p(\sigma_{\bar{\mu}}|\bar{\imath}\rightarrow\bar{f})\right)\right\}_{\bar{\imath},\bar{f}},
  \label{eq:compound-data}
\end{equation}
where the Pauli operators $\bar{\sigma}_{\mu}$ act only on the qubits
in the set $R$, see Eq.~(\ref{eq:qubits-remain}).

The simulator takes as input a quantum circuit (a sequential list of
operations of the four types above) and returns corresponding
experimental measurement samples.  The simulator is implemented as a
Markov chain whose internal states are specified by the leakage
space label vector $\bar a$ and (with $m=m_{\bar a}\le n$ devices in
the computational subspace) an $m$-qubit stabilizer state specified in
terms of $m$ Hermitian generators of the corresponding stabilizer
group ${\cal S}=\langle \hat G_1,\ldots,\hat G_m\rangle$, an abelian
subgroup of the $m$-qubit Pauli group ${\cal P}_m$ which does not include $-\hat I$.  Notice that it is
the stabilizer group but not the particular generators that represent
the state, thus at any point the set of generators can be changed by an arbitrary sequence of \emph{row transformations} $\hat G_i\to \hat G_i \hat G_j$, $j\neq i$.

To collect a single sample, the simulation begins by initializing all $n$
components of the leakage state label $\bar a$ to $c$ and the $m=n$ qubits in
the product state $\ket{00\ldots 0}$, by selecting the stabilizer
generators $\hat G_i=\sigma_z^{(i)}$, $i=1,2,\ldots, m$, where a
single-qubit state $\ket0$ is a $+1$ eigenstate of $\sigma_z$.
Subsequently, the simulator iterates over the operations in the
circuit, at each step updating the stabilizer generators and possibly
the leakage state label $\bar a$:
\begin{enumerate}
\item \texttt{clifford}: a Clifford unitary $\hat U$ acting on devices in
  the index set $J\subset \{1,2,\ldots,n\}$.
  \begin{enumerate}[label=(\roman*)]
  \item if any of the devices in $J$ are outside of the computational
    state, do nothing;
  \item otherwise, replace each stabilizer generator $\hat G_i$ with a
    Pauli operator $\hat G_i'=\hat U\hat G_i \hat U^\dagger$.
  \end{enumerate}
\item \texttt{measure}: an ideal computational-basis measurement of qubit $j$.
    \begin{enumerate}[label=(\roman*)]
    \item If the device $j$ is leaked, \texttt{return} the leakage
      label $a[j]$.
    \item Otherwise, use row transformations to select a set of
      generators with (a) only $\hat G_1$ supported on the qubit $j$
      if it is unentangled, or (b) only $\hat G_1$ and $\hat G_2$
      supported on $j$, where $\hat G_1[j]=\sigma_x$ and
      $\hat G_2[j]=\sigma_z$.
    \item In case (a), if $\hat G_1[j]\in \{\sigma_x,\sigma_y\}$, or in
      case (b), select the result $\alpha\in \{0,1\}$ randomly with
      equal probabilities and replace the 1st generator with
      $\hat G_1'=(-1)^\alpha \sigma_z^{(j)}$; \texttt{return} $\alpha$.
    \item Otherwise, in case (a) with $\hat G_1[j]= \sigma_z$, do full
      row reduction of the generator matrix to separate the
      independent generator
      $\hat G_1'=\pm \sigma_z\equiv (-1)^\alpha\sigma_z$ supported
      only on the qubit $j$, \texttt{return} the sign bit $\alpha$.
    \end{enumerate}
  \item \texttt{reset}: an ideal reset of device $j$ to the state
    $\ket0$.
    \begin{enumerate}[label=(\roman*)]
    \item If the device $j$ is leaked, set the label $a[j]=c$ and add
      an extra generator $\hat G_{m+1}=\sigma_z^{(j)}$; increment $m$.
    \item Otherwise, do partial row reduction as in step 2(ii) above,
      and replace the first generator with
      $\hat G_1'=+\sigma_z^{(j)}$.
    \end{enumerate}
    
  \item \texttt{paulichannel}: Generalized Pauli  (GP) channel acting on devices in the index set
    $J$.
  \begin{enumerate}[label=(\roman*)]
  \item Construct the initial subspace vector $\bar\imath=\bar a[J]$
    and choose the final subspace vector $\bar f$ according to the
    probability distribution $P(\bar \imath\to \bar f)$.
  \item For this outcome, select the set of devices $U$ transitioning
    upward from the computational subspace.  For each $j\in U$:
    \begin{enumerate*}[label=({\bf\alph*})]
    \item do partial row reduction as in step 2(ii);
    \item drop the first generator $G_1$ (swap it with $G_m$ and decrement $m$);
    \item if the qubit $j$ was entangled with the rest of the system,
      replace $G_2[j]=\sigma_z$ with the identity operator;
    \item with probability $p=1/2$, flip the sign of $G_2$;
    \item set leakage label $a[j]$ according to its value in $\bar f$.
    \end{enumerate*}
  \item Select the Pauli error operator $\hat E=\bar \sigma_\mu$
  according to the conditional distribution
  $p(\bar \sigma_\mu|\bar \imath\to \bar f)$ supported on the set $R$
  of devices remaining in the computational subspace, and flip the
  signs of all the generators $G_i$ which anticommute with the error
  $E$, $G_i'\equiv EG_iE^\dagger =\pm G_i$.
\item For every device $j\in D$ coming down from leakage,
      \begin{enumerate*}[label=({\bf\alph*})]
      \item add a generator $G_{m+1}=\pm\sigma_z^{(j)}$ with a random sign;
      \item set $a[j]:=c$; and \item increment $m$.
      \end{enumerate*}
    \end{enumerate}
  \end{enumerate}

Once the last operation in the circuit is reached, a record of all measurement
results is kept for each sample.

The validity of operations described follow from general theory
of entanglement for stabilizer
states\cite{Fattal-Cubitt-Yamamoto-Bravyi-Chuang-2004}.

\subsubsection{Generalized twirling approximation}
\label{sec:twirling}

Given an arbitrary quantum channel $\mathcal{E}$ with Kraus operators
$K_{j}$ acting on devices $J$, the Generalized Pauli Twirling approximation
allows us to define a corresponding GP channel.  
This mapping corresponds to the following physical process: 
\begin{enumerate}
\item For a given initial subspace $\mathcal{H}_{\bar{\imath}}$, double the number of degrees of freedom and prepare
a maximally entangled state over the product
$\mathcal{H}_{\bar{\imath}}\otimes\mathcal{H}_{\bar{\imath}}$, a direct product of EPR pairs. Namely, the
devices whose index in $\bar{\imath}$ match the computational subspace and their matching ancillary pairs
should be set as the $+1$ eigenstates of $\sigma_{x}\otimes\sigma_{x}$
and $\sigma_{z}\otimes\sigma_{z}$.
\item Apply $\mathcal{E}$ to the first subsystem.
\item Measure the leakage status of each device in the first subsystem, thereby determining the final subspace $\bar{f}$. This can be represented by the single-device projective measurement with elements 
\begin{eqnarray*}
  P_{c} & =&\ketbrad{0}+\ketbrad{1}\\
  P_{2} & =&\ketbrad{2}\\
  P_{3} & =&\ketbrad{3}
 \end{eqnarray*}
\item For the devices that remained in the computational subspace ($R$),
measure the commuting stabilizers $\sigma_{x}\otimes\sigma_{x}$ and
$\sigma_{z}\otimes\sigma_{z}$ acting on them and their entangled
pair. Associate the resulting stabilizer eigenvalue pairs $(1,1)$, $(1,-1)$, $(-1,1)$, $(-1,-1)$
with the Pauli operators $I$, $\sigma_{x}$, $\sigma_{z}$, $\sigma_{y}$, respectively.
The total error operator $\sigma_{\bar{\mu}}$ over all devices in
$R$ is the tensor product of all single qubit errors.
\end{enumerate}
The conditional probability distribution $P(\bar{i}\rightarrow\bar{f},\bar{\mu})$
is then defined as the probability of recording the outcomes $\bar{f},\bar{\mu}$
assuming an initial state preparation in subspace $\mathcal{H}_{\bar{\imath}}\otimes\mathcal{H}_{\bar{\imath}}$.
Accordingly, the extracted GP channel will have the exact same statistics
under the physical process represented above.

The probabilities $P(\bar{i}\rightarrow\bar{f},\bar{\mu})$ can be
determined directly from the Kraus operators $K_{j}$. For a given
transition $\bar{\imath}\rightarrow\bar{f}$, Kraus operator $K_{j}$
decomposed as a sum over tensor products for the devices in $U,R,D$,
and $L$.
\begin{eqnarray*}
&&P_{\bar{f}}K_{j}P_{\bar{\imath}}=\\
&&\sum_{\bar{u},\bar{d}}\biggl(K_{j}^{\bar{u},\bar{d},\bar{\imath}\rightarrow\bar{f}}\biggr)_{R}\otimes\ketbradt{\bar{f}[U]}{\bar{u}}_{U}\otimes\ketbradt{\bar{d}}{\bar{\imath}[D]}_{D}\otimes\ketbradt{\bar{f}[L]}{\bar{\imath}[L]}_{L}
\end{eqnarray*}
where $\ket{\bar{u}},\ket{\bar{d}}$ sum over the tensor product of
standard (normalized) basis vectors ($\ket{0},\ket{1}$) representing
an orthonormal basis for the computational subspaces of devices $U$
and $D$, respectively. The vectors $\ket{\bar{f}[U]}$, $\ket{\bar{\imath}[D]}$,
and $(\ket{\bar{\imath}[L]},\ket{\bar{f}[L]})$ correspond to the
leakage states of the devices which leaked, returned to the computational
subspace, and remained leaked, respectively. (Note that the last operator
in the tensor product can be used to represent transitions between
leaked states.) The sub-block Kraus operator $K_{j}^{\bar{u},\bar{d},\bar{f}[U+L],\bar{\imath}[D+L]}$
can correspondingly be written as
\begin{eqnarray*}
&&K_{j}^{\bar{u},\bar{d},\bar{\imath}\rightarrow\bar{f}}=\\
&&\hspace{3ex}\biggl(\bra{\bar{f}[U]}_{U}\otimes\bra{\bar{d}}_{D}\otimes\bra{\bar{f}[L]}_{L}\biggr)\\
&&\hspace{28ex}K_{j}\\
&&\hspace{28ex}\biggl(\ket{\bar{u}}_{U}\otimes\ket{\bar{\imath}[D]}_{D}\otimes\ket{\bar{\imath}[L]}_{L}\biggr)
\end{eqnarray*}
Given the above decomposition, we can further break up the operators
on the computational devices $R$ by decomposing them as a linear combination of Pauli operators,
\[
K_{j}^{\bar{u},\bar{d},\bar{\imath}\rightarrow\bar{f}}=\sum_{\bar{\mu}}c_{\bar{\mu}}^{j,\bar{u},\bar{d},\bar{\imath}\rightarrow\bar{f}}\sigma_{\bar{\mu}},
\]
where the sum is taken over the Pauli operators $\sigma_{\bar\mu}$ acting on the devices
in $R$; these operators are assumed to vanish on the leakage states $\ket{2},\ket{3}$.
Below we will shorten the notations by dropping the extra fixed indices $\bar{\imath}\rightarrow\bar{f}$. From this expansion, one may derive
the conditional distribution, cf.~Eq.~(\ref{eq:compound-data}):
\begin{equation}
P(\bar{\imath}\rightarrow\bar{f},\bar{\mu})\equiv P(\bar{\imath}\rightarrow\bar{f})P(\sigma_{\bar{\mu}}|\bar{\imath}\rightarrow\bar{f},\bar{\mu}) =\frac{1}{2^{|U|}}\sum_{\bar{u},\bar{d},j}\left|c_{\bar{\mu}}^{j,\bar{u},\bar{d}}\right|^{2}.\label{eq:MarkovProb}
\end{equation}
The first term denotes the fact that each state $\ket{\bar{u}}$ exists
in a computational subspace of dimension $2^{|U|}$ . Finally, the
sum itself is the average probability of applying $\sigma_{\bar{u}}$
on the qubits in $R$. In the case of no leakage transitions ($\bar{\imath}=\bar{f}$)
it corresponds to the typical Pauli Twirling Approximation. 

As a simple example, we apply the approximation to an error
channel representing coherent rotation between the $\ket{1}$ and
$\ket{2}$ states. The error channel is composed of a single unitary
Kraus operator
\begin{eqnarray*}
U= & \ketbrad{0}+\cos(\theta)\l(\ketbrad{1}+\ketbrad{2}\r)+\ketbrad{3}\\
 & +\sin(\theta)\l(\ketbradt{1}{2}-\ketbradt{2}{1}\r).
\end{eqnarray*}
Using the definitions of the single device projection operators, the
non-vanishing blocks of the Kraus operator are
\begin{eqnarray*}
P_{c}UP_{c} & =&\ketbrad{0}+\cos(\theta)\ketbrad{1}\\
 & =&\l(\cos^{2}(\theta/2)\sigma_{I}+\sin^{2}(\theta/2)\sigma_{z}\r)\cdot1_{U}\cdot1_{D}\cdot1_{L}\\
P_{c}UP_{2} & =&\sin(\theta)\times\l(1_{R}\cdot1_{U}\cdot\ketbradt{1}{2}_{D}\cdot1_{L}\r)\\
&& +0\times\l(1_{R}\cdot1_{U}\cdot\ketbradt{0}{2}_{D}\cdot1_{L}\r)\\
P_{2}UP_{c} & =&-\sin(\theta)\l(1_{R}\cdot\ketbradt{2}{1}_{U}\cdot1_{D}\cdot1_{L}\r)\\
&&+0\times\l(1_{R}\cdot\ketbradt{2}{0}_{U}\cdot1_{D}\cdot1_{L}\r)\\
P_{2}UP_{2} & =&\cos(\theta)1_{R}\cdot1_{U}\cdot1_{D}\cdot\ketbrad{2}_{L}\\
P_{3}UP_{3} & =&1_{R}\cdot1_{U}\cdot1_{D}\cdot\ketbrad{3}_{L}
\end{eqnarray*}
where we have added trivial terms with zero coefficients to emphasize that the subspace ${\cal H}_c$ it two-dimensional. From this decomposition we may
read off the non-vanishing Pauli coefficients :
\begin{eqnarray*}
c_{I}^{c\rightarrow c} & =\cos^{2}(\theta/2)&,\qquad c_{Z}^{c\rightarrow c}=\sin^{2}(\theta/2),\\
c_{0}^{u=1,c\rightarrow2} & = \sin(\theta)&=-c_{0}^{d=1,2\rightarrow c},\\
c_{0}^{2\rightarrow2} & =\cos(\theta)&,\\
c_{0}^{3\rightarrow3} & =1.&
\end{eqnarray*}
The conditional probabilities can now be read off using Eq.~(\ref{eq:MarkovProb}).  Evidently, for each input state $\imath$, the sum of output probabilities equals one.
\begin{center}
\begin{tabular}{|c|c|c|c|}
\hline 
$\imath$ & $f$ & $\mu$ & $P(\imath\rightarrow f,\mu)$\tabularnewline
\hline 
\hline 
$c$ & $c$ & $I$ & $\cos^{4}(\theta/2)$\tabularnewline
\hline 
$c$ & $c$ & $Z$ & $\sin^{4}(\theta/2)$\tabularnewline
\hline 
$c$ & $2$ & - & $\sin^{2}(\theta)/2$\tabularnewline
\hline 
$2$ & $c$ & - & $\sin^{2}(\theta)$\tabularnewline
\hline 
2 & 2 & - & $\cos^{2}(\theta)$\tabularnewline
\hline 
3 & 3 & - & 1\tabularnewline
\hline 
\end{tabular}
\par\end{center}

As expected, the probability to observe a transition down from state
$\ket{2}$ is just the modulus square of the transition amplitude,
$\sin^{2}(\theta)=|\bra{1}U\ket{2}|^{2}$. The reverse transition,
though, has half this probability because a device in the computational
subspace is (on average) in state $\ket{1}$ only half the time. A
higher order effect cause by this twirling is the dephasing observed
in the $c\rightarrow c$ transition. This can be interpreted as a
measurement ``back-action'': because the device was not observed
to leak and only state $\ket{1}$ can leak, the channel acts like
a weak measurement in the standard basis.

\subsection{Comparison of Pauli$+$ and quantum simulations}
\begin{figure}[htbp]
    \centering
     \includegraphics[width=\linewidth]{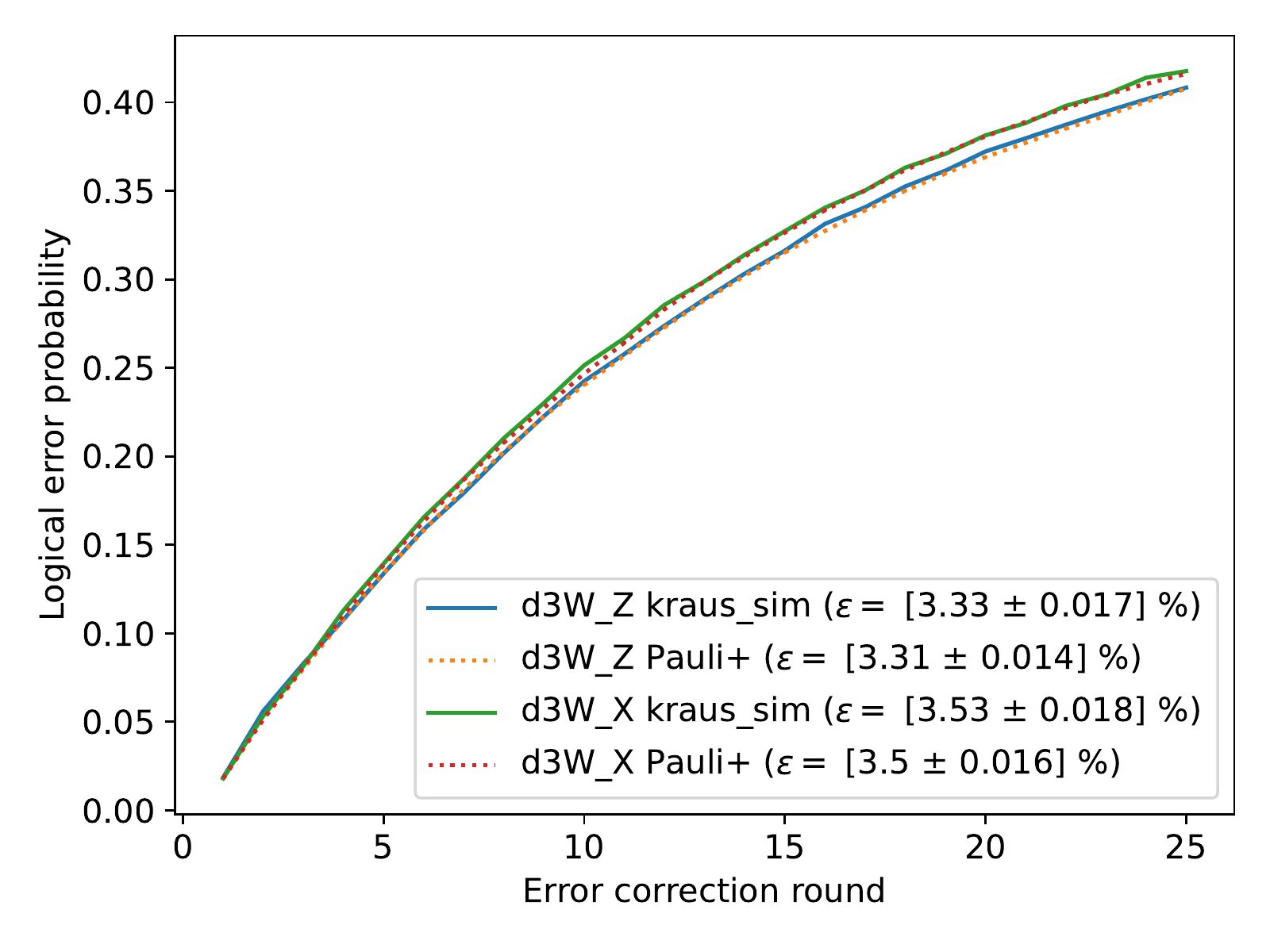}
    \caption{Comparison of simulated logical error probabilities for the distance-3 West, X and Z basis experiments. Each data point is the average over $N=2\times10^5$ samples, so that its standard deviation is bounded from above by $1/\sqrt{4N}\approx 0.11\%$.}
    \label{fig:quantum_classical_fidelities}

\end{figure}

\begin{figure}[htbp]
    \centering
    \includegraphics[width=\linewidth]{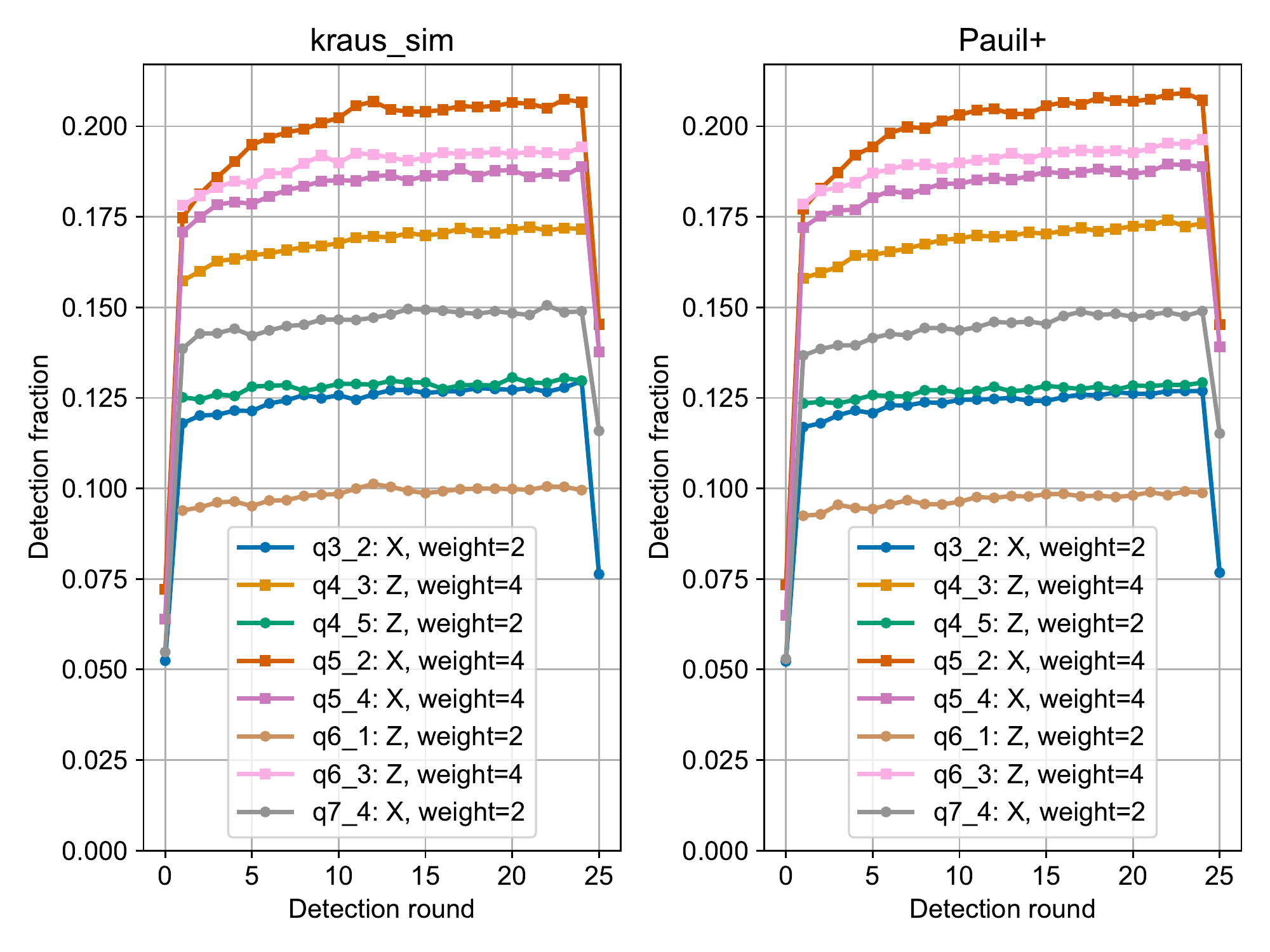}
    \caption{Comparison of simulated detection event fractions for the distance-3 West, X-basis experiment. As in the experimental data, we see a slow increase in detection event fraction as the number of rounds progresses. This can be attributed to leakage accumulation in the data qubits.}
    \label{fig:quantum_classical_defs}

\end{figure}

\begin{figure*}[htbp]
    \centering
    
    \includegraphics[width=\linewidth,trim={0 0 0 0},clip]{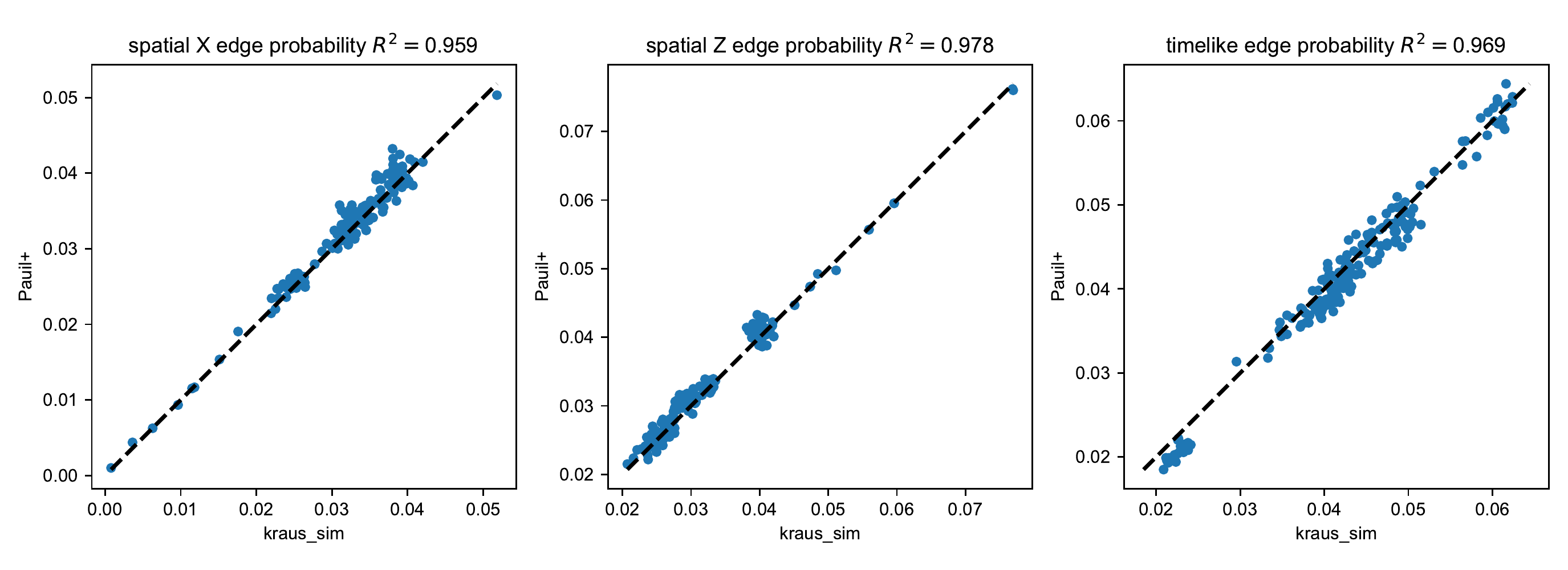}
    \caption{Comparison of simulated spacelike and timelike edge probabilities for the distance-3 West, X-basis experiment. From left to right, the scatter plots correspond to "X" spatial edges (between stabilizers in the initial state basis), "Z" spatial edges (between stabilizers with undetermined initial value), and timelike edges (between the stabilizers separated by one error correction round). The $R^2$ quantities reported above correspond to the coefficient of determination between the "observed" (kraus\_sim) and "predicted" (Pauli$+$) edge probabilities.}
    \label{fig:quantum_classical_scatter}
\end{figure*}

We tested the validity of the Generalized Twirling Approximation used by the \emph{Pauli$+$} model by doing a direct comparison with the equivalent kraus\_sim simulation. Specifically, we simulated the distance-3 $Z$-basis and $X$-basis experiments associated with the West device configuration. Each 25-round simulation corresponded to 200,000 samples. The classical \emph{Pauli$+$} simulation required about 160 seconds (distributed over 16 CPU cores), while the quantum simulation required ~11 hours\footnote{Note that the qsim backend for the kraus\_sim simulator has not been optimized for performance on these system sizes.}. Decoding was implemented using the algorithm in Ref.~\cite{fowler2013optimal}, with detection cluster probabilities inferred from the $p_{ij}$ matrix.

As can be seen in Figures \ref{fig:quantum_classical_fidelities}-\ref{fig:quantum_classical_scatter}, there is generally quite good agreement between both simulations with respect to logical error and typical detection event statistics. It is not entirely clear why this is the case.  Generally, the surface code measurement circuit implies a stabilizer-state projection at the end of each cycle; this has the effect of "twirling" the quantum noise, so that for a single data qubit error in the absence of leakage the approximation is exact\cite{cai2019constructing}.  The same is true for errors in non-neighboring circuit locations.  With the net error rates $\sim 1\%$ per gate and total of $n\sim 20$ qubits, we expect around one error per cycle. Moreover, some of the modeled error channels, (such as dephasing), are described by a Pauli channel exactly. 

On the other hand, our simulations also include some unitary rotation errors (due to crosstalk) which may add coherently in a QMC simulation\cite{Bravyi-Englbrecht-Koenig-Peard-2018}.  Early work\cite{katabarwa2015logical} simulating smaller distance codes suggested that Twirling Approximation under both decoherence and unitary gate errors strongly over predicted the expected logical error (by factors of 2 or more). Interestingly, an early numerical test of the distance-3 surface code \cite{tomita_low-distance_2014} found good agreement with the Pauli Twirling Approximation under amplitude damping ($T_1$ decay) and (white noise) dephasing, but only in the logical $Z$ basis (twirling over-predicted the error for the logical $X$).  
There are several factors that may have contributed to reducing these effects in our simulations.  First, the Hadamard operations in our experiments are chosen so that the stabilizer generators describing our system are all of mixed $ZXXZ$ type\cite{wen2003quantum, bonilla2021xzzx}. This has the effect of approximately symmetrizing all bit-flip and phase-flip errors.  Second, we deliberately insert $X$ gates whenever a qubit is idle, which is also known to reduce coherent errors.  With these measures in place, a good agreement between an exact QMC and a classical simulations indicates that the net contribution of coherently-added errors is not large enough to see.

\begin{table}[b!]
\centering
    \begin{tabular}{|l|c|}
    \hline
      \multicolumn{1}{|c|}{Component}   & Error probability\\ \hline
      SQ gates   & $1.09\times 10^{-3}$\\
      CZ gates & $6.05\times 10^{-3}$ \\
      Data qubit idle & $2.46\times10^{-2}$ \\
      Reset & $1.86\times 10^{-3}$ \\
      Readout & $1.96\times 10^{-2}$\\
      CZ leakage & $2.0\times10^{-4}$\\
      Leakage from heating& $6.4\times10^{-4}$\\
      CZ crosstalk & $9.5\times 10^{-4}$ \\
      \hline
    \end{tabular}
    \caption{Average probabilities of main errors at the operation point of $d=5$ surface code experiments.}
    \label{tab:table-expt-operation-point}
\end{table}

\section{Sensitivity of logical error per cycle to physical errors}
In this section we provide the sensitivity coefficients of surface code logical error per cycle ($\varepsilon_L$) to errors in the main components of the device; namely, control-Z (CZ) gates, single qubit (SQ) gates, data qubit idle (this includes dynamical decoupling sequence during readout and reset time), readout and reset. Additionally, we also present $\varepsilon_L$ sensitivity coefficients to leakage and crosstalk. The results are obtained using the Pauli+ simulator and the correlated minimum-weight perfect matching decoder. We note that this analysis is separate from the $1/\Lambda$ analysis presented in Fig.~4a of the main text. Specifically, in Fig.~4a, we used the methodology of Ref.~\cite{chen2021exponential} to estimate the $1/\Lambda$ budget with sensitivities evaluated at the half-operation point (i.e., component errors being 0.5 times the component errors shown in Table~\ref{tab:table-expt-operation-point}, used in the main simulation).

For a certain component error (e.g., SQ gate error), we evaluate the sensitivity coefficient for the logical error per cycle $\varepsilon_{\rm L}$  using the formula
\begin{align}
    \nu_{\rm error} = \frac{\delta \varepsilon_{\rm L}}{\delta p_{\rm error}}, 
    \label{eq:sensitivity_formula}
\end{align}
where $\delta \varepsilon_{\rm L}$ is the change of the logical error per cycle due to a variation $\delta p_{\rm error}$ of the error probability of all components of the same type (e.g., when calculating the $\varepsilon_{\rm L}$ sensitivity to SQ gate errors, $\delta p_{\rm error}$ is the error change of all SQ gates, excluding dynamical decoupling gates).

The logical error per cycle $\varepsilon_{\rm L}$ is a nonlinear function of the component error probabilities, so the sensitivity coefficients depend on the operation point (i.e., linearization point). We are particularly interested in evaluating the $\varepsilon_{\rm L}$ sensitivities at the experimental operation point, discussed below.

\subsection{Experimental operation point}
The experimental operation point of the $d=5$ surface code experiment (used in the simulator) is specified by the error probabilities of all components in the surface code circuit. We independently measure the component error probabilities to be used in the simulator. Table~\ref{tab:table-expt-operation-point} shows Pauli error probabilities averaged over all components of the same type, while individual values for every circuit element are used in the simulator. 

In particular, the SQ gate errors are measured by RB for each qubit and the CZ gate errors are measured by XEB for each CZ. Data qubit idle errors are measured individually by interleaved RB, which includes dynamical decoupling sequence used during readout and reset. The readout and reset errors are also measured independently. 

Another imperfection present in our device is leakage into non-computational states (mainly to state $|2\rangle$). We assume that leakage is generated at the high frequency qubit of a CZ gate (e.g., due to qubit dephasing during CZ operation) with a nominal probability of \begin{equation} p_2^{\rm CZ} = 0.25\, p_t = 2.0\times10^{-4}, \label{eq:p2_CZ}\end{equation} where we assume parameter $p_t=8\times10^{-4}$ for all CZs (see Sec.~\ref{sec:leakage}) and the factor 0.25 is the probability of state $|11\rangle$ before a CZ gate during surface code operation. Moreover, leakage can be generated from a non-equilibrium ``heating'' process at each qubit, for which we use a probability per QEC cycle of 
\begin{equation}p_2^{\rm heating}=0.5\, \Gamma_{1\to2}\, t_{\rm cycle} = 6.4\times10^{-4},\label{eq:p2_heating}\end{equation} where the factor 0.5 is the probability that the qubit is in state $|1\rangle$ during surface code operation, the QEC cycle duration is $t_{\rm cycle}=896$~ns, and we assume the experimentally estimated heating rate of $\Gamma_{1\to2}=1/(700\,{\rm \mu s})$ being the same for all qubits. Note that in the simulator the possibility of heating is considered after each gate or idling, so the actual parameter is $\Gamma_{1\to2}$, but we still characterize the heating strength by $p_2^{\rm heating}$. 

Finally, another imperfection in the device is the presence of unwanted interactions (crosstalk) between simultaneous gates (either single qubit idle or CZ), see Sec.~\ref{sec:x-talk}. This error channel is modelled individually for all elements and 
leads to CZ errors with a Pauli error probability (averaged over all CZs) of 
\begin{align} p_{\rm CZ} ^{\rm crosstalk} = 9.5\times10^{-4},\end{align} which is roughly 15\% of the observed total CZ error. Note that in the simulator this contribution is subtracted from the total CZ error to obtain the depolarizing part of the CZ error. 

\begin{table}[t!]
\centering
    \begin{tabular}{|l|c|c|}
    \hline
      \multicolumn{1}{|c|}{Component}   &  sensitivity (d3) & sensitivity (d5) \\   \hline
      SQ gates   & $6.2 \pm 0.3 $ & $8.4\pm0.2$ \\
      CZ gates & $2.13\pm0.08$ & $4.0\pm 0.2$ \\
      Data qubit idle & $0.91\pm0.02$ & $1.05\pm0.03$ \\
      Readout or reset & $0.26\pm 0.01$ & $0.41\pm0.02$\\
      Leakage (heating) & $2.4\pm0.3$ & $4.0\pm0.8$\\
      CZ crosstalk & $5.0\pm0.5$& $6.5\pm0.6$ \\
      \hline
    \end{tabular}
    \caption{Logical error rate sensitivities at the experimental operation point.}
    \label{tab:LER_sensitivities_expt_op_point}
\end{table}

\subsection{Logical error rate sensitivities for $d = 3$, 5 surface codes}
Table~\ref{tab:LER_sensitivities_expt_op_point} shows the sensitivity coefficients of $d=3$ and $d=5$ logical error per cycle to errors in the main components of the surface code circuit, and also the sensitivities to leakage and crosstalk errors. The main takeaway from this table is that $d=5$ surface code is more sensitive to component errors than $d=3$ surface code.  

\begin{figure}[t!]
    \centering
    \includegraphics[width=\linewidth, trim =0cm 0cm 1.5cm 2cm,clip=true]{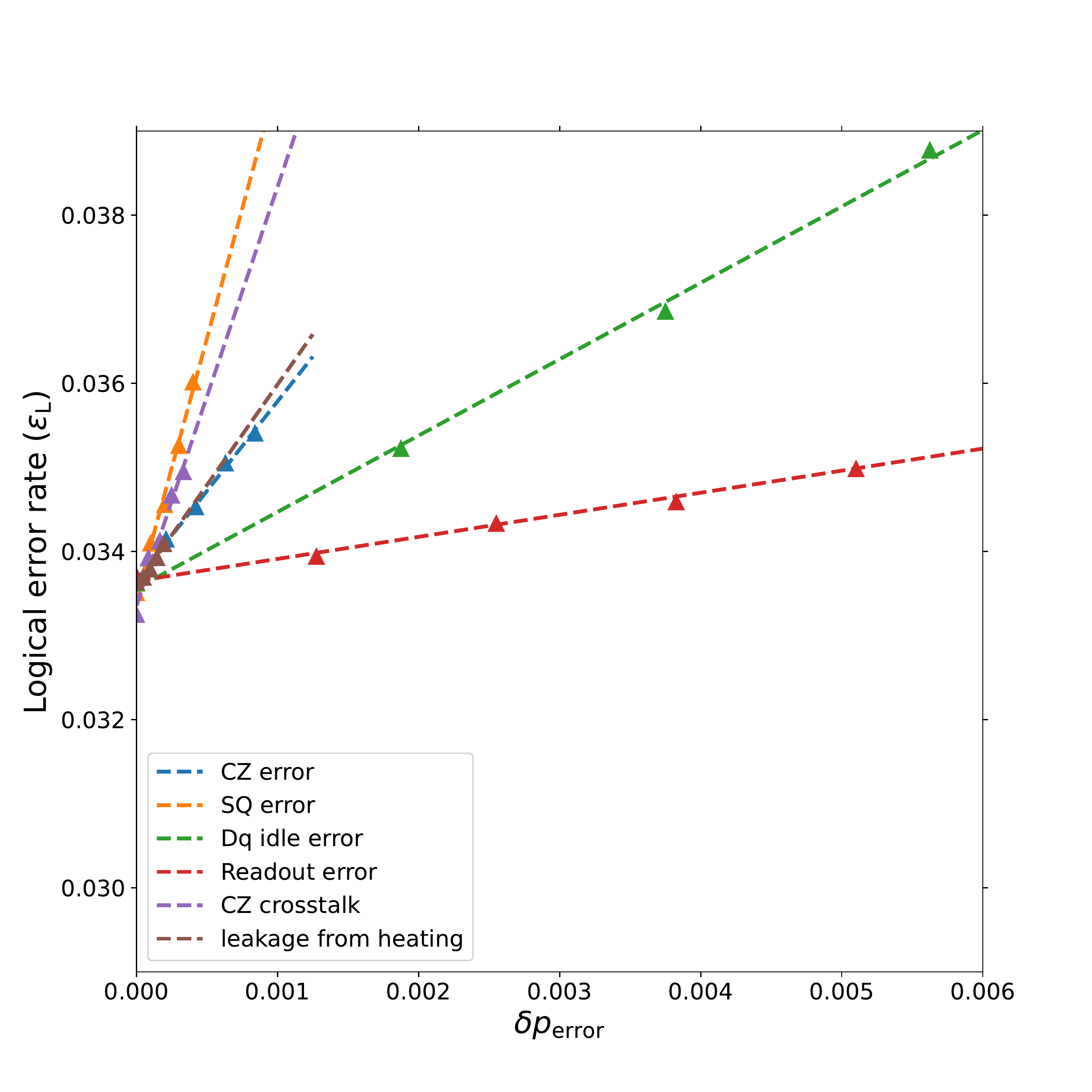}
    \caption{Calculation of logical error rate sensitivities for $d=3$ surface code. The depicted logical error rates are averaged values over the four $d=3$ surface code grids. The variable $\delta p_{\rm error}$ is the variation of the corresponding component error probability. Dashed lines are linear fits to the simulation data (markers) and their slopes are the sensitivities  reported in Table~\ref{tab:LER_sensitivities_expt_op_point}.
    }
    \label{fig:sensitivities_d3}
\end{figure}

\begin{figure}[t!]
    \centering
    \includegraphics[width=\linewidth, trim =0cm 0cm 1.5cm 2cm,clip=true]{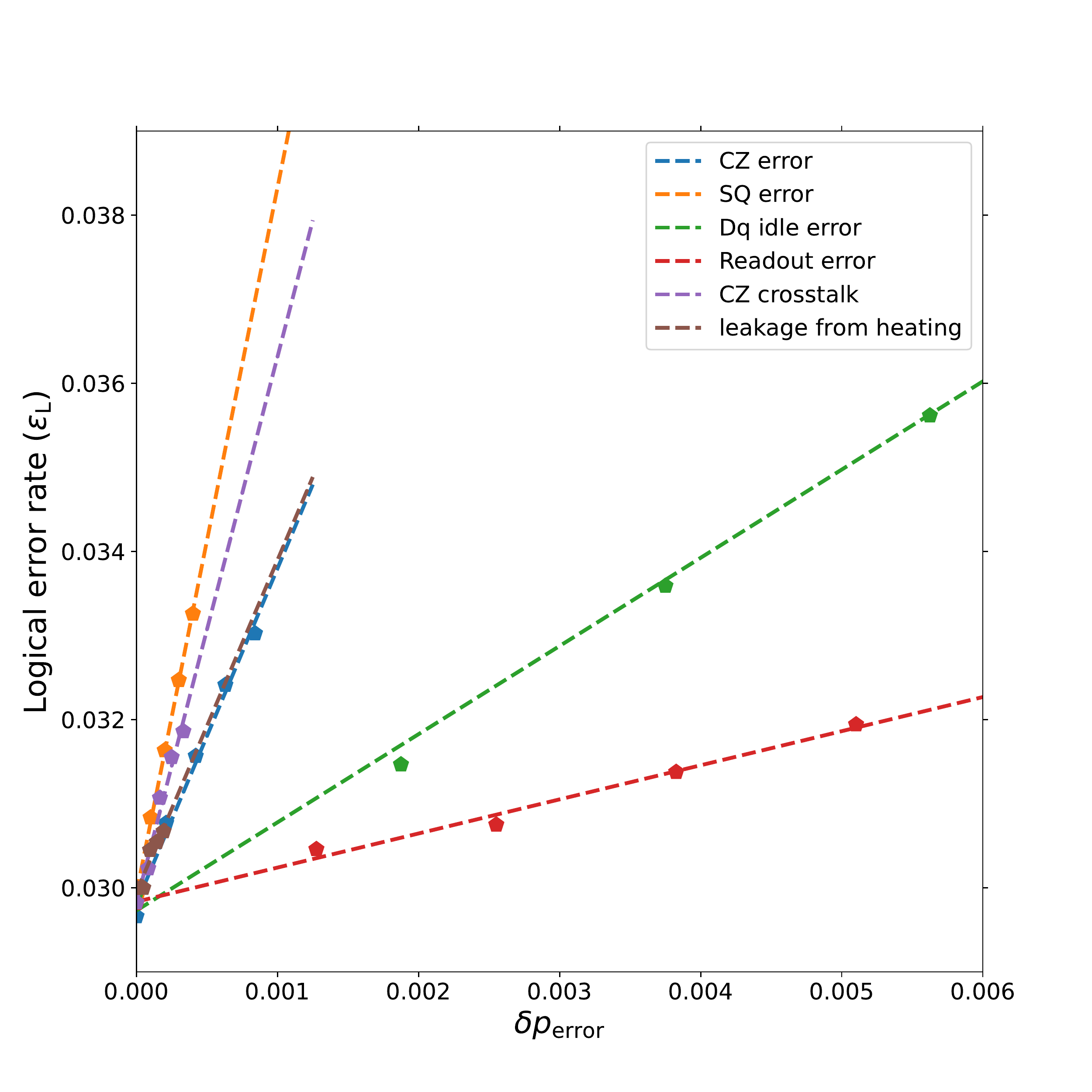}
    \caption{Calculation of logical error rate sensitivities for $d=5$ surface code. Similarly to Fig.~\ref{fig:sensitivities_d3}, dashed lines are linear fits to the simulation data (markers) and their slopes are the sensitivities  reported in  Table~\ref{tab:LER_sensitivities_expt_op_point}.}
    \label{fig:sensitivities_d5}
\end{figure}

We have computed these sensitivity coefficients at the experimental operation point. For instance, for SQ gate errors, $\delta p_{\rm error}$ in Eq.~\eqref{eq:sensitivity_formula} is the variation of the SQ gate error probability $\delta p_{\rm sq}$, which is assumed to be the same for all SQ gates (excluding those in dynamical decoupling sequence). As shown by the orange markers in Figs.~\ref{fig:sensitivities_d3} and~\ref{fig:sensitivities_d5}, the dependence of the change of logical error per cycle $\delta \varepsilon_{\rm L}$ on the variation of the SQ gate error $\delta p_{\rm sq}$ is fairly linear. From a linear fit (orange dashed lines in  Figs.~\ref{fig:sensitivities_d3} and~\ref{fig:sensitivities_d5}) to the simulation data  we obtain the sensitivity coefficients for single qubit gate errors: $\nu_{\rm sq} = 6.2\pm0.3$ ($d=3$) and $8.4\pm0.2$ ($d=5$). A similar procedure is used to obtain the $\varepsilon_{\rm L}$ sensitivity coefficients for the component errors indicated in Table~\ref{tab:LER_sensitivities_expt_op_point}. 

For leakage and crosstalk, there are subtleties in this procedure. For leakage from heating, we actually vary the heating rate $\Gamma_{1\to 2}$ applied to all qubits, while in Eq.~\eqref{eq:sensitivity_formula} we use $\delta p_{\rm error} = 0.5\, \delta \Gamma_{1\to 2} \, t_{\rm cycle}$. For crosstalk, we scale all crosstalk unitaries as $U_I \to U_I^\alpha$ (with the power $\alpha=1$ corresponding to the operation point) and use $\delta p_{\rm error} = (\alpha^2-1) p_{\rm CZ}^{\rm crosstalk}$ in Eq.~\eqref{eq:sensitivity_formula} with the nominal crosstalk error value $p_{\rm CZ}^{\rm crosstalk} = 9.5\times10^{-4}$.

\begin{figure}[t!]
    \centering
    \includegraphics[width=1.1\linewidth, trim =0.6cm 9.5cm 0.8cm 0.5cm,clip=true]{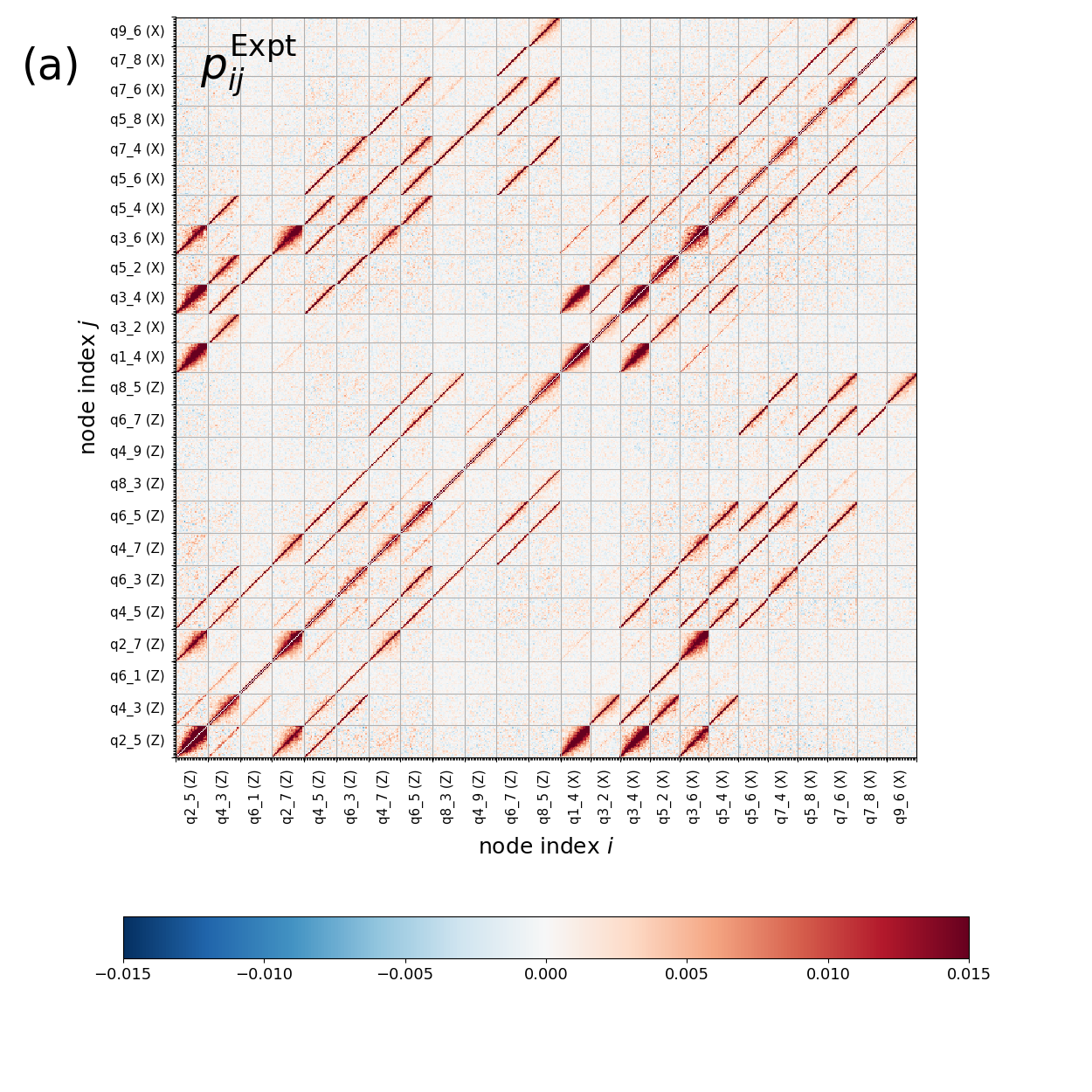}
    \includegraphics[width=1.1\linewidth, trim =0.6cm 3cm 0.8cm 0.5cm,clip=true]{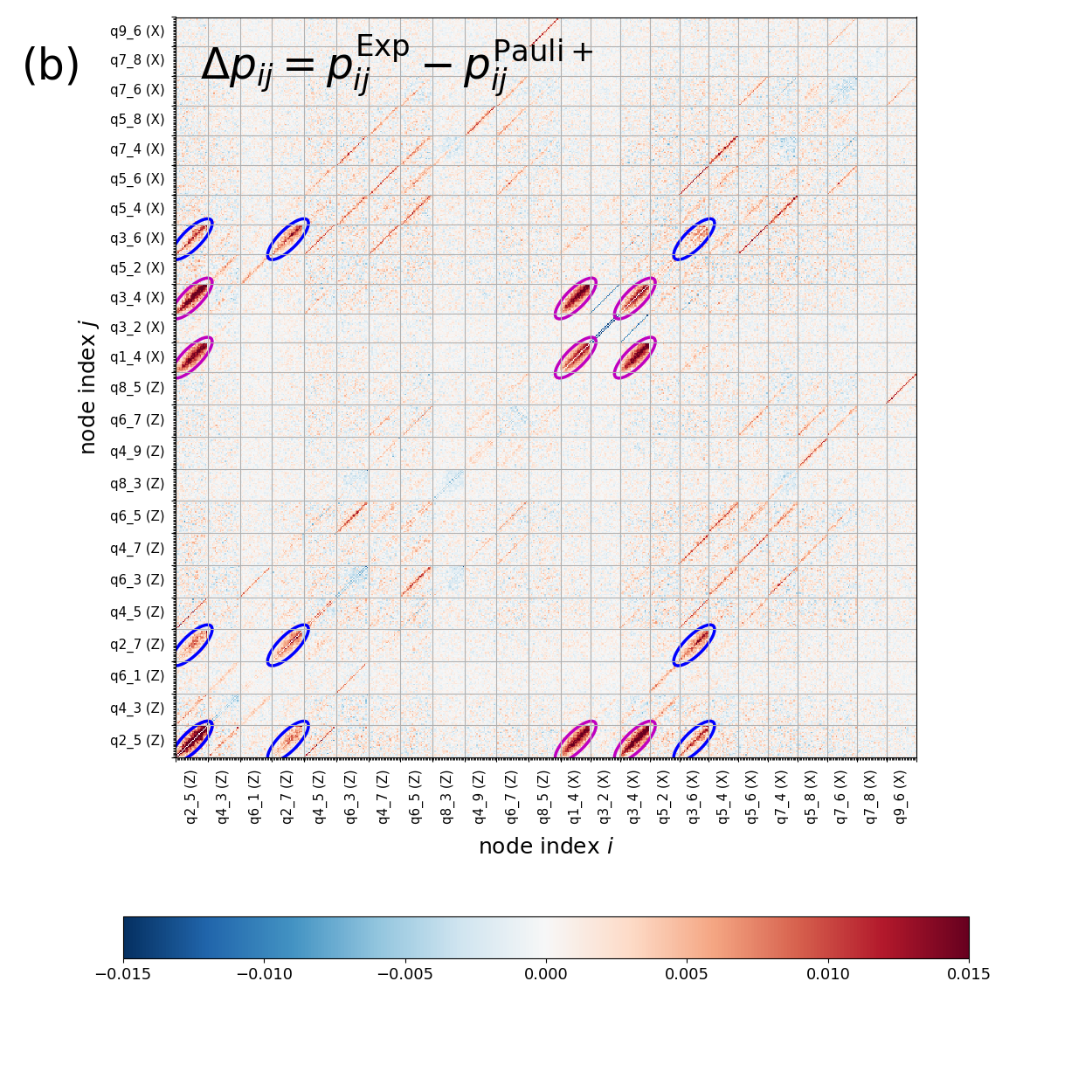}
    \caption{Correlation matrix $p_{ij}$. Panel (a) illustrates the $600\times600$ symmetric correlation matrix $p_{ij}$ for a 25 rounds $d=5$ surface code experiment with data qubits prepared in the $Z$-basis. Each detection node $i = (t, s)$ has time coordinate  $t$ indicated by the minor ticks and a spatial coordinate $s$ indicated by the measure qubit labels ($s=$ q2\_5, q4\_3, ... q9\_6). Panel (b) illustrates the difference of the $p_{ij}$ correlation matrices from experimental and simulation Pauli$+$ data. The ellipses indicate the main discrepancies between the experiment and simulation, which are due to excess leakage accumulation in a pair of data qubits (2\_6 and 2\_4) during the surface code experiment.}
    \label{fig:pij_matrix}
\end{figure}

\section{$p_{ij}$ diagnostics of surface codes}
We analyze the pairwise detection event correlations in surface code data using the $p_{ij}$ {\it correlation matrix method} that was introduced in Ref.~\cite{chen2021exponential}. This method quantifies the probability $p_{ij}$ of simultaneous triggering of two detection events at the error detection nodes $i$ and $j$. The $p_{ij}$ method can provide the correlation probability for any pair $i$-$j$ (edge), but we will focus here on the dominant pairwise correlations that we observe in the data; namely, timelike, spacelike and spacetimelike (referred to as diagonal in the main text) pairs. Hereon, we will refer to pairs as edges. We also compare the experiment {\it vs.} Pauli$+$ simulation $p_{ij}$ correlation matrices. The experimental data that is used to compare against simulations is from the 25 rounds, $d=5$ surface code experiment with data qubits prepared in the Z basis. 

\subsection{Experiment {\it vs.} Pauli$+$ correlation matrices $p_{ij}$}
Panel (a) of Fig.~\ref{fig:pij_matrix} is a graphical representation of the $600\times600$ symmetric correlation matrix $p_{ij}$ for the considered $d=5$ surface code experiment with 25 QEC rounds and initial data qubit states prepared in the $Z$-basis. The number of detection nodes $i$ is $12\times 26 +12\times 24 = 600$, since for the twelve measure qubits that measure the $Z$ stabilisers, we have 26 detection rounds (an additional detection round comes from the final data qubit measurements), while for the remaining twelve measure qubits that measure the X stabilisers, we have 24 detection rounds (two detection rounds less since the values of the X-stabilisers are fully random at the start and there is no $X$-measurement at the end of the Z-basis surface code experiment). The detection nodes $i=(t, s)$ have a time coordinate $t$ (spanning from $t=0$ to 25 for the considered experiment) as well as a space coordinate $s$ associated to the measure qubits. In Fig.~\ref{fig:pij_matrix}(a), the time coordinates are indicated by the minor ticks and the space coordinates are indicated by the qubit labels such as $s={\rm q}2\_5({\rm Z})$ for the measure qubit 2\_5 that measures a Z stabiliser. Note that even though we actually use the $ZXXZ$ variant of the surface code, all notations for stabilizers here are as if we were using the traditional surface code.  

There are two distinct types of features in the $p_{ij}$ correlation matrix illustrated in Fig.~\ref{fig:pij_matrix}(a). Namely, we see some narrow reddish features (e.g., the one inside the qubit-qubit block q8\_5-q6\_5) that can be explained in terms of conventional Pauli errors in the surface code circuit (such conventional errors manifest in the $p_{ij}$ matrix as one-pixel-wide reddish features since these errors produce detection events separated in time by 0 or 1 round). In addition, we also see some broader reddish features (e.g., the one inside the qubit-qubit block q1\_4-q2\_5) that indicate the presence of long-lived time correlations in the error detection events. Such correlations are due to leakage accumulation in the data qubits, see Ref~\cite{chen2021exponential}. 

Panel (b) of Fig.~\ref{fig:pij_matrix} shows the difference between the $p_{ij}$ correlation matrices from experimental data and Pauli$+$ simulation data (used in the main text). We see that the $p_{ij}$ matrix difference looks much cleaner (with fainter reddish regions than the experimental $p_{ij}$ matrix shown in Fig.~\ref{fig:pij_matrix}(a)). This indicates that Pauli$+$ simulation does a good job in capturing most of the detection event correlations that are present in the experiment. There are, however, some spatial/temporal correlations that are significantly stronger in the experiment than in the simulation, indicated by the ellipses in the $p_{ij}$ matrix difference. We attribute these features to  stronger-than-usual leakage in two data qubits: 2\_4 and 2\_6. Leakage accumulation in these data qubits leads to detection events that are correlated locally in space (with detection events at measure qubits that are neighbors to the leaky data qubits 2\_4 and 2\_6) and correlated nonlocally in time (with a correlation time of roughly $T_1/2\approx 10\,\mu$s). These correlations produce the reddish regions indicated by the magenta and blue ellipses for leakage accumulation in the data qubits 2\_4 and 2\_6, respectively. We stress that Pauli$+$ simulation can in principle capture these missing correlations by properly adjusting the simulation parameters related to leakage generation in the data qubits 2\_4 and 2\_6 (e.g., by increasing the heating rates $\Gamma_{1\to2}$ or by increasing the leakage generation probabilities $p_2^{\rm CZ}$ of the CZs that act on these leaky data qubits). Besides the broad reddish regions indicated by the ellipses in Fig.~\ref{fig:pij_matrix}(b), we also notice some faint narrow reddish features in the $p_{ij}$ matrix difference. These features indicate short temporal correlations of detection events separated in time by 0 or 1 round, and they can be understood as a discrepancy between the actual component error probabilities during the surface code experiment and the component error probabilities  assumed in the simulation. To quantify the discrepancy between the component error probabilities, we need to compare the $p_{ij}$ probabilities for the edges of the surface code error graph, which we discuss below.

\begin{figure}[t!]
    \centering
    \includegraphics[width=0.9\linewidth, trim =7cm 5cm 7cm 3cm,clip=true]{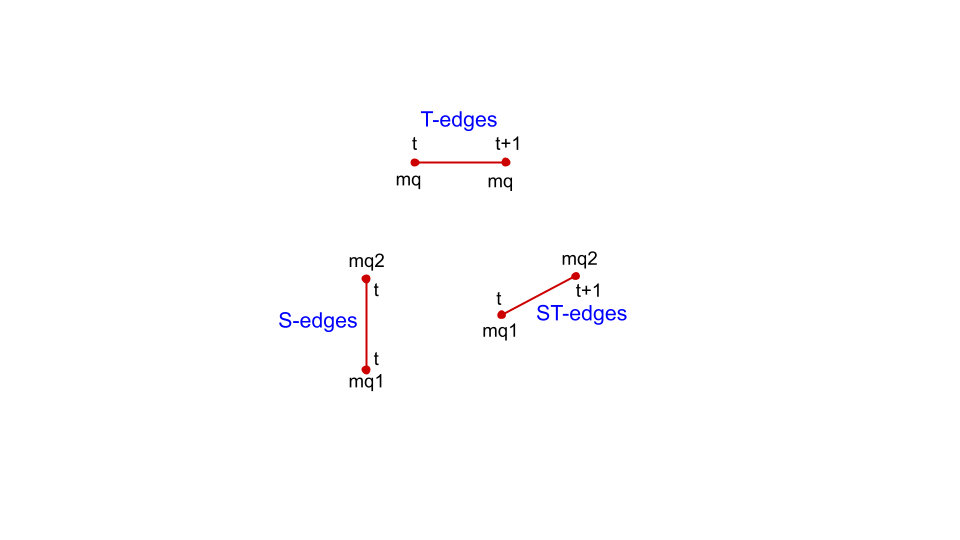}
    \caption{Main pairwise detection event correlations in the surface code. The red dots indicate detection events. The considered pairwise correlations (edges) are timelike (T) edges, spacelike (S) and  spacetimelike (ST) edges. These correlations are mainly from readout/reset errors, idle data qubit errors and CZ errors, respectively.}
    \label{fig:edges}
\end{figure}

\subsection{$p_{ij}$ probability of main error edges in the surface code}
In this section we examine the probabilities of the main pairwise correlations (error edges) in the $d=5$ surface code. These error edges are depicted in Fig.~\ref{fig:edges}. Timelike edges (T-edges) indicate correlations of detection events occurring at the same measure qubit and separated in time by 1 round. Physical errors that give rise to T-edges mainly include readout/reset errors. Spacelike edges (S-edges) indicate correlations of detection events occurring at the same round  and at two measure qubits of the same type (X or Z). Physical errors that give rise to spacelike edges mainly include bit-flip errors (from $T_1$ qubit decay) and phase-flip errors (from qubit dephasing) at the data qubits during measurement and reset time (idle time). Finally, spacetimelike edges indicate correlations of detection events separated in time by 1 round and occurring at different measure qubits. Spacetimelike edges that come from Pauli CZ errors are denoted as ST edges, while other spacetimelike edges that are not expected from CZ errors are denoted as ST' edges. The latter can be due to leakage or crosstalk. 

\subsubsection{Timelike edges}
\begin{figure}[t!]
    \centering
    \includegraphics[width=0.9\linewidth, trim =0cm 1.6cm 0cm 0cm,clip=true]{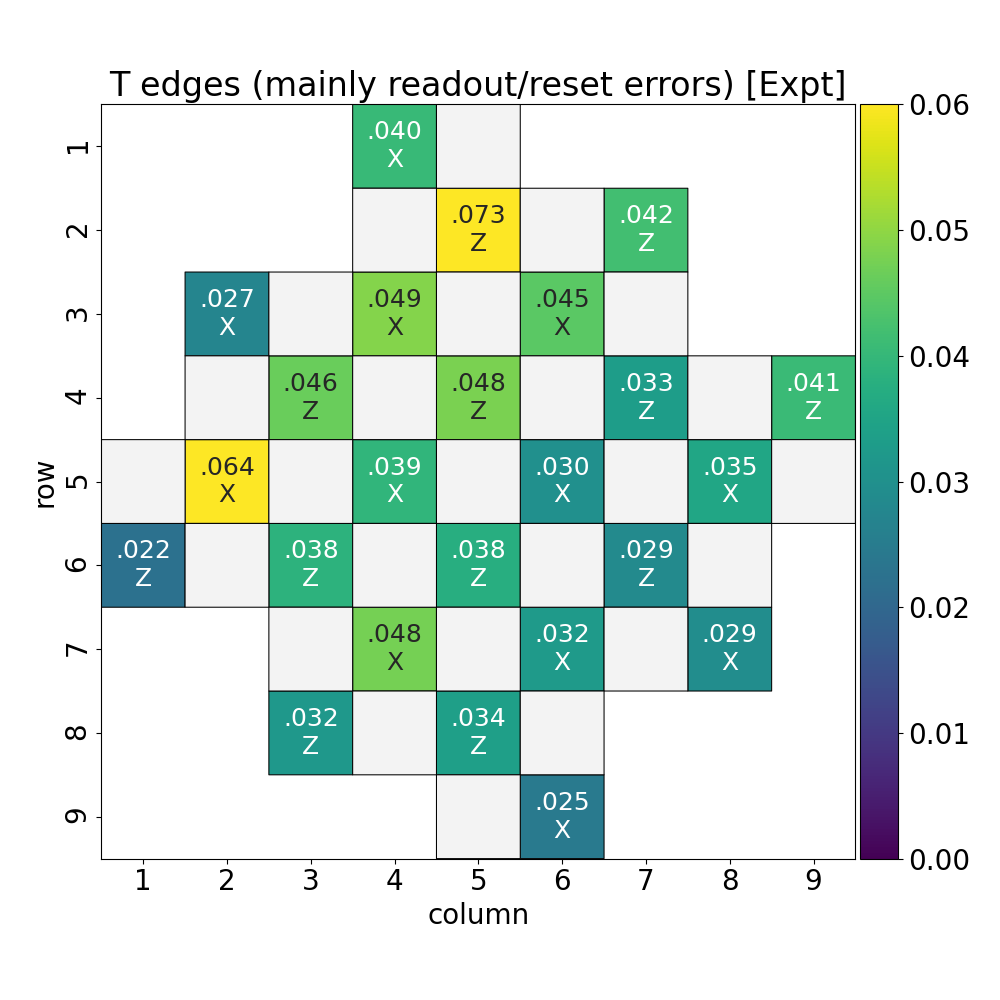}
    \includegraphics[width=0.95\linewidth, trim =0cm 2.4cm 0cm 1cm,clip=true]{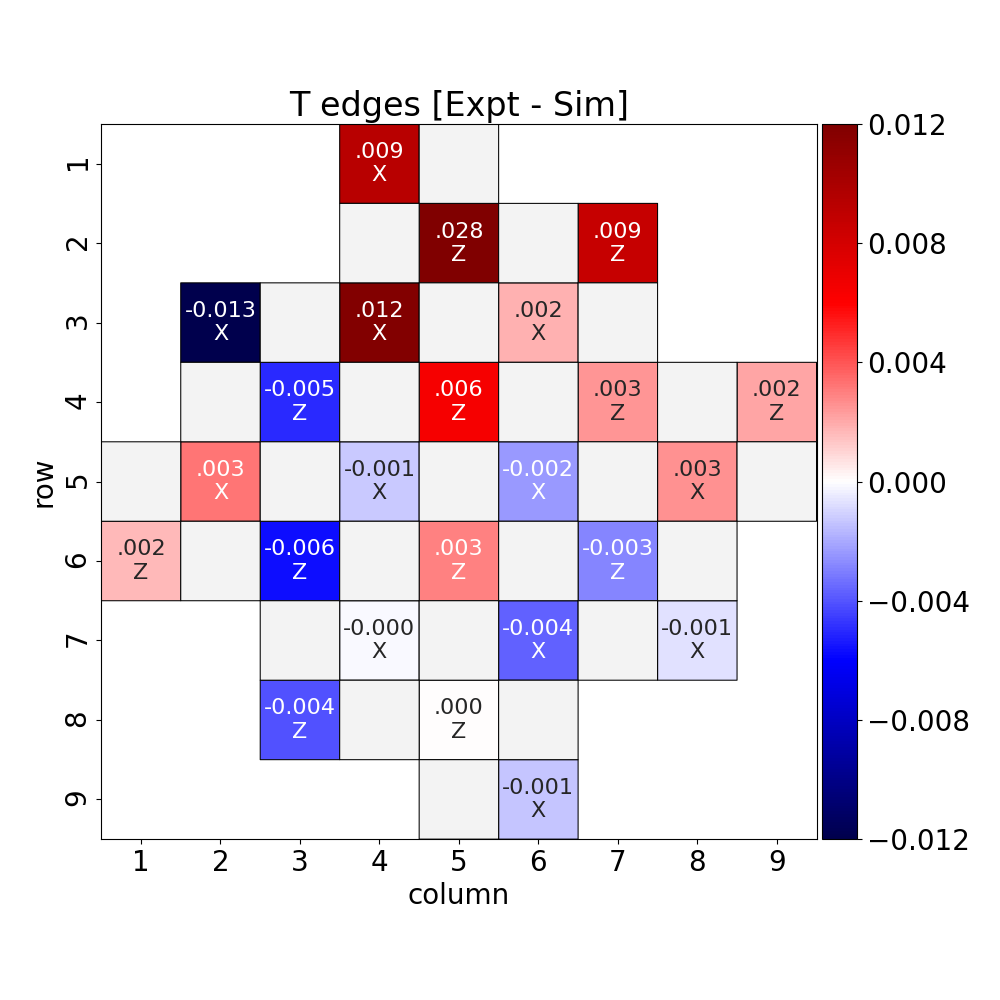}
    \caption{Top panel: timelike edge probabilities. Each number in the heatmaps is the median of the T-edge probabilities of a measure qubit. Measure qubits that measure X or Z stabilisers are indicated by X or Z. Top panel shows the time-averaged T-edge probabilities for the 25 rounds $d=5$ surface code experiment with data qubits initialized in the Z-basis. Bottom panel shows the difference of T-edge probabilities between experiment and  Pauli$+$ simulation.}
    \label{fig:t_edges}
\end{figure}

The probability of T-edges are obtained from the $p_{ij}$ correlation matrix with $i=(t, s)$ and $j=(t+1, s)$, where the space coordinate $s$ refers to one of the measure qubits. It is convenient to average over time $t$ and present the time-averaged T-edge probabilities as shown at the top panel heatmap of Fig.~\ref{fig:t_edges}. This way of presenting the $p_{ij}$ data for the T-edges is particularly useful to spot the measure qubits with good or bad readout/reset errors. We point out that, although readout/reset errors are the main contributors to the T-edges in our experiment, other physical errors (e.g., leakage and CZ errors) can also contribute to the T-edge probabilities. 

The mean T-edge probability (averaging over all qubits and all rounds) is 
\begin{align}
    p_{ij}^{{\rm T}-{\rm edge},\,{\rm mean}} = 3.85\times 10^{-2}\; [{\rm Pauli\!\!+\, sim}=&\; 3.7\times10^{-2}, \nonumber \\
     {\rm Pauli\; sim} =&\; 3.0\times 10^{-2}
    ],\nonumber
\end{align}
where we have also included the averaged T-edge probabilities from Pauli$+$ and Pauli simulations.

The lower panel heatmap of Fig.~\ref{fig:t_edges} shows the difference of the time-averaged T-edge probabilities between experiment and Pauli$+$ simulation. We see that the difference is relatively small, indicating a good agreement between experiment and simulation. The biggest discrepancy between experiment and simulation is for T-edges of the measure qubits 1\_4, 3\_4, 2\_5, 2\_9, which are spatially close to the data qubits 2\_4 and 2\_6 that exhibit more leakage in the experiment than in the Pauli$+$ simulation. Thus the reason for this discrepancy is most likely leakage accumulation in these data qubits.

\subsubsection{Spacelike edges}
\label{sec:space_edges}
\begin{figure}[t!]
    \centering
    \includegraphics[width=0.9\linewidth, trim =0cm 1.6cm 0cm 0cm,clip=true]{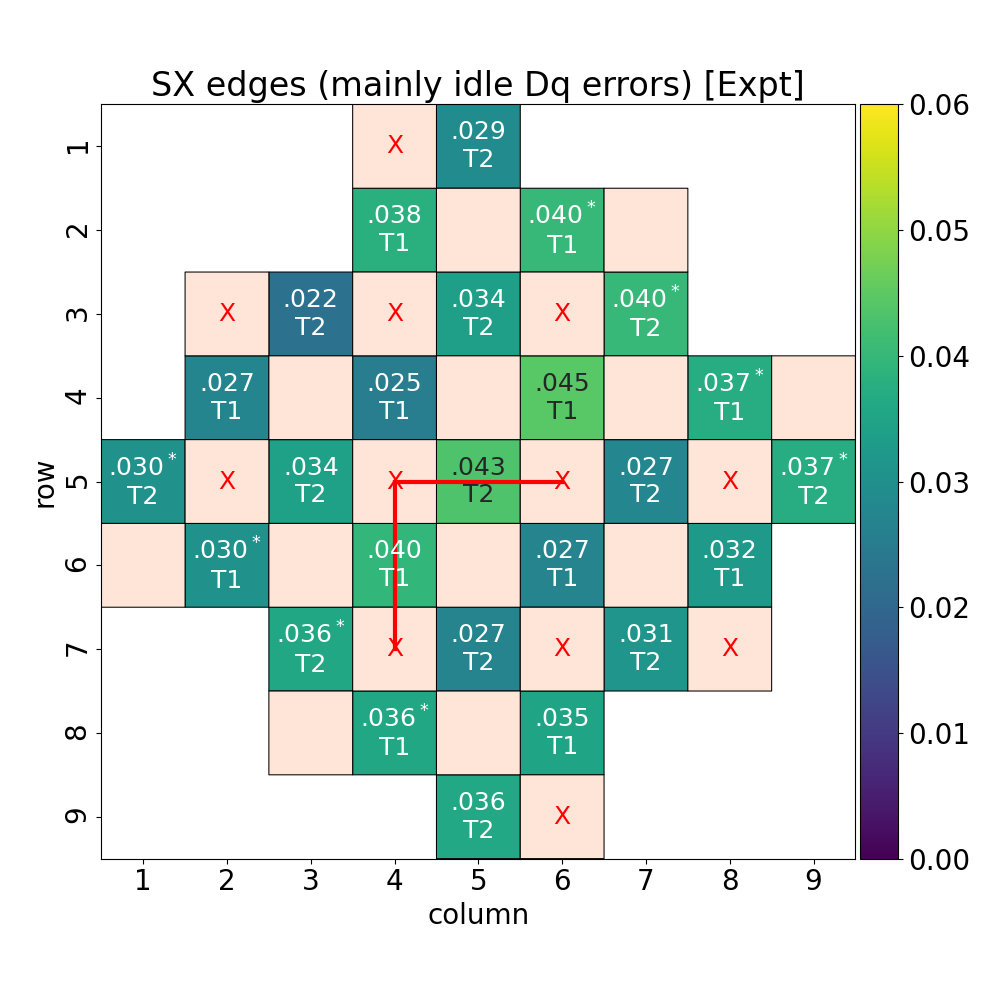}
    \includegraphics[width=0.95\linewidth, trim =0cm 2.4cm 0cm 1cm,clip=true]{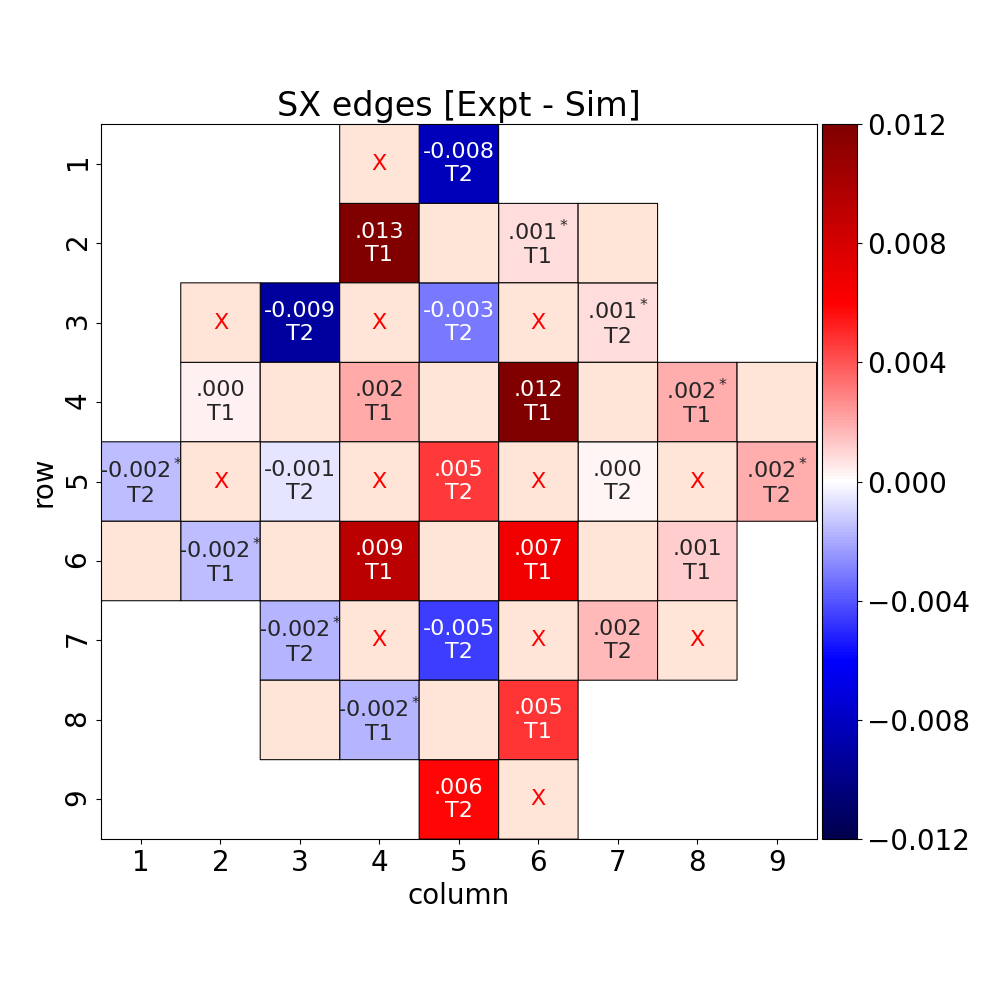}
    \caption{Top panel: time-averaged $p_{ij}$ probabilities of SX edges. Pink tiles with an X indicate the measure qubits that measure X stabilisers and the empty pink tiles indicate the other (Z) measure qubits. The red horizontal edge indicates SX edges between the X measure qubits 5\_4 and 5\_6. The probabilities of these SX edges are obtained from the $p_{ij}$ matrix with $i=(s,t)$ and $j=(s',t)$, and $s={\rm q}5\_4$ and $s'={\rm q}5\_6$. After averaging over time $t$, we write the result in the middle data qubit tile (5,5) since phase-flip errors in this data qubit would produce such SX edges. Asterisks indicate ambiguous data qubits (e.g., the pair 5\_1 and 6\_2) in the sense that idle errors at those data qubits produce the same boundary edge (see Sec.~\ref{sec:space_edges}). Lower panel shows the difference of the time-averaged SX edge probabilities between experiment and Pauli$+$ simulation. Labels T1 and T2 at the tiles indicate the type of idle data qubit error (bit-flip and phase-flip errors, respectively) that gives rise to the SX edges (see Sec.~\ref{sec:space_edges}).
    }
    \label{fig:sx_edges}
\end{figure}

The heatmap at the top panel of Fig.~\ref{fig:sx_edges} shows the time-averaged probabilities of SX edges (spacelike edges between two X measure qubits) for the $d=5$ surface code experiment in the Z-basis. To explain the information displayed in this heatmap, let us consider the middle data qubit 5\_5. A phase-flip error in this data qubit during idle time produces two detection events at the neighboring X measure qubits 5\_4 and 5\_6 (see horizontal red edge in Fig.~\ref{fig:sx_edges}) and at some round $t$. The number 0.043 written in the middle tile (5,5) is the average (specifically, the median) of the $p_{ij}$ probabilities of all SX edges between these two X measure qubits. Idle Y errors in the data qubit 5\_5 also contributes to the displayed probability 0.043; for this reason, we write 'T2' below 0.043 to remind us that this SX edge probability comes from phase-flip or Y idle data qubit errors (in reality, other errors such as leakage and CZ errors can also contribute to the SX edge probabilities). Let us consider another example: an idle bit-flip error (due to T1 qubit decay) in the data qubit 6\_4 produces a SX edge between the X measure qubits 5\_4 and 7\_4 (vertical red edge in Fig.~\ref{fig:sx_edges}). One might expect that all SX edges are produced by phase-flip idle data qubit errors. This is not the case for our surface code experiments since we implement the ZXXZ variant of the surface code circuit (see main text). \newline Idle errors in the boundary data qubits of the surface code grid sometimes  produce just one detection event (i.e., a boundary edge). For instance, this is the case of an idle phase-flip error in the data qubit 1\_5 that produces only one detection event at the X measure qubit 1\_4 and at some time $t$. Finally, we also have the situation where idle errors in two different boundary data qubits produce the same detection event or boundary edge. For instance, a bit-flip error in data qubit 6\_2 and a phase-flip error in data qubit 5\_1 both produce the same boundary edge at the X measure qubit 5\_2. In this ambiguous situation, we split equally the $p_{ij}$ probability of the boundary edge among these data qubits [in this example 0.03 in the tiles (5,1) and (6,2)] and write asterisks at the corresponding data qubit tiles. 

\begin{figure}[t!]
    \centering
    \includegraphics[width=0.9\linewidth, trim =0cm 1.6cm 0cm 0cm,clip=true]{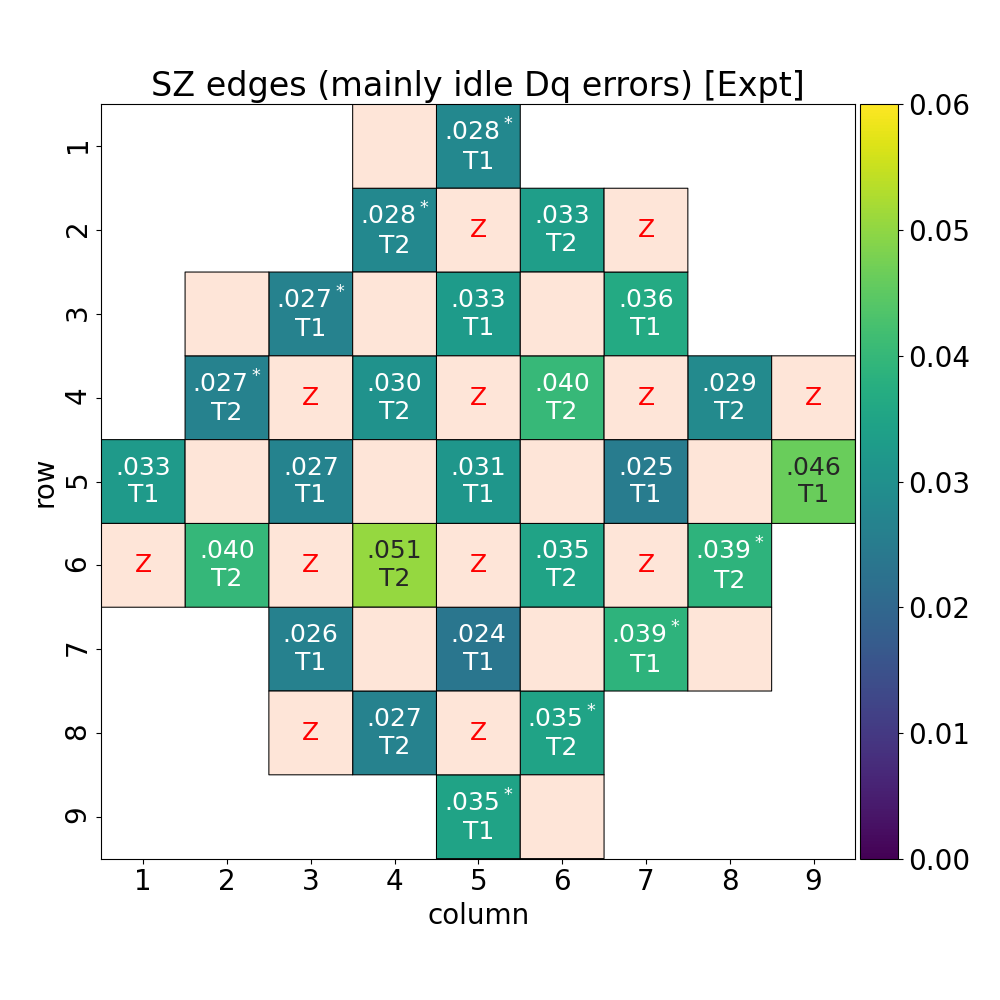}
    \includegraphics[width=0.95\linewidth, trim =0cm 2.4cm 0cm 1cm,clip=true]{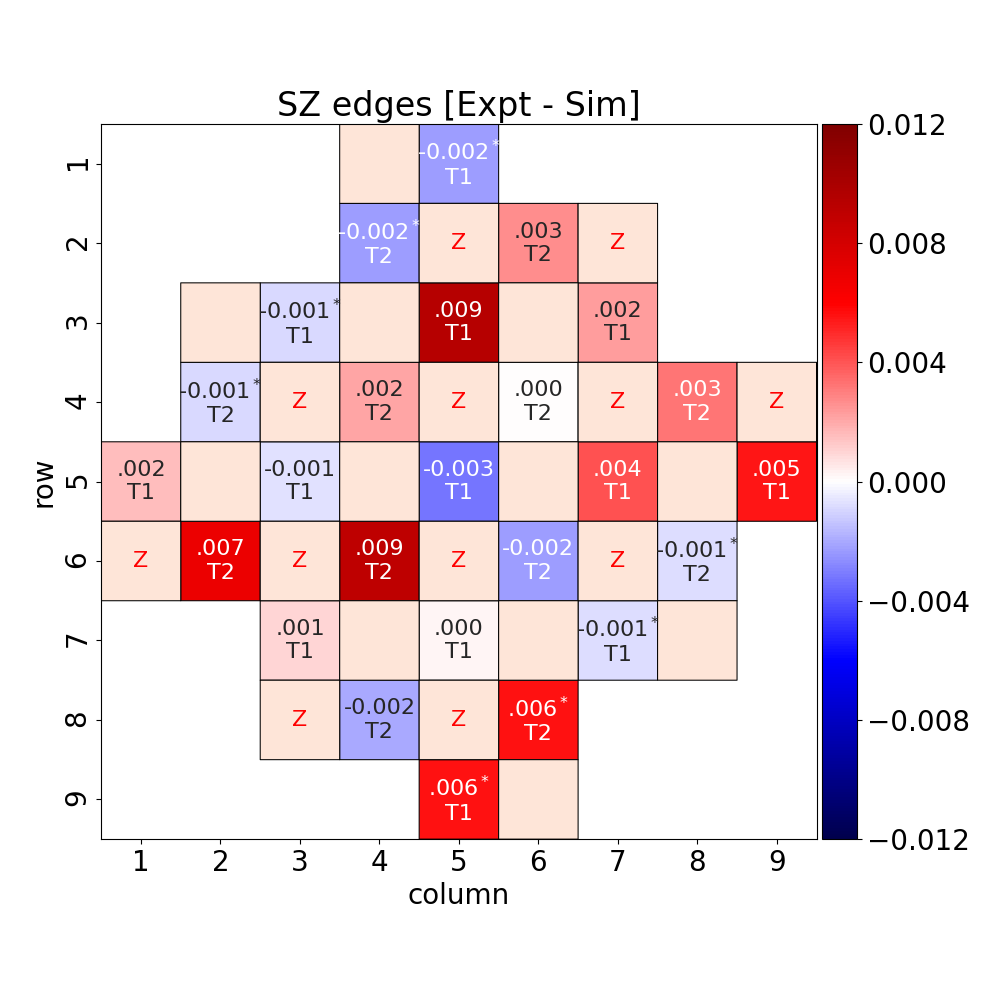}
    \caption{Top panel: time-averaged $p_{ij}$ probabilities of SZ edges. Description of the information displayed at the top panel is similar to that of Fig.~\ref{fig:sx_edges}. Lower panel shows the difference of the time-averaged SZ edge probabilities between experiment and Pauli$+$ simulation.}
    \label{fig:sz_edges}
\end{figure}

Figure~\ref{fig:sz_edges} is complementary to Fig.~\ref{fig:sx_edges}, and it displays the time-averaged SZ edge probabilities. For instance, for the middle data qubit 5\_5, we read a SX edge probability of 0.043 from Fig.~\ref{fig:sx_edges}, and a SZ edge probability of 0.031 from Fig.~\ref{fig:sz_edges}. The bottom panels of Figs.~\ref{fig:sx_edges} and \ref{fig:sz_edges} show the difference between the experimental and simulated values, which is relatively small. 

The mean spacelike edge probabilities are 
\begin{align}
    p_{ij}^{{\rm SX}-{\rm edge},\,{\rm mean}} = 3.4\times 10^{-2}\; [&{\rm Pauli\!\!+ sim}= 3.3\times10^{-2}, \nonumber \\
    &{\rm Pauli\; sim}= 3.1\times10^{-2}
    ], \nonumber
\end{align}
and 
\begin{align}
    p_{ij}^{{\rm SZ}-{\rm edge},\,{\rm mean}} = 3.2\times 10^{-2}\; [&{\rm Pauli\!\!+ sim}=  3.0\times10^{-2} \nonumber \\
    &{\rm Pauli\; sim}= 2.9\times10^{-2}
    ], \nonumber
\end{align}
where we have also included the results from Pauli$+$ and Pauli simulations.

\begin{figure}[t!]
    \centering
    \includegraphics[width=\linewidth, trim =0.75cm 1.6cm 0cm 0cm,clip=true]{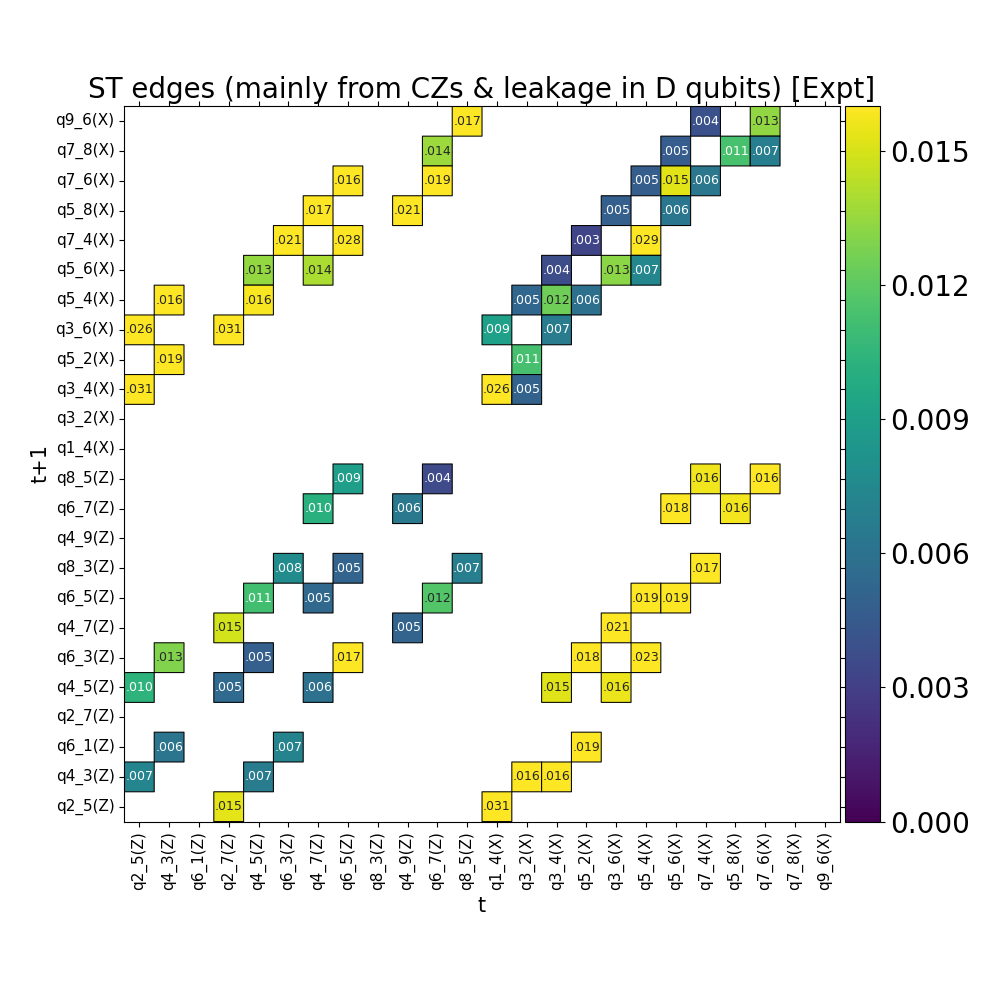}
    \includegraphics[width=1.05\linewidth, trim =0.75cm 2.4cm 0cm 1cm,clip=true]{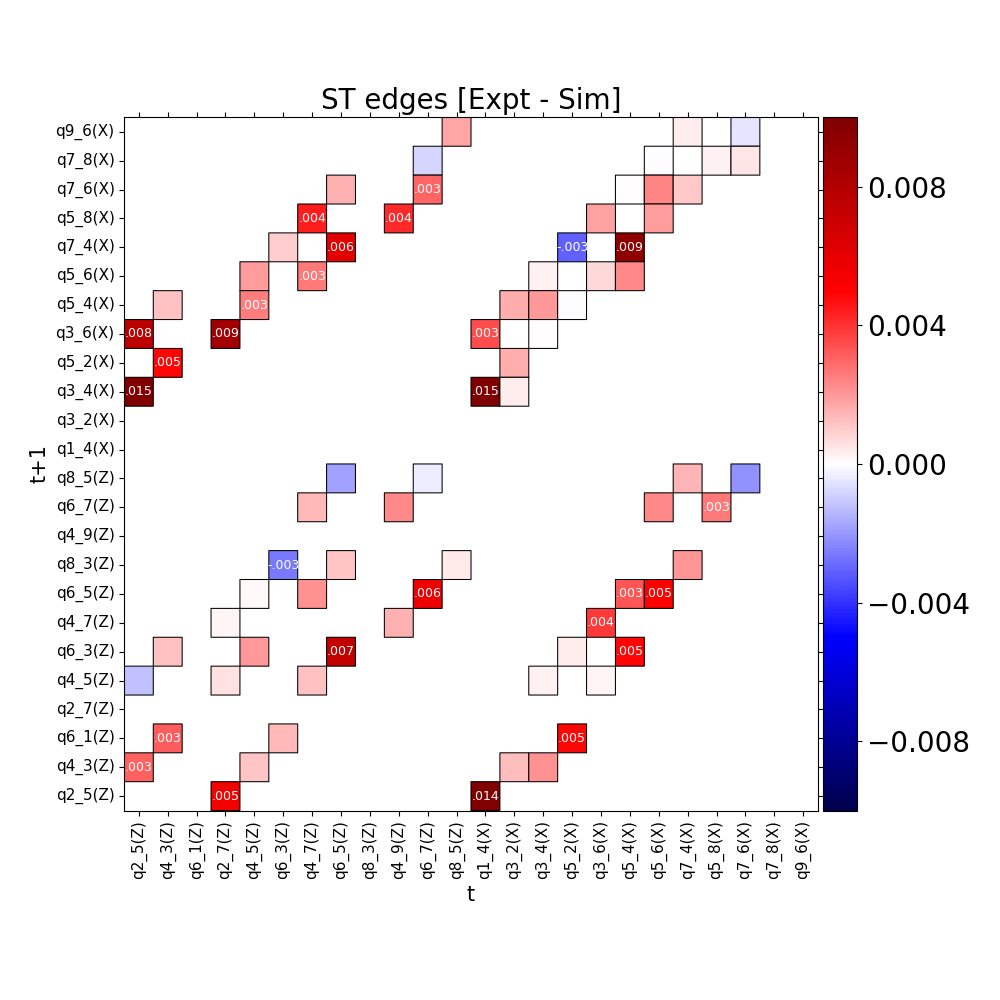}
    \caption{Time-averaged $p_{ij}$ probabilities of ST-edges. At the top panel, each number is the time average of ST edges from $i=(t,s_i)$ [$s_i$ are the qubit labels indicated at the x-axis] to $j=(t+1, s_j)$ [$s_j$ are the qubit labels indicated at the y-axis]. In this figure we only include pairwise correlations $i$-$j$ that are expected from CZ errors. At the lower panel, we show the difference of the time-averaged ST-edges between experiment and Pauli$+$ simulation. At the lower panel, numbers that are smaller in magnitude than $2.5\times 10^{-3}$ are not shown. 
    } 
    \label{fig:st_edges}
\end{figure}

\begin{figure}[t!]
    \centering
    \includegraphics[width=\linewidth, trim =0.5cm 1.6cm 0cm 0cm,clip=true]{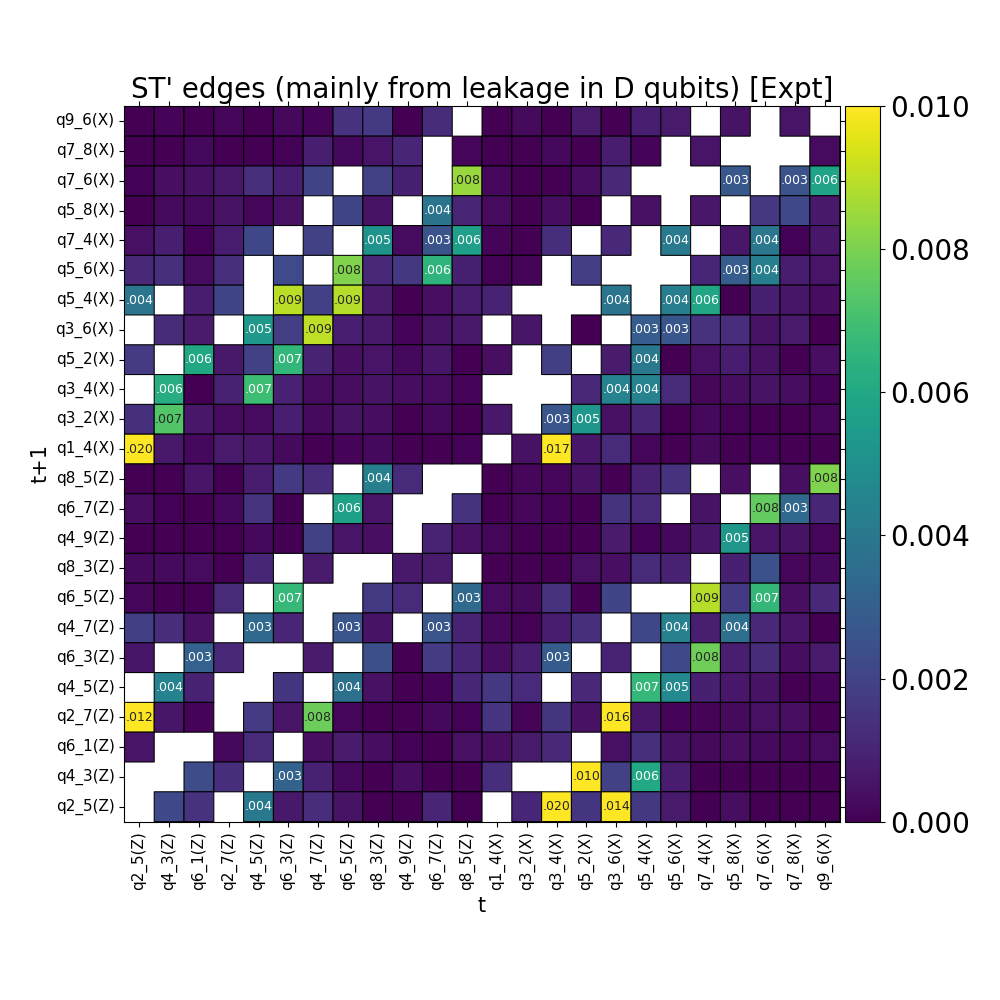}
    \includegraphics[width=1.05\linewidth, trim =0.5cm 2.4cm 0cm 1cm,clip=true]{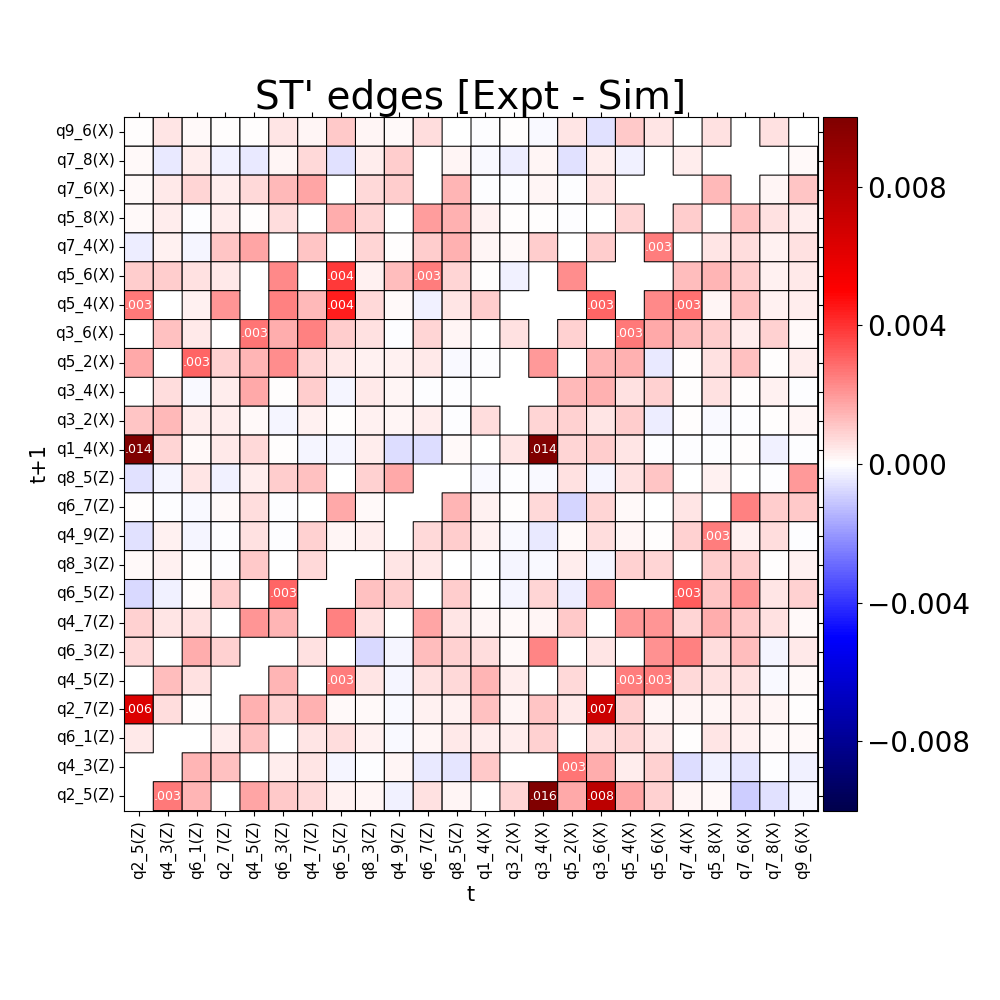}
    \caption{Time-averaged $p_{ij}$ probabilities of spacetimelike edges not expected from CZ errors (ST' edges). These unexpected pairwise correlations are mainly from leakage in data qubits. Top panel shows the ST'-edge probabilities from the $d=5$ surface code experiment, and the lower panel shows the discrepancy between experiment and Pauli$+$ simulation. Numbers that are smaller in magnitude than $2.5\times10^{-3}$ are not shown.}
    \label{fig:st_prime_edges}
\end{figure}

\subsubsection{Spacetimelike edges}
Within the conventional Pauli error model, spacetimelike edges are due to CZ Pauli errors. Those spacetimelike edges are referred to as ST-edges. In the experiment and in the Pauli$+$ simulations, we also find other unexpected spacetimelike edges that we denote as ST' edges, which are mainly due to leakage accumulation in the data qubits. We expect that leakage in the data qubits should increase the ST- and ST'-edges by roughly the same amount, and this is indeed validated by the experimental and simulation data, as shown below. 

Top panel of Fig.~\ref{fig:st_edges} shows the time-averaged ST-edge probabilities for all ST edges (expected due to CZ errors) for experimental data. The lower panel shows the discrepancy between experiment and Pauli$+$ simulation. 

The mean ST-edge probability for the $d=5$ surface code experiment is
\begin{align}
    p_{ij}^{{\rm ST}-{\rm edge},\,{\rm mean}} = 1.3\times 10^{-2}\; [{\rm Pauli\!\!+ sim}=&\; 1.1\times10^{-2},\nonumber \\
    {\rm Pauli\; sim} =&\; 0.8\times10^{-2}
    ]. \nonumber
\end{align}
We see that the ST-edge probability that is expected from the conventional Pauli error model is $0.8\times10^{-2}$; however, the actual ST-edge probability is higher by $0.5\times10^{-2}$ in the experiment and by $0.3\times 10^{-2}$ in the Pauli$+$ simulation. This excess of ST-edge probability turns out to be similar to the average probability of the unexpected ST'-edges: 

\begin{align}
    p_{ij}^{{\rm ST'}-{\rm edge},\,{\rm mean}} = 0.6\times 10^{-2}\; [&{\rm Pauli\!\!+ sim}= 0.35\times10^{-2}]. \nonumber
\end{align}
A significant difference here indicates that we most likely underestimate leakage in the simulation. 

The time-averaged probability of all ST' edges for the experimental data is shown at the top panel of Fig.~\ref{fig:st_prime_edges}, and the lower panel shows the discrepancy between experiment and simulation. We notice that the largest discrepancies are present in ST'-edges between measure qubits that are spatially close to the data qubits 2\_4 and 2\_6 which exhibit more leakage accumulation in the experiment than in the simulation, see lower panel of Fig.~\ref{fig:pij_matrix}.

\subsection{Probability of clusters of detection events using generalized $p_{ij}$}
In the surface code, there are physical errors (e.g., Y errors in the data qubits during measurement and reset time) that can produce more than two detection events. To characterize such physical processes using experimental error detection data, we use a generalized version of the $p_{ij}$ correlation matrix method~\cite{chen2021exponential} that allows us to calculate the probability of clusters of detection events. To be specific, let us assume that we are interested in finding out the probability $p_{123}$ of a physical process that can simultaneously flip the state $x_i$ ($x_i=1$ if there is an error detection event, otherwise $x_i=0$) of three error detection nodes $i=1,2,3$. To do this, we describe the statistics of the error detection events at these three nodes in terms of seven independent physical processes: three 1-body processes that separately flip the state of the nodes $i$ with probabilities $p_i$; three 2-body processes where each of them simultaneously flips the states of two nodes $i,j$ with probabilities $p_{ij}$, and one 3-body process that simultaneously flips the states of the three considered nodes with the sought probability $p_{123}$. The idea to find $p_{123}$ is to express the following  seven experimental averages: $\langle x_1\rangle$, $\langle x_2 \rangle$, $\langle x_3\rangle$, $\langle x_1x_2\rangle$, $\langle x_1x_3 \rangle$, $\langle x_2x_3\rangle$ and $\langle x_1x_2x_3\rangle$ in terms of the probabilities of the seven processes $p_1$, $p_2$, $p_3$, $p_{12}$, $p_{23}$, $p_{13}$ and $p_{123}$. Then solving numerically this nonlinear systems of seven equations with seven unknowns, we obtain the sought probability $p_{123}$. This generalized $p_{ij}$ method can be applied to analyze processes that can produce arbitrary number $n$ of detection events; in this case, we would need to solve $2^n-1$ nonlinear equations with the same number of unknown probabilities (one of which is the probability of the $n$-body process we are interested to find). 

We have used the generalized $p_{ij}$ method to do error diagnostics of higher order processes (e.g., Y data qubit errors) using the experimental error detection data. We also use the cluster probabilities provided by the generalized $p_{ij}$ method to set the weights of decoders, discussed in Section~\ref{sec:decoding}.

\begin{table}[t!]
\centering
    \begin{tabular}{|l|c|c|c|}
    \hline
      \multicolumn{1}{|c|}{Component}   & $p_{\rm expt}^{(i)}$ & $w_i$ & $1/\Lambda$ contrib.\\   \hline
      SQ gates & $1.09\!\times\!10^{-3}$ & 78.7 & 0.086 (9.6\%)  \\
      CZ gates & $4.85\!\times\!10^{-3}$ & 54.5 & 0.264 (29.4\%)  \\
      Data qubit idle & $2.46\!\times\!10^{-2}$ &7.0 & 0.172 (19.2\%) \\
      Readout & $1.96\!\times\!10^{-2}$ & 5.6 & 0.11 (12.2\%) \\
      Reset & $1.86\!\times\!10^{-3}$ & 5.6 & 0.0104 (1.2\%) \\
      Leakage (heating) & $6.4\!\times\!10^{-4}$ & 125 & 0.08 (8.9\%)\\
      CZ leakage & $2.0\!\times\!10^{-4}$ & 125* & 0.025 (2.8\%)\\
      CZ crosstalk & $9.5\!\times\!10^{-4}$& 158 & 0.15 (16.7\%)\\ 
      \hline
    \end{tabular}
    \caption{Calculation of the error budget for $1/\Lambda_{3/5}$. We multiply the component errors $p_{\rm expt}^{(i)}$ by the sensitivities (weights) $w_i$ at the half-operation point, resulting in the contributions to $1/\Lambda_{3/5}$ shown in the right column.} 
    \label{tab:1_Lambda_contributions}
\end{table}

\section{Surface code $1/\Lambda$ error budget}
\label{sec:surface_code_1_over_Lambda_budget}
In this section we discuss the procedure to obtain the surface code error budget for $1/\Lambda_{3/5}$, which is presented as the bar chart in Fig.~4a of the main text. Following Ref.~\cite{chen2021exponential}, we write $1/\Lambda_{3/5}$ as a sum of contributions from each error channel $i$ that is considered in the Pauli+ simulation, 
\begin{align}
(\Lambda_{3/5})^{-1} = \sum_i w_i \, p_{\rm expt}^{(i)} = w\,p_{\rm expt}
\label{eq:1_Lambda_budget_formula}, 
\end{align}
where $p_{\rm expt}$ is the vector of component error probabilities at the experimental operation point and $w$ is the vector of weights. The latter is obtained from the gradient (sensitivities) of the nonlinear function $[1/\Lambda_{3/5}](p)$ at half experimental operation point (i.e., at $p=p_{\rm expt}/2$), which for simplicity is referred to as "half-operation point". 

The half-operation point is defined as half of the experimentally measured component error probabilities for SQ gates, CZ gates, data qubit idle, reset and readout -- see Table \ref{tab:table-expt-operation-point}. Additionally, the simulation leakage parameters become $1/(1.4\, {\rm ms})$ for the heating rate and $1.0\times 10^{-4}$ for the probability of leakage generation from qubit dephasing during CZs (CZ leakage), and the crosstalk unitaries are rescaled from $U_I$ to $(U_{I})^{1/\sqrt{2}}$ at half-operation point. 

The third column in Table~\ref{tab:1_Lambda_contributions} shows the sensitivities of $1/\Lambda_{3/5}$ at half-operation point for the eight error channels that are considered in the simulations. These sensitivities are then used as the weight parameters $w_i$ in Eq.~\eqref{eq:1_Lambda_budget_formula} to obtain the contribution of each error channel to $1/\Lambda_{3/5}$. The resulting contributions are shown in the right column of Table~\ref{tab:1_Lambda_contributions}. Note that for the CZ error channel, we use error probability $4.85\times10^{-3}$ instead of $6.05\times 10^{-3}$ in Table \ref{tab:table-expt-operation-point} to avoid double count of the CZ error coming from crosstalk and CZ leakage: we subtract the contribution of $9.5\times10^{-4}$ from crosstalk and the contribution of $1.25\times 2\times10^{-4}$ from CZ leakage. Also note that for the sensitivity to CZ leakage we use the same value as for the sensitivity to the leakage from heating. 

The values from the right column of Table~\ref{tab:1_Lambda_contributions} are used in Fig.\ 4a of the main text. The sum of these values is $0.90$, which is very close to the inverse of the simulated value $\Lambda_{3/5}=1.10$. 

\clearpage

\onecolumngrid
\section{Overview of error correction experiments}

\begin{table*}[hb]
\caption{\label{table:experiments}Summary of various error correction and error detection experiments.
    Experiments using ``classical" codes (i.e. codes that only detect one type of error e.g. only phase flips or only bit flips) use $[n,k,d]$ code notation instead of quantum $[[n,k,d]]$ code notation.
    Entries with an N/A are experiments related to embedding error correction into the physical qubits as opposed to layering the error correction on top of the physical qubits.}
\resizebox{\linewidth}{!}{

\begin{tabular}{cccccccccc}
\hline\hline
Paper & Year & Code name & [[\#data,\#logical,distance]] & Physical qubits & Rounds & Physical qubit type \\
\hline
\cite{cory1998experimental} & 1998 & Repetition Code & [3,1,3] & 3 & single shot & NMR \\
\cite{knill2001benchmarking} & 2001 & Perfect Code & [[5,1,3]] & 5 & single shot & NMR \\
\cite{schindler2011experimental} & 2011 & Repetition Code & [3,1,3] & 3 & 3 & Ion trap \\
\cite{moussa2011demonstration} & 2011 & Repetition Code & [3,1,3] & 3 & 2 & NMR \\
\cite{zhang2011experimental} & 2011 & Repetition Code & [3,1,3] & 3 & single shot & NMR \\
\cite{reed2012realization} & 2012 & Repetition Code & [3,1,3] & 3 & single shot & Superconducting \\
\cite{zhang2012experimental} & 2012 & Perfect Code & [[5,1,3]] & 5 & single shot & NMR \\
\cite{bell2014experimental} & 2014 & Surface Code & [[4,1,2]] & 4 & single shot & Photons \\
\cite{kelly2015state} & 2014 & Repetition Code & [3,1,3]-[5,1,5] & 9 & 8 & Superconducting \\
\cite{nigg2014quantum} & 2014 & Color Code & [[7,1,3]] & 7 & single shot & Ion trap \\
\cite{waldherr2014quantum} & 2014 & Repetition Code & [[3,1,3]] & 4 & single shot & NV center \\
\cite{riste2015detecting} & 2015 & Repetition Code & [3,1,3] & 5 & single shot & Superconducting \\
\cite{corcoles2015demonstration} & 2015 & Bell State & [[2,0,2]] & 4 & single shot & Superconducting \\
\cite{cramer2016repeated} & 2016 & Repetition Code & [3,1,3] & 4 & 1-3 & Superconducting \\
\cite{ofek2016extending} & 2016 & Cat States & N/A & 1 & 1-6 & 3D cavity \\
\cite{takita2017experimental} & 2017 & Color Code & [[4,2,2]] & 5 & single shot & Superconducting \\
\cite{linke2017fault} & 2017 & Color Code & [[4,2,2]] & 5 & single shot & Ion trap \\
\cite{wootton2018repetition} & 2018 & Repetition Code & [3,1,3]-[8,1,8] & 15 & single shot & Superconducting \\
\cite{andersen2019entanglement} & 2019 & Bell State & [[2,0,2]] & 3 & 1-12 & Superconducting \\
\cite{gong2019experimental} & 2019 & Perfect Code & [[5,1,3]] & 5 & single shot & Superconducting \\
\cite{hu2019quantum} & 2019 & Binomial Bosonic States & N/A & 1 & 1-19 & 3D cavity \\
\cite{wootton2020benchmarking} & 2020 & Repetition Code & [3,1,3]-[22,1,22] & 5-43 & single shot & Superconducting \\
\cite{andersen2020} & 2020 & Surface Code & [[4,1,2]] & 7 & 1-11 & Superconducting \\
\cite{bultink2020protecting} & 2020 & Bell State & [[2,0,2]] & 3 & 1-26 & Superconducting \\
\cite{egan2021fault} & 2020 & Bacon-Shor Code & [[9,1,3]] & 15 & single shot & Ion trap \\
\cite{luo2020quantum} & 2020 & Bacon-Shor Code & [[9,1,3]] & 11 & single shot & Photons \\
\cite{campagne2020quantum}& 2020 & GKP States & N/A & 1 & 1-200 & 3D cavity \\
\cite{chen2021exponential} & 2020 & Repetition Code & [3,1,3]-[11,1,11], [[4,1,2]] & 5-21 & 1-50 & Superconducting \\
\cite{ryan2021realization} & 2021 & Color Code & [[7,1,3]] & 10 & 2-12 & Ion trap \\
\cite{zhao2021realizing} & 2021 & Surface Code & [[9,1,3]] & 17 & 1-11 & Superconducting \\
\cite{abobeih2022fault} & 2022 & Perfect Code & [[5,1,3]] & 7 & 1-11 & NV center \\
\cite{krinner2022realizing} & 2022 & Surface Code & [[9,1,3]] & 17 & 1-16 & Superconducting \\
\cite{marques2022logical} & 2022 & Surface Code & [[4,1,2]] & 7 & 1-15 & Superconducting \\
\cite{sundaresan2022matching} & 2022 & Subsystem Code & [[9,1,3]] & 23 & 1-10 & Superconducting \\
\cite{bluvstein2022quantum} & 2022 & Surface Code & [[13, 1, 3]] & 19 & 1 & Rydberg \\
\cite{bluvstein2022quantum} & 2022 & Toric Code & [[16, 2, 2]] & 24 & 1 & Rydberg \\
This work & 2022 & Repetition Code & [3,1,3]-[25,1,25] & 5-49 & 50 & Superconducting \\
This work & 2022 & Surface Code & [[9,1,3]]-[[25,1,5]] & 17-49 & 1-25 & Superconducting \\
\hline\hline
\end{tabular}

}
\end{table*}

\newpage

\section{Additional plots of surface code data}
\begin{figure*}[htbp]
    \centering
    
    \includegraphics[height=7.8in]{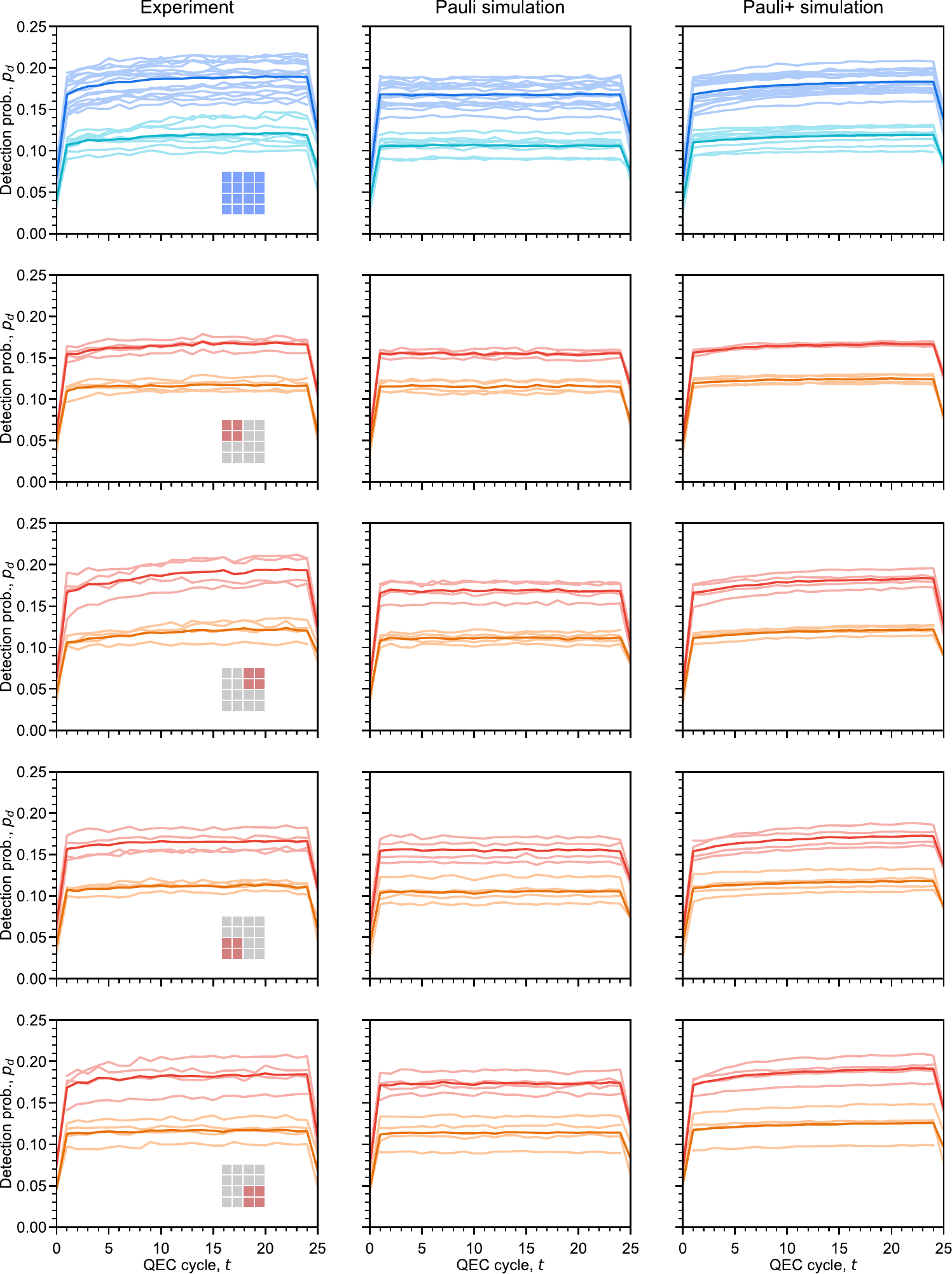}
    \caption{Detection probabilities for individual codes.}
    \label{fig:defs}
\end{figure*}

\begin{figure*}[htbp]
    \centering
    
    \includegraphics[width=7.2in]{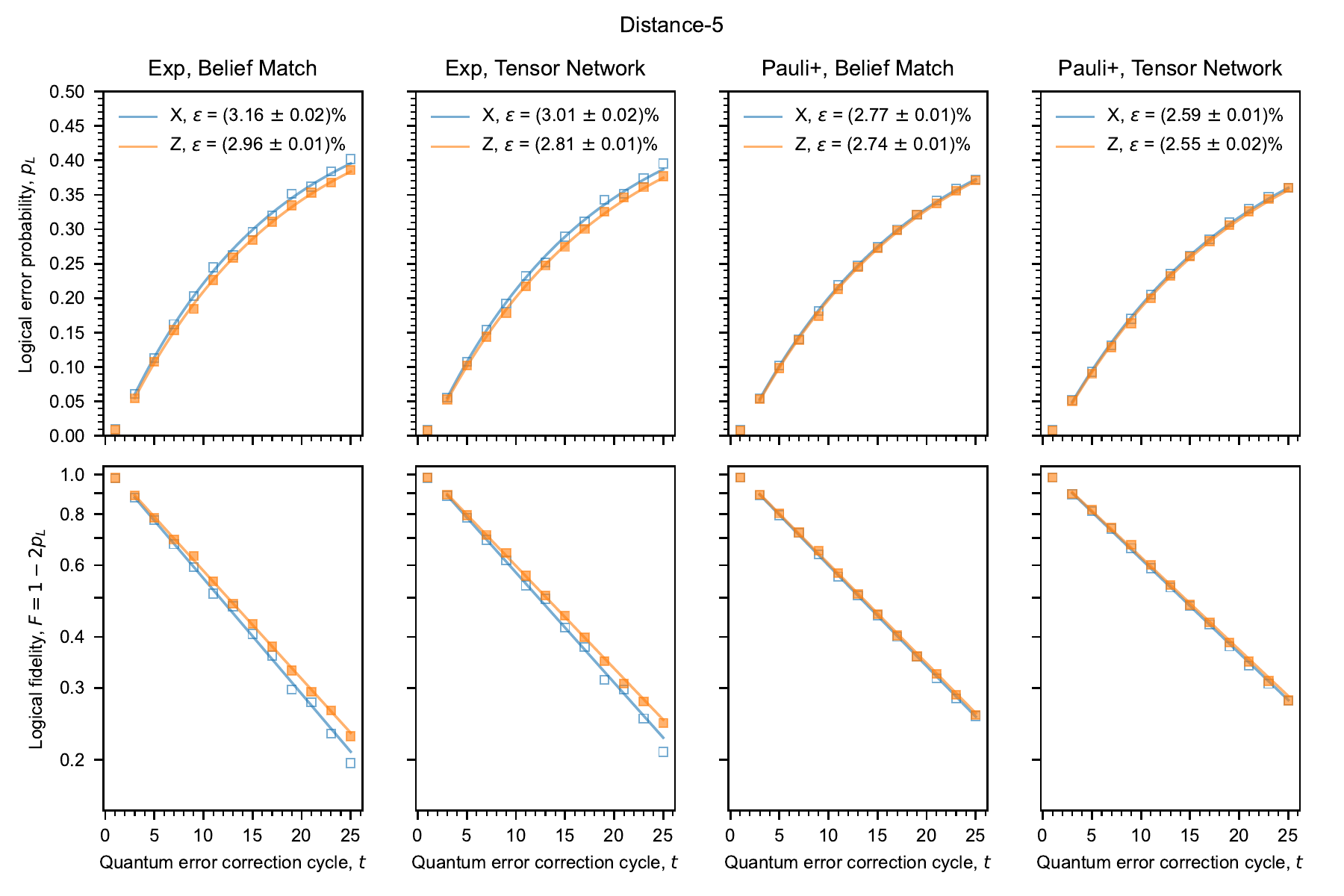}
    \caption{Logical error probabilities vs quantum error correction cycle for the distance-5 code for four scenarios, from left to right: experimental data with belief matching, experimental data with tensor network decoding, Pauli$+$ simulation with belief matching, and Pauli$+$ with tensor network decoding. Top row: logical errors plotted on normal scale. Bottom row: logical fidelities plotted on logscale.}
    \label{fig:d5-lers}
\end{figure*}

\begin{figure*}[htbp]
    \centering
    
    \includegraphics[width=7.2in]{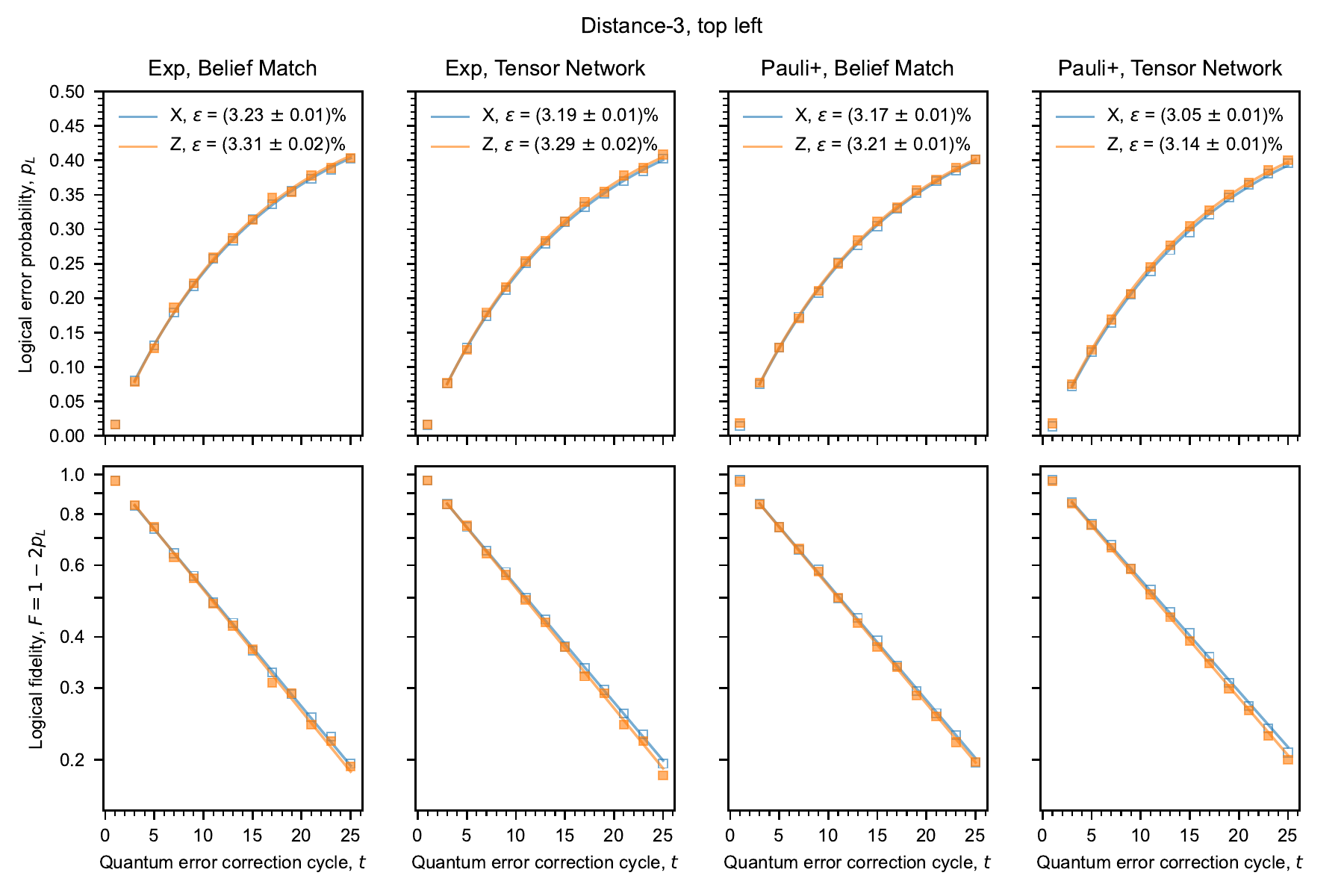}
    \caption{Logical error probabilities vs quantum error correction cycle for the top left distance-3 code for four scenarios, from left to right: experimental data with belief matching, experimental data with tensor network decoding, Pauli$+$ simulation with belief matching, and Pauli$+$ with tensor network decoding. Top row: logical errors plotted on normal scale. Bottom row: logical fidelities plotted on logscale.}
    \label{fig:d3-tl-lers}
\end{figure*}

\begin{figure*}[htbp]
    \centering
    
    \includegraphics[width=7.2in]{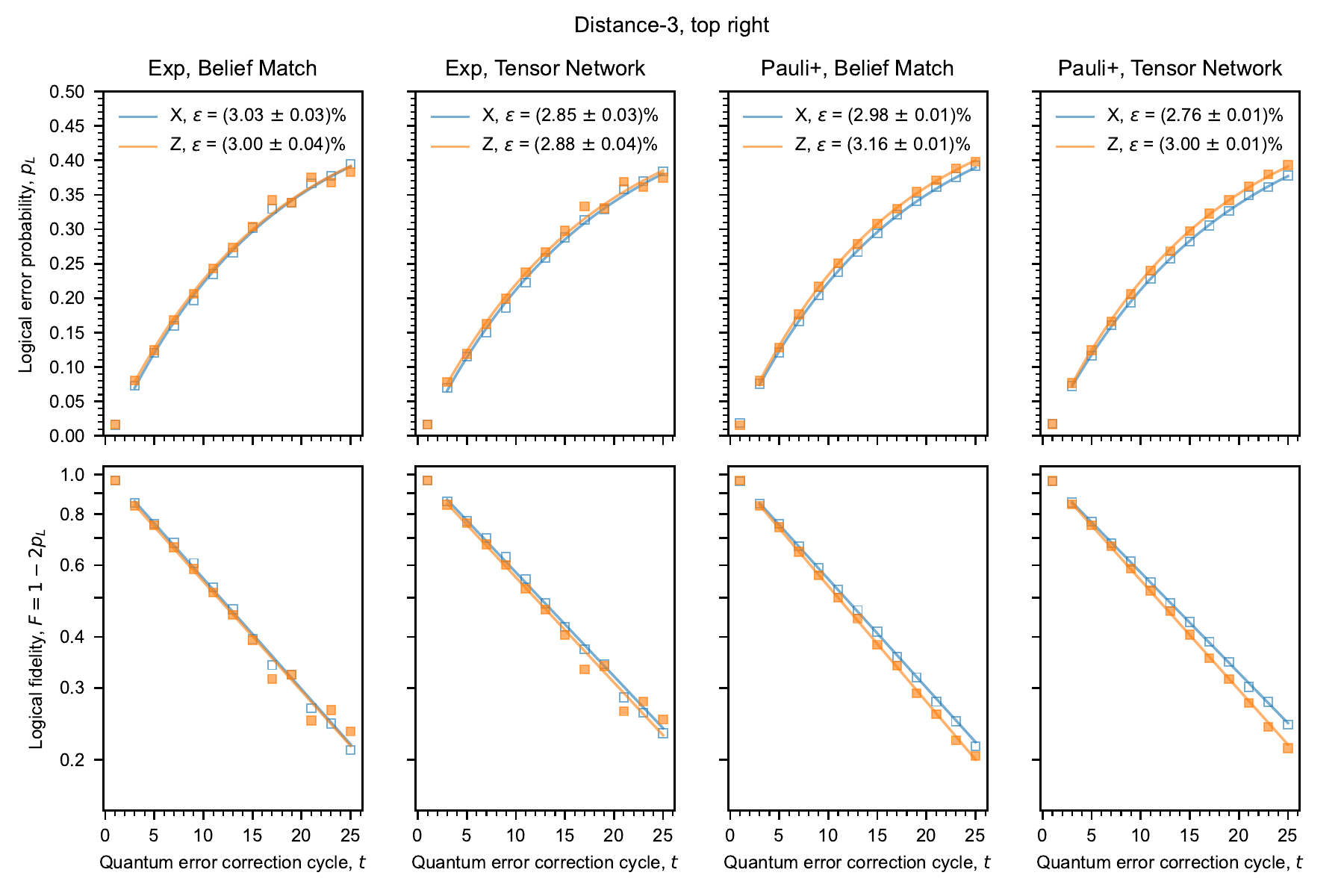}
    \caption{Logical error probabilities vs quantum error correction cycle for the top right distance-3 code for four scenarios, from left to right: experimental data with belief matching, experimental data with tensor network decoding, Pauli$+$ simulation with belief matching, and Pauli$+$ with tensor network decoding. Top row: logical errors plotted on normal scale. Bottom row: logical fidelities plotted on logscale.}
    \label{fig:d3-tr-lers}
\end{figure*}

\begin{figure*}[htbp]
    \centering
    
    \includegraphics[width=7.2in]{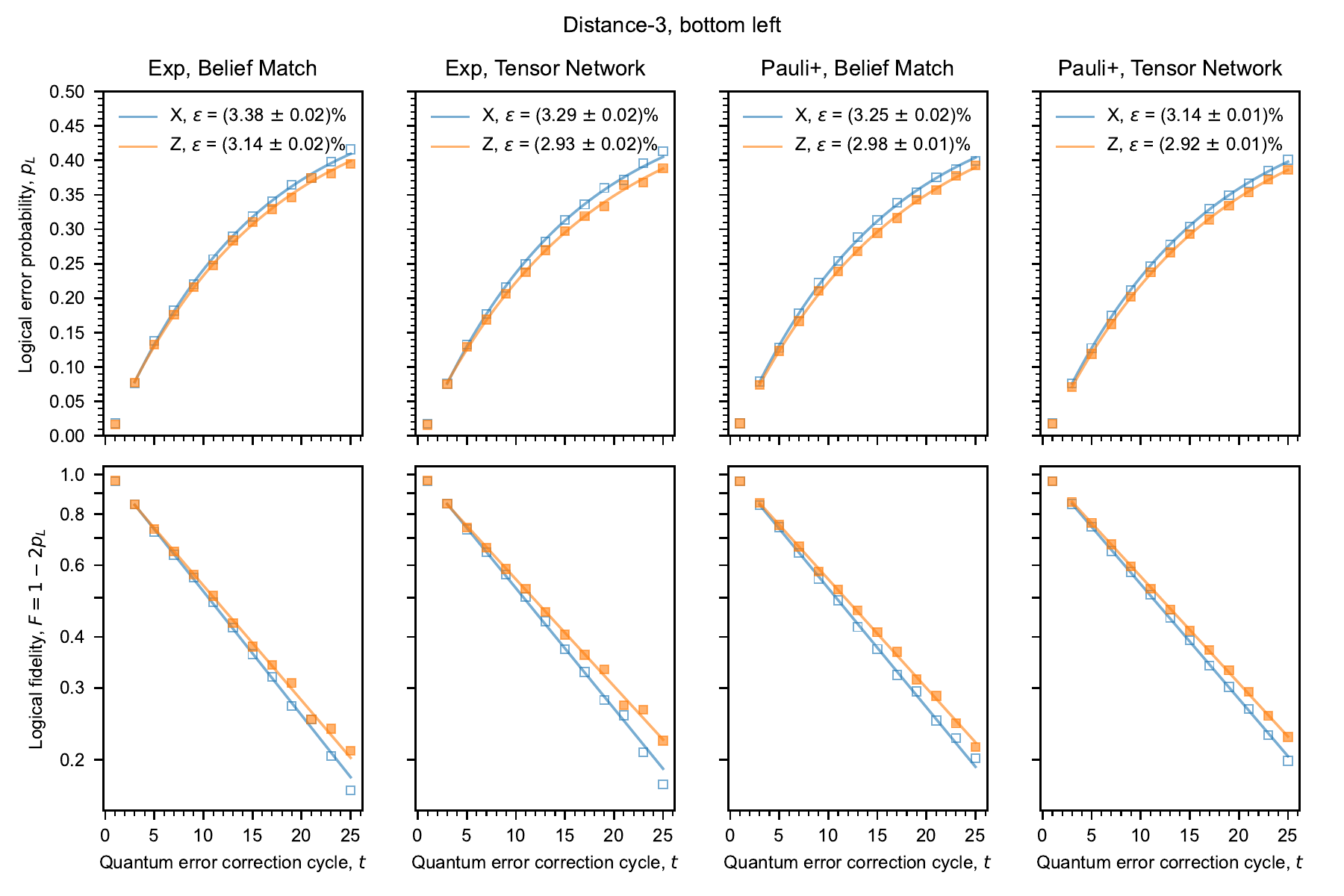}
    \caption{Logical error probabilities vs quantum error correction cycle for the bottom left distance-3 code for four scenarios, from left to right: experimental data with belief matching, experimental data with tensor network decoding, Pauli$+$ simulation with belief matching, and Pauli$+$ with tensor network decoding. Top row: logical errors plotted on normal scale. Bottom row: logical fidelities plotted on logscale.}
    \label{fig:d3-bl-lers}
\end{figure*}

\begin{figure*}[htbp]
    \centering
    
    \includegraphics[width=7.2in]{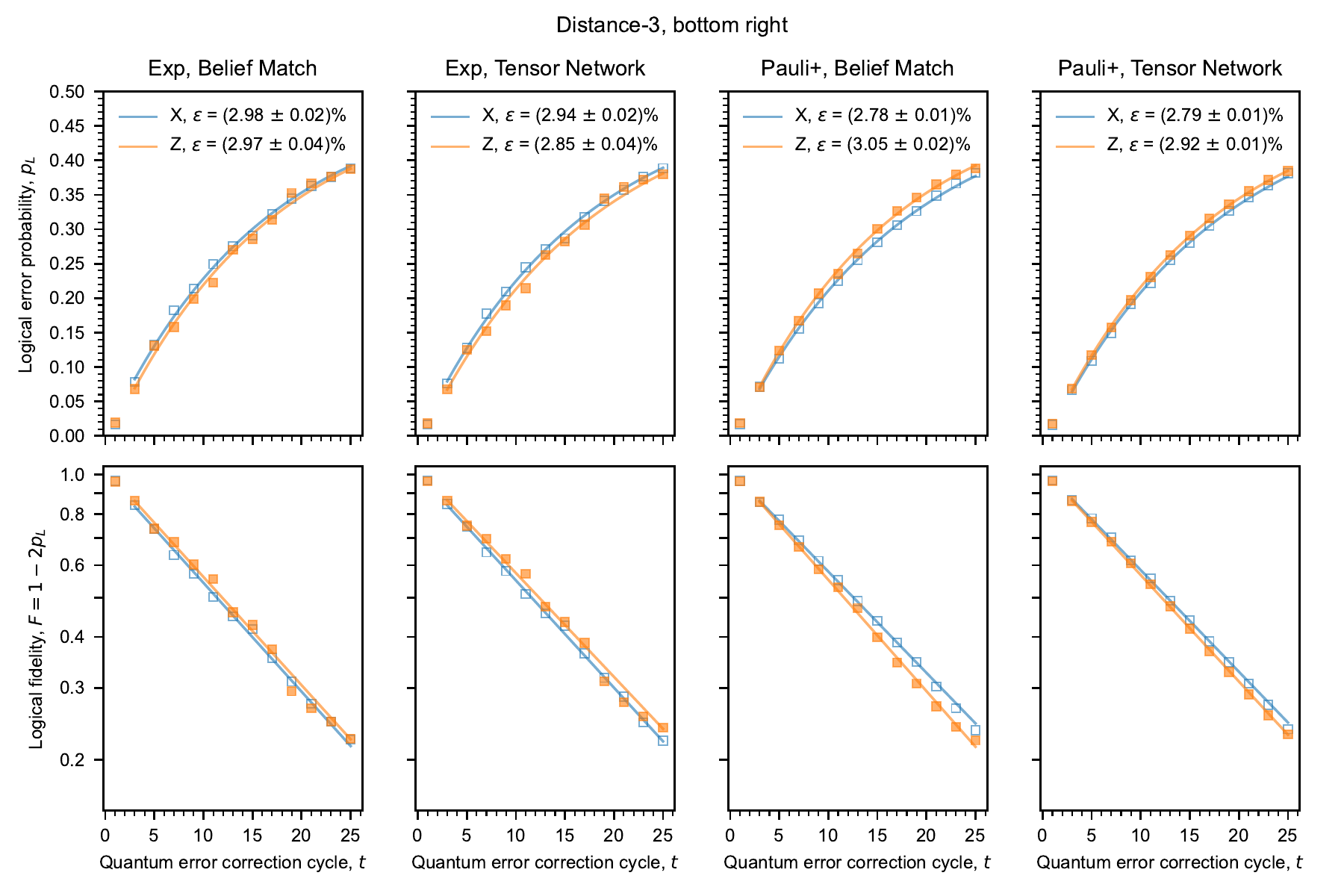}
    \caption{Logical error probabilities vs quantum error correction cycle for the bottom right distance-3 code for four scenarios, from left to right: experimental data with belief matching, experimental data with tensor network decoding, Pauli$+$ simulation with belief matching, and Pauli$+$ with tensor network decoding. Top row: logical errors plotted on normal scale. Bottom row: logical fidelities plotted on logscale.}
    \label{fig:d3-br-lers}
\end{figure*}

\clearpage

\end{document}